\documentclass{aa} 
\usepackage[dvips]{graphicx}
\usepackage{amsmath}
\usepackage{natbib}
\usepackage{txfonts}
\usepackage{color}

\begin{document}
   \title{Radiative hydrodynamics simulations of red supergiant stars. IV gray versus non-gray opacities}
 \titlerunning{Numerical simulations of red supergiant stars with CO$^5$BOLD}
 \authorrunning{Chiavassa et al.}
   \author{A. Chiavassa
          \inst{1}
          B. Freytag
          \inst{2,3,4}
          T. Masseron
          \inst{1}
          B. Plez
          \inst{5}
          }

   \institute{Institut d'Astronomie et d'Astrophysique, Universit\'e Libre de Bruxelles, CP. 226, Boulevard du Triomphe, B-1050 Bruxelles, Belgium\\
              \email{achiavas@ulb.ac.be}
         \and
             Centre de Recherche Astrophysique de Lyon,
          UMR 5574: CNRS, Universit\'e de Lyon,
          \'Ecole Normale Sup\'erieure de Lyon,
          46 all\'ee d'Italie, F-69364 Lyon Cedex 07, France
          \and
          Department of Physics and Astronomy,
          Division of Astronomy and Space Physics,
          Uppsala University,
          Box 515, S-751~ 20 Uppsala,
          Sweden
          \and
          Istituto Nazionale di Astrofisica, Osservatorio Astronomico
          di Capodimonte, Via Moiariello 16, I-80131 Naples, Italy
          \and
          LUPM, Laboratoire Univers et Particules, Universit\'{e} de Montpellier II, CNRS, Place Eug\'{e}ne Bataillon
         	 34095 Montpellier Cedex 05, France
                         }

   \date{Received ; accepted }

  \abstract
   {Red supergiants are massive evolved stars that contribute extensively to the chemical enrichment of our Galaxy. It has been shown that convection in those stars gives rise to
large granules that cause surface inhomogeneities and shock waves in the photosphere. The understanding of their dynamics is crucial to unveil the unknown mass-loss mechanism, their chemical composition and stellar parameters.}
   {We present a new generation of red supergiants simulations with a more sophisticated opacity treatment done with 3D radiative-hydrodynamics CO5BOLD.}
   {In the code, the coupled
equations of compressible hydrodynamics and non-local radiation transport are solved in the presence of a spherical potential.
The stellar core is replaced by a special spherical inner boundary condition, where the gravitational potential is smoothed
and the energy production by fusion is mimicked by a simply producing heat corresponding to the stellar luminosity. All outer
boundaries are transmitting for matter and light. The
   post-processing radiative transfer code OPTIM3D is used to extract spectroscopic and interferometric observables.}
   {We show that the relaxation of the assumption of frequency-independent opacities shows a steeper mean thermal gradient in the optical thin region that affect strongly the atomic strengths and the spectral energy distribution. Moreover, the weaker temperature fluctuations reduce the incertitude on the radius determination with interferometry. We show that 1D models of red supergiants must include a turbulent velocity calibrated on 3D simulations to obtain the effective surface gravity that mimic the effect of turbulent pressure on the stellar atmosphere. We provide an empirical calibration of the ad-hoc micro- and macroturbulence parameters for 1D models using the 3D simulations: we find that there is not a clear distinction between the different macroturbulent profiles needed in 1D models to fit 3D synthetic lines.}
   {}

\keywords{stars: atmospheres --
 		stars: supergiants --
                hydrodynamics --
                radiative transfer --
                Methods: numerical
               }
               
  \maketitle

%

\section{Introduction}

The dynamical nature of the solar-surface layers, manifested
for instance in granules and sunspots, has been known for a
long time. With every improvement of ground-based or spaceborne
instruments the complexity of the observed processes increased.
Red supergiant (RSG) stars are among the largest stars in the universe and the brightest in the optical and
near infrared. They are massive stars with masses between roughly 10 and 25~$M_{\odot}$ with effective temperatures ranging from
     3\,450 to 4\,100\,K, luminosities of 20\,000 to 300\,000\,$L_\odot$,
     and radii up to 1\,500\,$R_\odot$ \citep{2005ApJ...628..973L}. These stars exhibit variations in integrated brightness,
surface features, and the depths, shapes, and Doppler shifts of
spectral lines; as a consequence, stellar parameters and abundances are difficult to determine. Progress has been done using 1D hydrostatic models revising the $T_{\rm eff}$-scale and reddening both at solar and Magellanic Clouds metallicities \citep{2005ApJ...628..973L,2006ApJ...645.1102L,2007ApJ...660..301M,2007ApJ...667..202L,2010NewAR..54....1L} but problems still remain, e.g. the blue-UV excess that may be due to scattering by circumstellar dust or to an insufficiency in the models, and the visual-infrared effective temperature mismatch \citep{2006ApJ...645.1102L}. Finally, RSGs eject massive amounts of mass back to the interstellar medium with an unidentified process that may be related to acoustic waves and radiation pressure on molecules \citep{2007A&A...469..671J}, or to the dissipation of Alfv\'en waves from magnetic field, recently discovered on RSGs \citep{2010A&A...516L...2A,2010MNRAS.408.2290G}, as early suggested by \cite{1984ApJ...284..238H, 1989A&A...209..198P, 1997A&A...325..709C}.

The dynamical convective pattern of RSGs is then crucial for the understanding of the physics of these stars that contribute extensively to the chemical enrichment of the Galaxy. There is a number of multiwavelength imaging examples of RSGs (e.g. $\alpha$~Ori) because of their high luminosity and large angular diameter. Concerning $\alpha$~Ori, \cite{1990MNRAS.245P...7B, 1992MNRAS.257..369W, 1997MNRAS.285..529T,
  1997MNRAS.291..819W, 2000MNRAS.315..635Y, 2004young} detected the presence of time-variable inhomogeneities on his surface with WHT and COAST; \cite{2009A&A...508..923H} reported a reconstructed image in the H band with two large spots; \cite{2009A&A...503..183O,2011A&A...529A.163O} attributed motions detected by VLTI/AMBER observations to convection; \cite{2009A&A...504..115K} resolved $\alpha$~Ori using diffraction-limited adaptive optics in the near-infrared and found an asymmetric envelope around the star with a
bright plume extending in the southwestern region. \cite{1997MNRAS.285..529T} the presence of bright spots on the surface of the supergiants $\alpha$~Her and $\alpha$~SCO using WHT and \cite{2010A&A...511A..51C} on VX~Sgr using VLTI/AMBER.

The effects of convection and non-radial waves can be represented
by numerical multi-dimensional time-dependent radiation
hydrodynamics (RHD) simulations with realistic input
physics. Three-dimensional radiative hydrodynamics simulations are no longer restricted to the Sun \citep[for a review on the Sun models see][]{2009LRSP....6....2N} but cover a substantial portion of the Hertzsprung-Russell diagram \citep{2009MmSAI..80..711L}.  Moreover, they have been already extensively employed to study the effects of
photospheric inhomogeneities and velocity fields on the formation of
spectral lines and on interferometric observables in a number of cases, including the Sun, dwarfs and
subgiants \citep[e.g. ][]{1999A&A...346L..17A, 2001A&A...372..601A,
 2009ARA&A..47..481A,2010SoPh..tmp...66C,2010A&A...513A..72B,2010A&A...522A..26S}, red giants \citep[e.g.][]{2010A&A...524A..93C,2007A&A...469..687C,
 2009MmSAI..80..719C,2009A&A...508.1429W}, asymptotic giant branch stars \citep{2010A&A...511A..51C}, and red supergiant stars \citep{2009A&A...506.1351C, 2010A&A...515A..12C,2011A&A...528A.120C}. In particular, the presence and the characterization of the size of convective cells on $\alpha$~Ori has been showed by \cite{2010A&A...515A..12C} by comparing a large set of interferometric observations ranging from the optical to the infrared wavelengths.
 
This paper is the fourth in a series aimed to present the numerical simulations of red supergiant stars with CO$^5$BOLD and to introduce the new generation of RSG simulations with a more sophisticated opacity treatment. Such simulations are of great importance for an accurate quantitative analysis of observed data. 

 \begin{figure*}
   \centering
   \includegraphics[width=1.\textwidth]{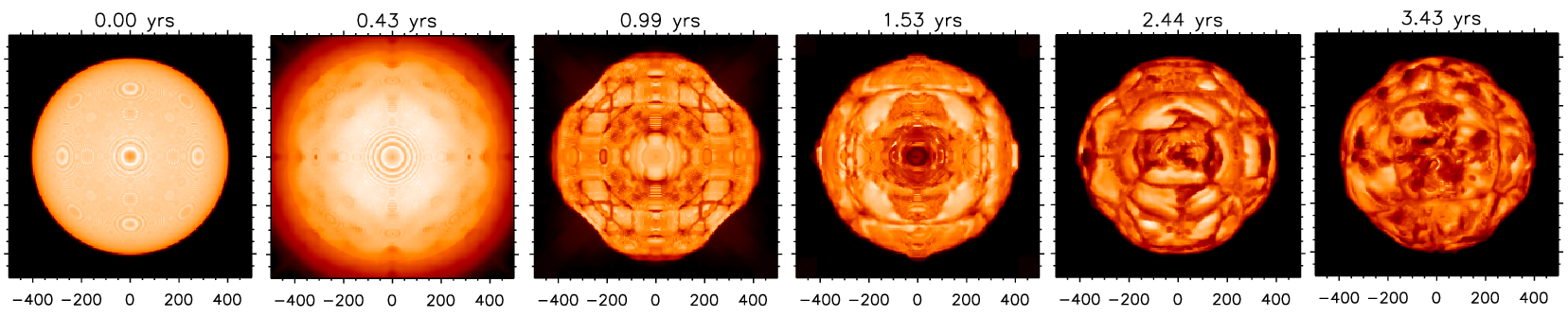}
   \caption{Gray intensity on one side of the computational cube from the initial sequence of the model st35gm00n05 in Table~\ref{simus}. The axes are in solar radii. The artifacts due to the mismatch between the spherical object and the Cartesian grid become less evident with time passing.}
              \label{figstarting}%
    \end{figure*}
    
\section{3D numerical simulations with CO$^5$BOLD}

\subsection{Basic equations}

The code solves the coupled equations of compressible hydrodynamics
and non-local radiation transport

\begin{eqnarray}
 \frac{\partial\rho}{\partial t}  + \nabla\cdot\left(\rho \vec{v}\right)      =   	 0    \\
\frac{\partial\rho\vec{v}}{\partial t}+\rho\left(\vec{v}\cdot\nabla\right)\vec{v}+\nabla P     	= 	-\rho \vec{\nabla}\Phi  \\
\frac{\partial\rho e_{\mathit{ikg}}}{\partial t}+\nabla\cdot\left(\left[\rho e_{\mathit{ikg}}+P\right]\vec{v}+F\mathbf{_{\rm{rad}}}\right)          =           0 
\end{eqnarray}

They describe the inviscid flow of density $\rho$, momentum $\rho \mathbf{v}$,
and total energy $e_{\mathit{ikg}}$ (including internal, kinetic, and gravitational potential
energy). Further quantities are the velocity vector $\mathbf{v}$, pressure
$P$, and radiative energy flux $F_\mathrm{_{rad}}$. The latter is computed from the frequency-integrated intensity $I$ for the gray treatment of opacity and for each wavelength group in the non-gray approach (Sect.~\ref{radsection}).

The gravitational potential is spherical,
\begin{equation}
\Phi=-GM_{\mathrm{pot}}\left(r_0^4+r^4/\sqrt{1+\left(r/r_1\right)}\right)^{-1/4},\label{eqgrav}
\end{equation}
where $M_{\mathrm{pot}}$ is the mass of the star to be modeled (see Fig.~\ref{1dquantities}, bottom right panel); $r_0$ and $r_1$ are free smoothing parameters: when $r_0^4 \ll r^4 \ll r_1^4$, $\Phi=-\frac{GM_{\mathrm{pot}}}{r}$; while when $r\rightarrow 0$, $\Phi\rightarrow \Phi=-\frac{GM_{\mathrm{pot}}}{r_0}$; and for $r\rightarrow\infty$, $\Phi\rightarrow \Phi=-\frac{GM_{\mathrm{pot}}}{r_1}$. Typically, $r_0\approx0.2$ $R_\star$ and $r_1\approx1.2$ $R_\star$ for RSG simulations \citep{2002AN....323..213F}.

\subsection{The code}

The numerical simulations described here are performed with ÒCO$^5$BOLDÓ (ÒCOnservative COde for the COmputation of COmpressible COnvection in a BOx of $L$ Dimensions, $L=2,3$Ó). It uses operator splitting \citep{1968SJNA....5..506S} to separate the various (explicit) operators: the hydrodynamics, the optional tensor viscosity, and the radiation transport.

The hydrodynamics module is based on a finite volume approach
and relies on directional splitting to reduce the 2D or
3D problem to one dimension. In the 1D steps an approximate
Riemann solver of Roe-type \citep{1986AnRFM..18..337R} is applied, modified
to account for a realistic equation of state, a non-equidistant
Cartesian grid, and the presence of source terms due to an external
gravity field. In addition to the stabilizing mechanism
inherent in an upwind-scheme with a monotonic reconstruction
method (typically a piecewise-linear van Leer interpolation), a
2D or 3D tensor viscosity can be activated. This step eliminates
certain errors of Godunov-type methods dealing with strong
velocity fields aligned with the grid \citep{1994IJNMF..18..555Q}.

The equation of state uses pre-tabulated values as functions
of density and internal energy $\left(\rho,e_i\rightarrow P,\Gamma_1,T,s\right)$. It accounts for H{\sc{I}}, H{\sc{II}}, H$_{2}$, He{\sc{I}}, He{\sc{II}}, He{\sc{III}} and a representative metal for any prescribed chemical composition. The equation of state
does not account for the ionization states of metals, but it uses
only one neutral element to achieve the appropriate atomic weight
(in the neutral case) for a given composition.
Two different geometries can be used with CO$^{5}$BOLD, that are characterized by different gravitational potentials,
boundary conditions, and modules for the radiation transport:
\begin{itemize}
\item The \textsc{box-in-a-star} setup is used to model a statistically representative volume of the stellar atmosphere with a constant gravitation, where the lateral boundaries are periodic,
and the radiation transport module relies on a Feautrier
scheme applied to a system of long rays \citep[\cite{2002AN....323..213F}, ][ and for analysis, e.g, \citeauthor{2010SoPh..tmp...66C}  \citeyear{2010SoPh..tmp...66C}]{2004A&A...414.1121W}.
\item The \textsc{star-in-a-box} setup is used to model RSG stars of this work. The computational domain is a
cubic grid equidistant in all directions, and the same
open boundary condition is employed for all side of the computational box.
\end{itemize}

As the outer boundaries are usually either hit at some angle by an outgoing
shockwave, or let material fall back (mostly with supersonic velocities),
there is not much point in tuning the formulation for
an optimum transmission of small-amplitude waves. Instead, a
simple and stable prescription, that lets the shocks pass, is sufficient.
It is implemented by filling typically two layers of ghost cells where the velocity components and the internal energy
are kept constant. The density is assumed to decrease exponentially
in the ghost layers, with a scale height set to a controllable
fraction of the local hydrostatic pressure scale height. The control
parameter allows to account for the fact that the turbulent
pressure plays a significant role for the average pressure stratification. The acceleration
due to gravity is derived from Eq.~\ref{eqgrav}. Within a radius $r_0$ the potential is smoothed (Eq.~\ref{eqgrav}). In this sphere a source term to the internal energy
provides the stellar luminosity. Motions in the core are damped
by a drag force to suppress dipolar oscillations. The hydrodynamics and the radiation transport scheme ignore
the core completely and integrate right through it.

The code is parallelized with Open Multi-Processing (OpenMP) directives.

\subsection{Radiation transport and opacities tables}\label{radsection}

The radiation transport step for RSGs' simulations uses a short-characteristics
method. To account for the short radiative time scale several
(typically 6 to 9) radiative sub$-$steps are performed per
global step. Each sub-step uses only 3 rays (e.g. in the directions
{(1,1,0), (1,$-$1,0), (0,0,1)} or {(1,0,1), (1,0,$-$1), (0,1,0)} or
{(0,1,1), (0,1,$-$1), (1,0,0)} or 4 rays (along the space diagonals
{($\pm$1,$\pm$1,$\pm$1)}). The different orientation sets are used cyclically.
The radiation transport is solved for a given
direction for both orientations. More
Òirregular directionsÓ (and more rays) are possible but avoided
in the models presented here to save computational time. The
radiation transport operator is constructed to be stable in the
presence of strong jumps in opacity and temperature (even on
a coarse grid).

The frequency dependance of the radiation field in the CO$^5$BOLD models can be calculated with two approaches:
\begin{itemize}
\item  the gray approximation, that completely ignores the frequency dependence, is justified only
in the stellar interior and it is inaccurate in the optically
thin layers. The Rosseland mean opacities are calculated as a function of pressure and temperature ($T,P\rightarrow \kappa_{\mathrm{Ross}}$) and available in a 2D table. The necessary values are found by interpolation in
a 2D table. It has been merged at around 12 000K from
high-temperature OPAL \citep{1992ApJ...397..717I} data and low-temperature
PHOENIX \citep{1997ApJ...483..390H} data by Hans-G\"unter Ludwig. 
The models which use theses opacities are reported in Table~\ref{simus}.
\item the more elaborate scheme accounting for non-gray effects which is based on the idea of
\emph{opacity binning} \citep{1982A&A...107....1N,1990A&A...228..155N}. The basic approximation
is the so called multi-group scheme \citep{1994A&A...284..105L,2004A&A...421..741V}. In this scheme, the frequencies which reach monochromatic optical
depth unity within a certain depth range of the model atmosphere will
be put into one frequency group. The opacity table used in this work is sorted in 5 wavelengths groups accordingly to the run of the monochromatic optical depth in
a corresponding MARCS \citep{2008A&A...486..951G} 1D model. The corresponding logarithmic Rosseland optical depth are $+\infty, 0.0, -1.5, -3.0, -4.5, -\infty$. In each group
there is a smooth transition from a Rosseland average in the optically
thick regime to a Planck average in the optically thin regime, except
for the group representing the largest opacities, where the
Rosseland average is used throughout. The implementation of non-gray opacities tables has been carried out by Hans-G\"unter Ludwig and we computed one non-gray model (Table~\ref{simus}).
\end{itemize}

\section{Modelling RSG stars}

The most important parameters that determine the type of the modeled star are: 
\begin{itemize}
\item the input luminosity into the core
\item the stellar mass that enters in Eq.~\ref{eqgrav} for the gravitational potential
\item the abundances that are used to create the
tables for the equation-of-state and the opacities
\end{itemize}
In addition, there are a number of parameters that influence the outcome
of a simulation to some extent, i.e. the detailed formulation
of the boundary conditions, the smoothing parameters of the
potential, the numerical resolution, detailed settings of the hydrodynamics
scheme and the additional tensor viscosity, choice
of ray directions for the radiation transport, and the time-step
control. The model presented in \cite{2002AN....323..213F} relies on the same assumptions
as the current ones. In the meantime, modifications of the code
(e.g. restrictions to rays only along axes and diagonals avoiding
some interpolations in the radiation transport module) and
faster computers allow models with higher resolution,
frequency-dependent opacity tables, and various stellar parameters. These new models show a significant increase in the number of small convective cells on the stellar
surface. 

\begin{table*}
\centering
\begin{minipage}[t]{\textwidth}
\caption{Simulations of red supergiant stars used in this work. }             
\label{simus}      
\renewcommand{\footnoterule}{} 
\begin{tabular}{c c c c c c c c c c c}        
\hline\hline                 
model & Simulated & grid &  grid & $M_{\mathrm{pot}}$ & $M_{\mathrm{env}}$ & $L$ &  $T_{\rm{eff}}$ & $R_{\star}$ &  $\log g$ & comment \\
            & $\!\!\!$relaxed$\!$ &  $\!\!\!$[grid points]$\!$ & $\!\!\!$[$R_\odot$]$\!$ & $\!\!\!$[$M_\odot$]$\!$ & $\!\!\!$[$M_\odot$]$\!$ & $\!\!\!$[$L_\odot$]$\!$ & $\!\!\!$$[\rm{K}]$$\!$ & $\!\!\!$[$R_\odot$]$\!$ &  $\!\!\!$[cgs]$\!$  & \\
	  & $\!$time [years]$\!$ &&&&&&&&& \\
\hline
st35gm03n07\footnote{Used and deeply analyzed in \cite{2009A&A...506.1351C,2010A&A...515A..12C}. Note that a sequence of 1.5 stellar years has been added with respect to the paper.}& 5.0 &235$^3$ & 8.6 & 12 & 3.0 & 91\ 932$\pm$1400 & 3487$\pm$12 & 830.0$\pm$2.0 & $-$0.335$\pm$0.002 & gray \\
st35gm03n13& 7.0 & 235$^3$ & 8.6 & 12 & 3.0 & 89\ 477$\pm$857 & 3430$\pm$8 & 846.0$\pm$1.1 & $-$0.354$\pm$0.001 & non-gray \\
   st36gm00n04 &    6.4   & 255$^3$ & 3.9 & 6 & 0.4 & 24\ 211$\pm$369& 3663$\pm$14& 386.2$\pm$0.4& 0.023$\pm$0.001& gray \\
     st36gm00n05 &      3.0        & 401$^3$ & 2.5 & 6 & 0.4 & 24\ 233$\pm$535& 3710$\pm$20& 376.7$\pm$0.5& 0.047$\pm$0.001& gray,  \\
    & & & & & & & & &  & highest\\
	& & & & & & & & &  & resolution\\
\hline\hline                          
\end{tabular}
\end{minipage}
\tablefoot{All models have solar chemical composition. The table reports: the simulated time used to average the stellar parameters; the numerical resolution in grid points and $R_\odot$; the mass used to compute the potential in Eq.~\ref{eqgrav}; total mass actually included in the model (the rest is assumed to sit
in the unresolved stellar core); the luminosity, effective temperature, radius and surface gravity averaged over both
spherical shells and simulated time (2$^{\mathrm{nd}}$ column), the errors are
one standard deviation fluctuations with respect to the average
over time; comment.}
\end{table*}

The initial model is produced starting from a sphere in hydrostatic equilibrium with a small velocity field inherited from a previous model with different stellar parameters (top left panel of Fig.~\ref{figstarting}). The temperature
in the photosphere follows a gray $T\left(\tau\right)$ relation. In the
optically thick layers it is chosen to be adiabatic. The first frame of Fig.~\ref{figstarting}, taken just after one snapshot, displays the limb-darkened surface without any convective signature but with some regular patterns due to the
coarse numerical grid and the correspondingly poor sampling
of the sharp temperature jump at the bottom of the photosphere. The central spot, quite evident at the beginning of the simulation, vanishes completely when convection becomes strong. A regular pattern of small-scale convection cells develops
initially and then, as the cells merge, the average structure becomes big and the regularity
is lost. The intensity contrast grows. After several years, the state is relaxed and the
pattern becomes completely irregular. All memory from the
initial symmetry is lost. The influence of the cubical box and the Cartesian grid is
strongest in the initial phase onset of convection. At later
stages, there is no alignment of structures with the grid nor a tendency
of the shape of the star to become cubical.

\begin{figure*}
   \centering
   \begin{tabular}{cccc}
   \includegraphics[width=0.25\hsize]{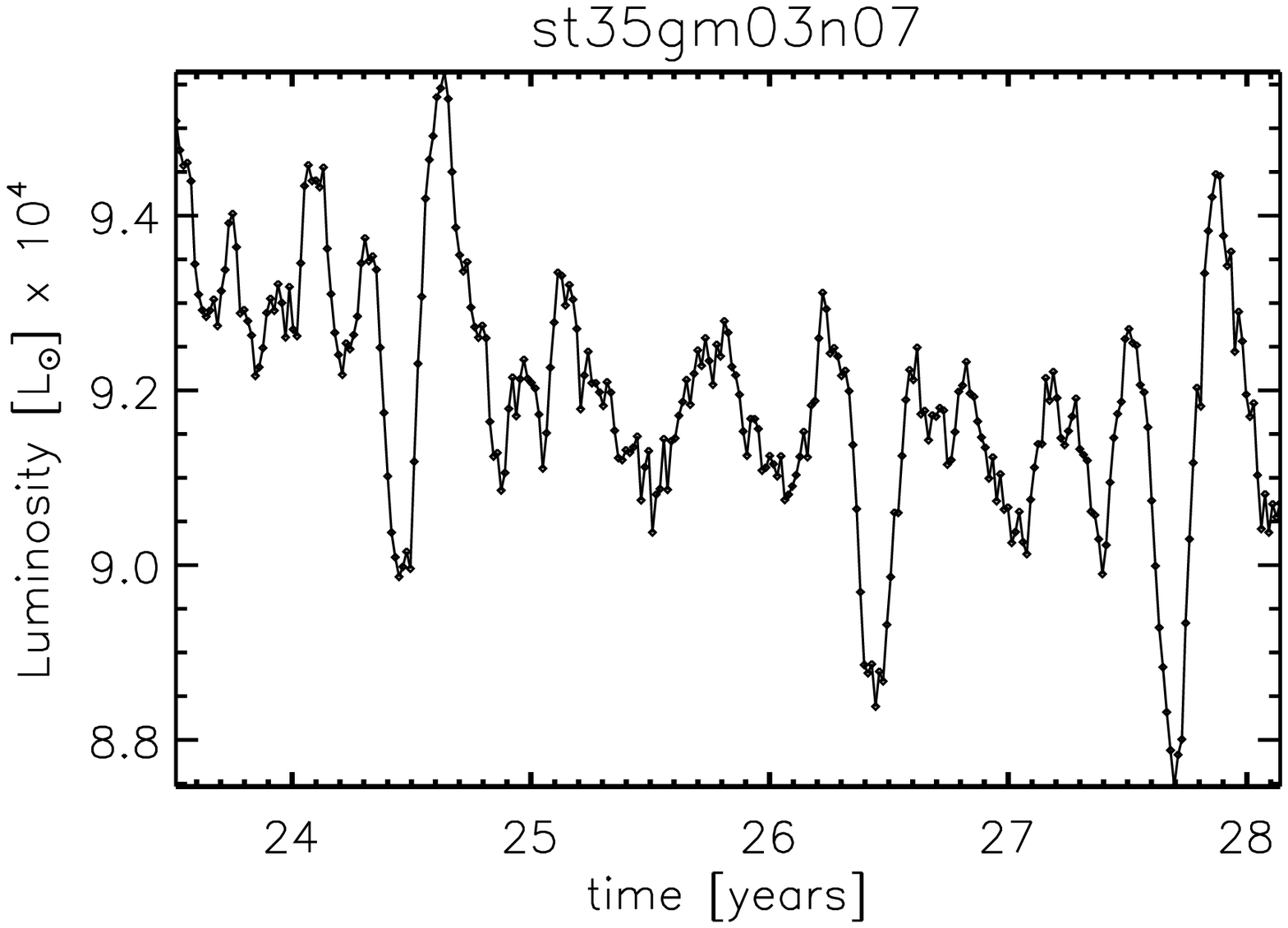}
       \includegraphics[width=0.25\hsize]{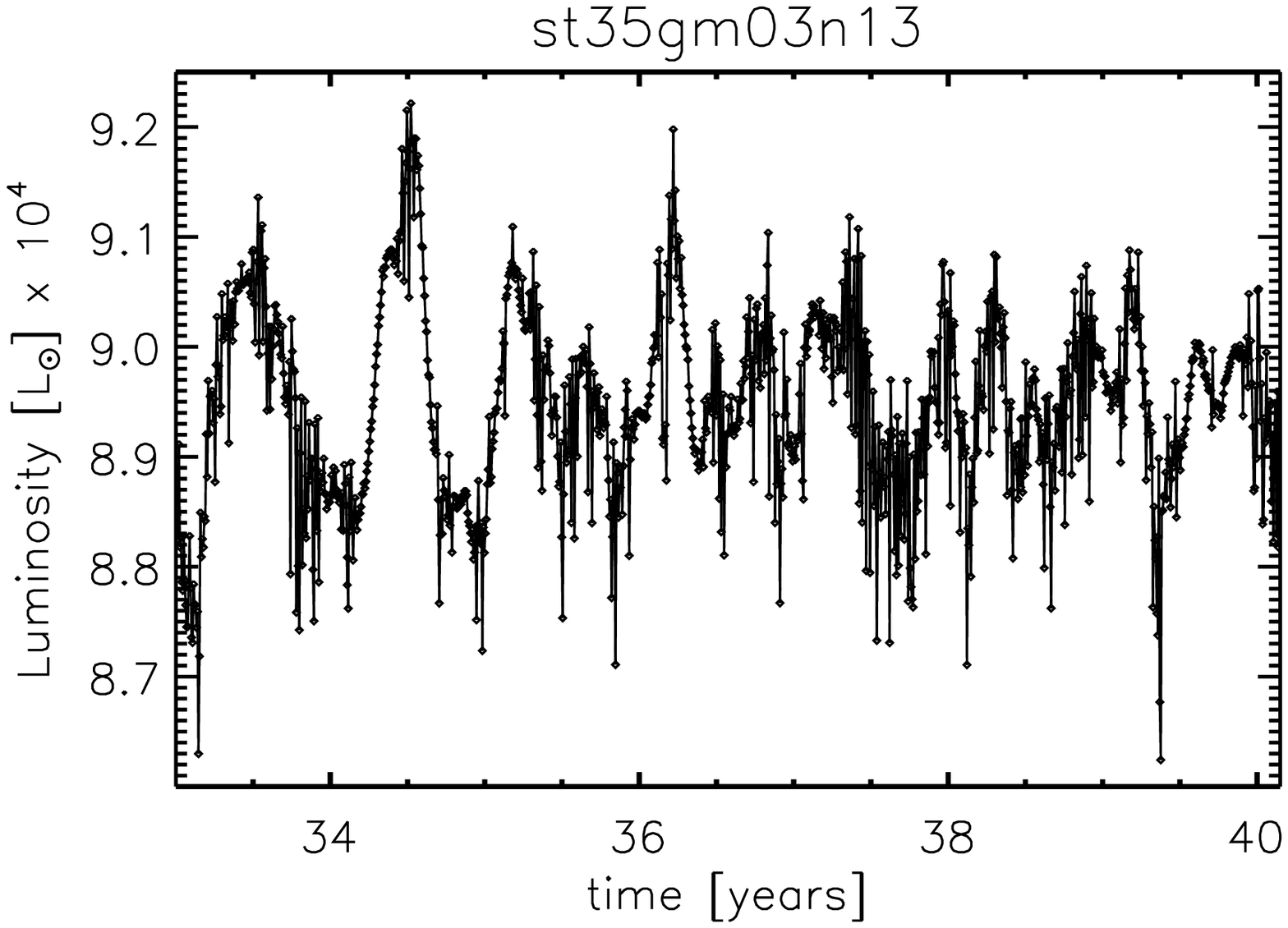}
           \includegraphics[width=0.25\hsize]{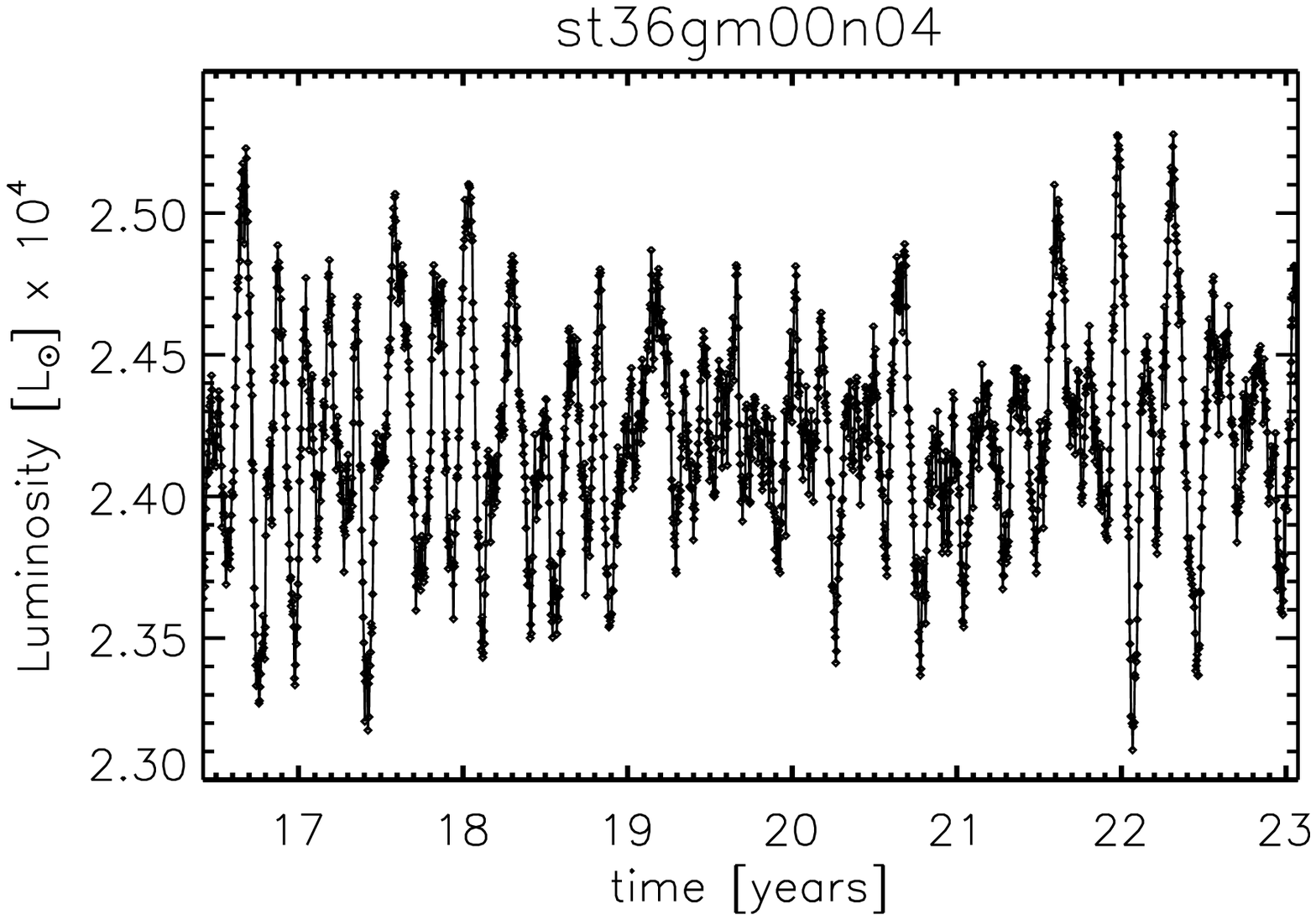}
               \includegraphics[width=0.25\hsize]{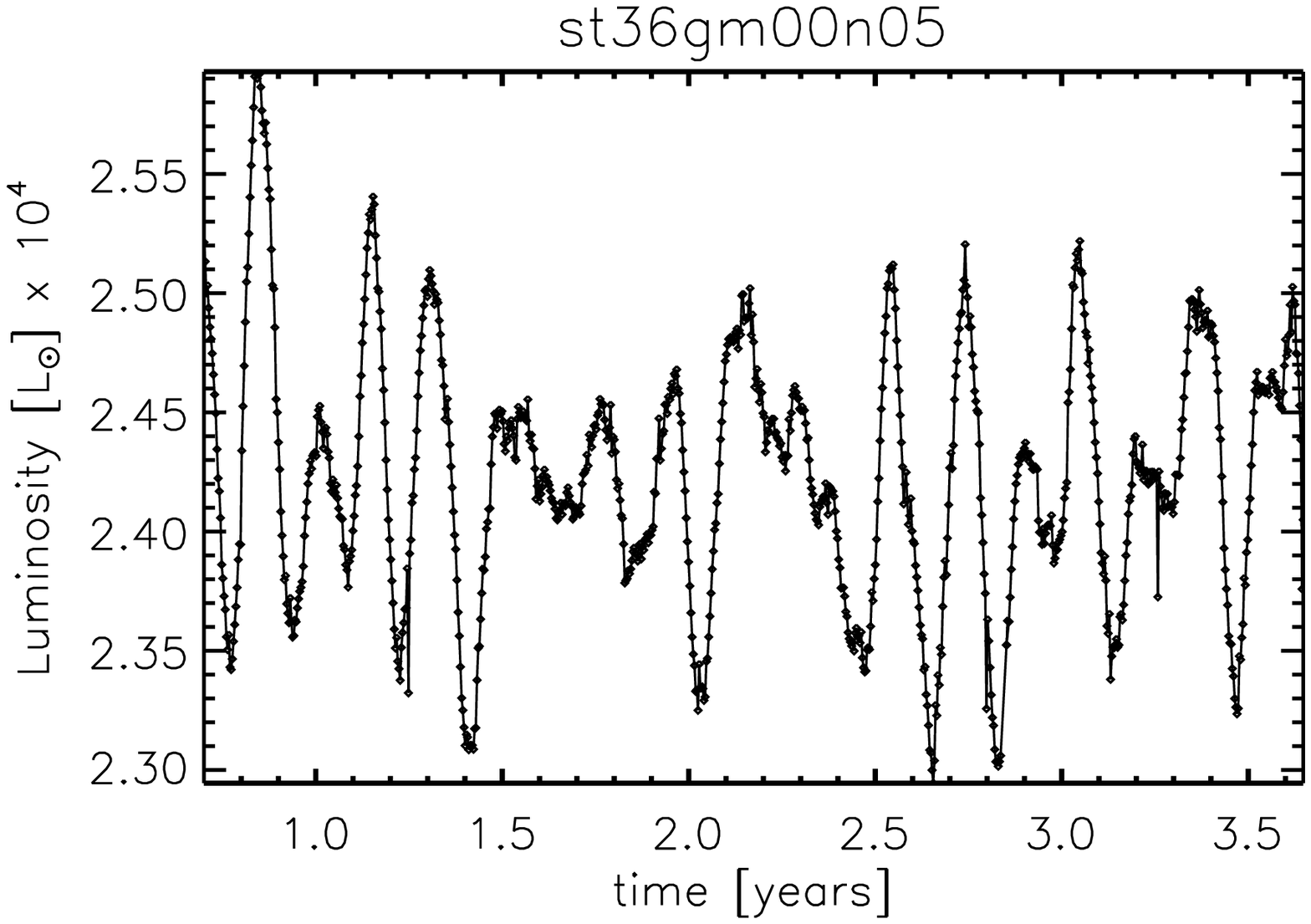}\\
    \includegraphics[width=0.25\hsize]{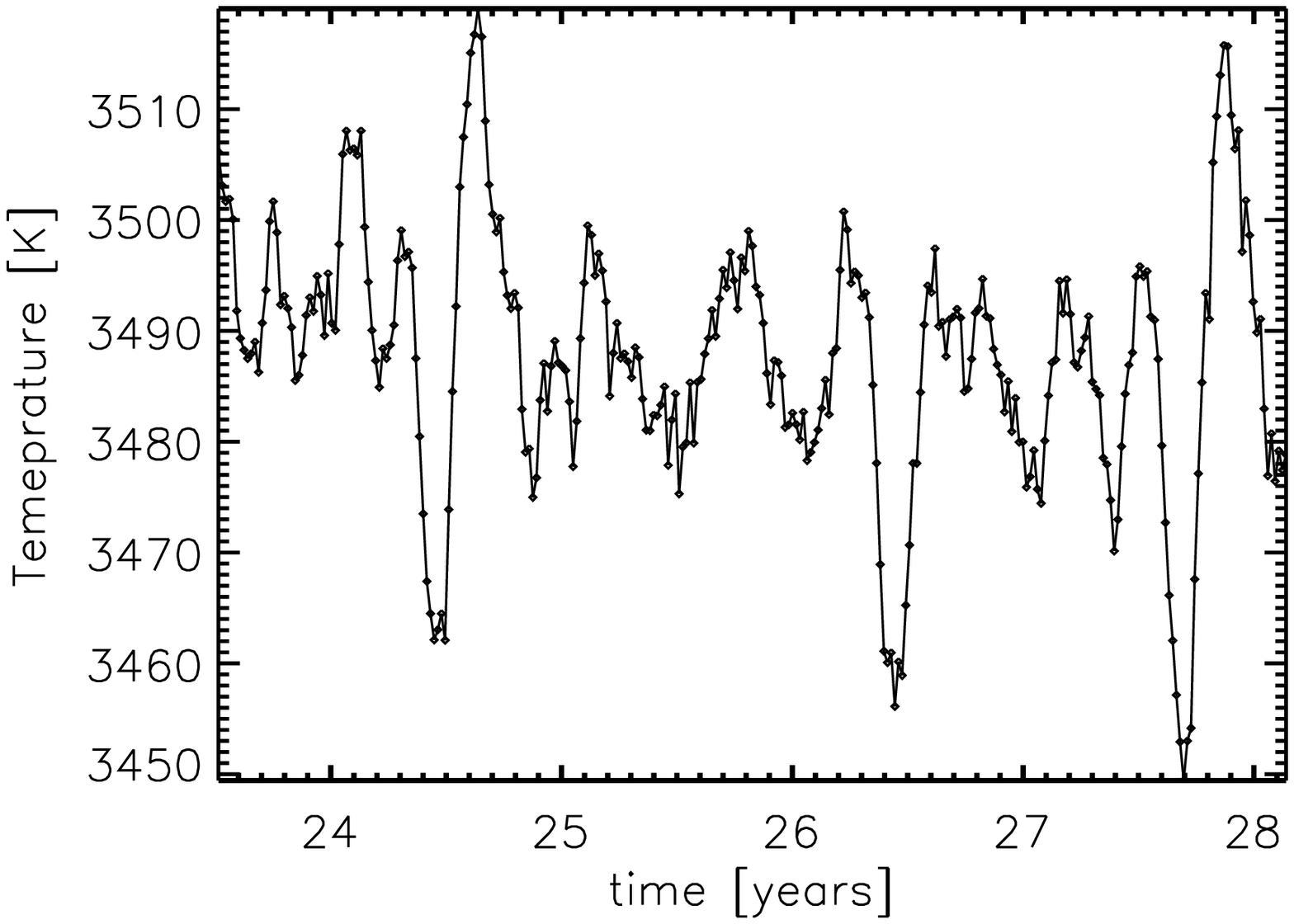}
       \includegraphics[width=0.25\hsize]{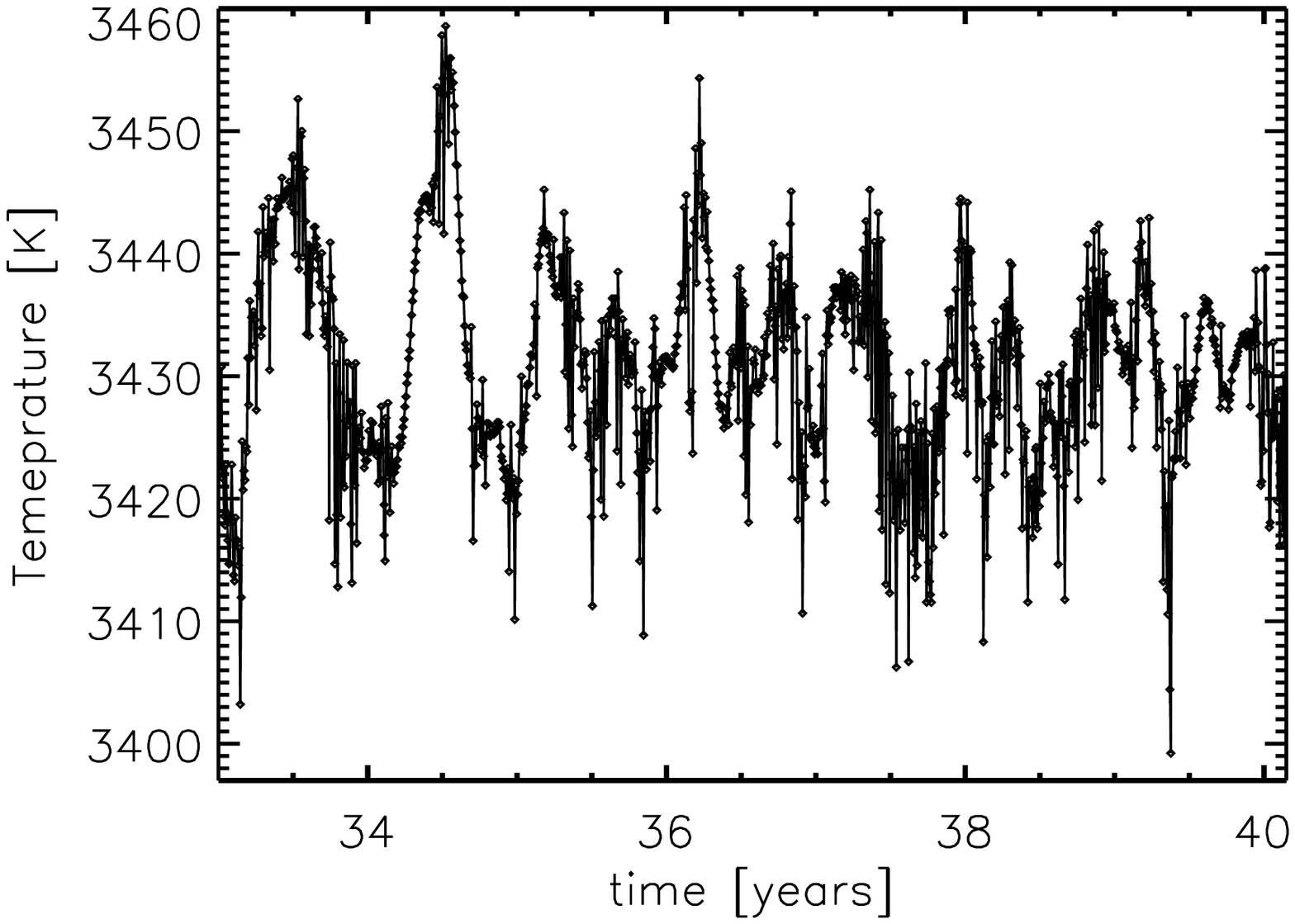}
           \includegraphics[width=0.25\hsize]{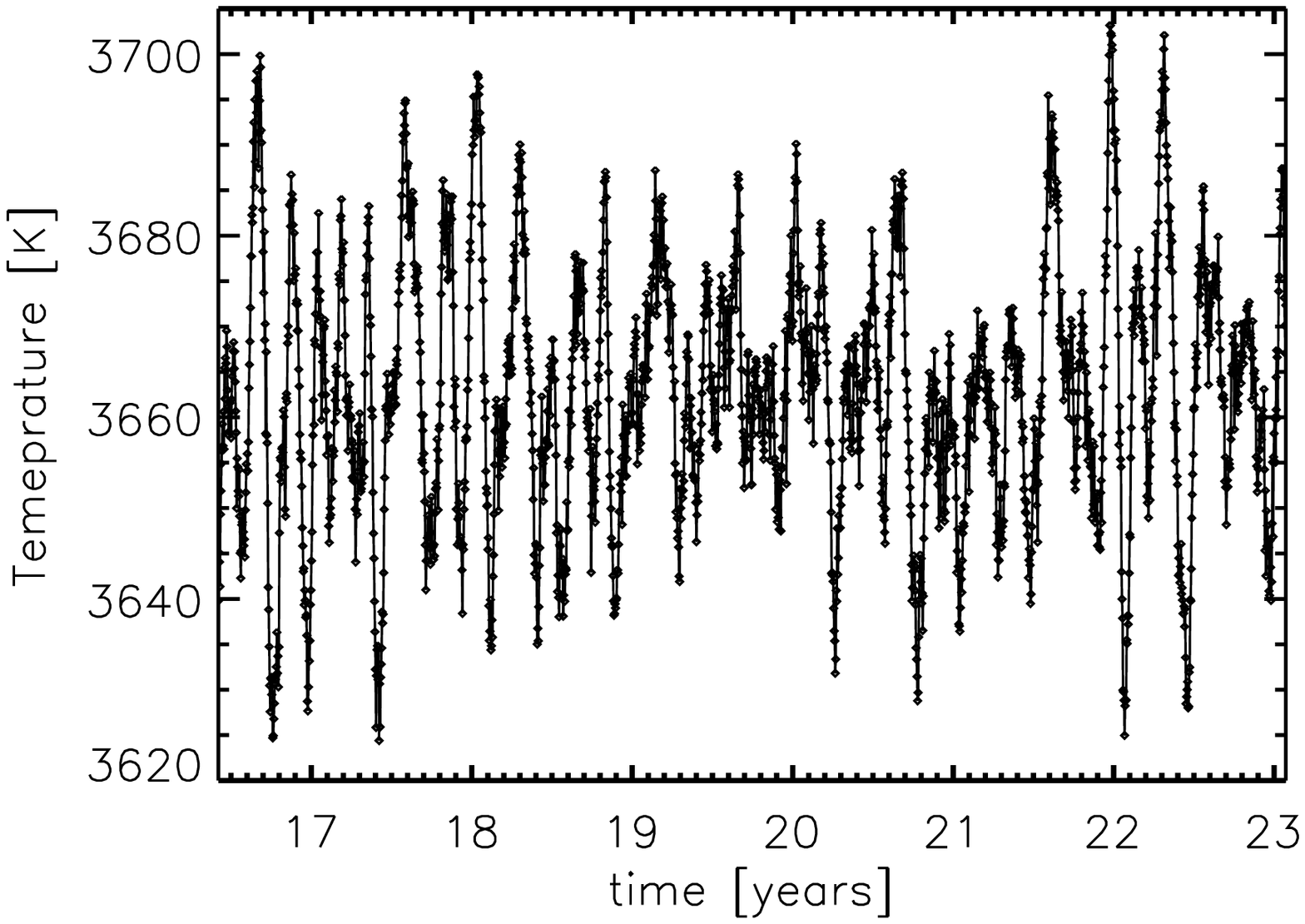}
               \includegraphics[width=0.25\hsize]{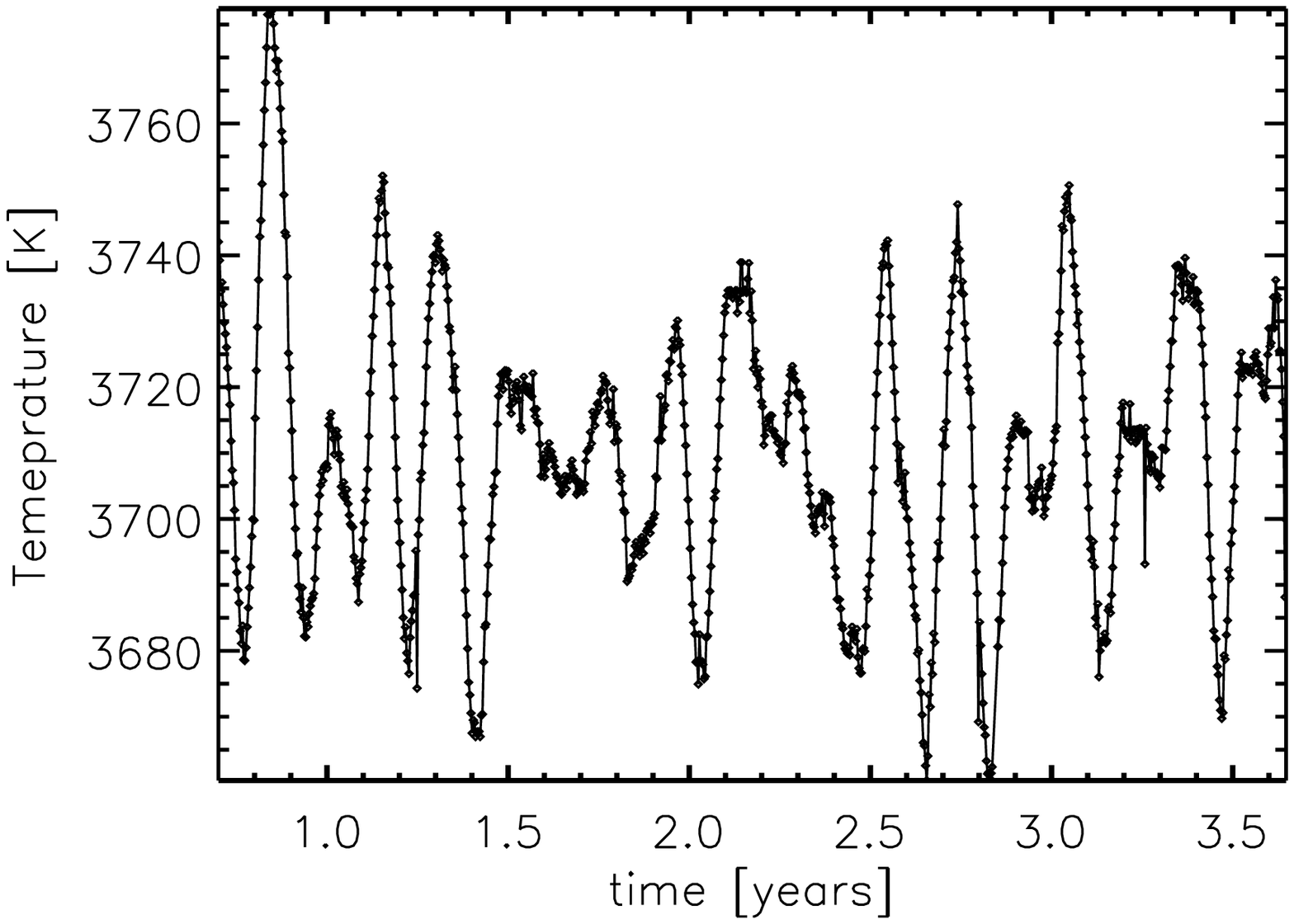}\\
    \includegraphics[width=0.25\hsize]{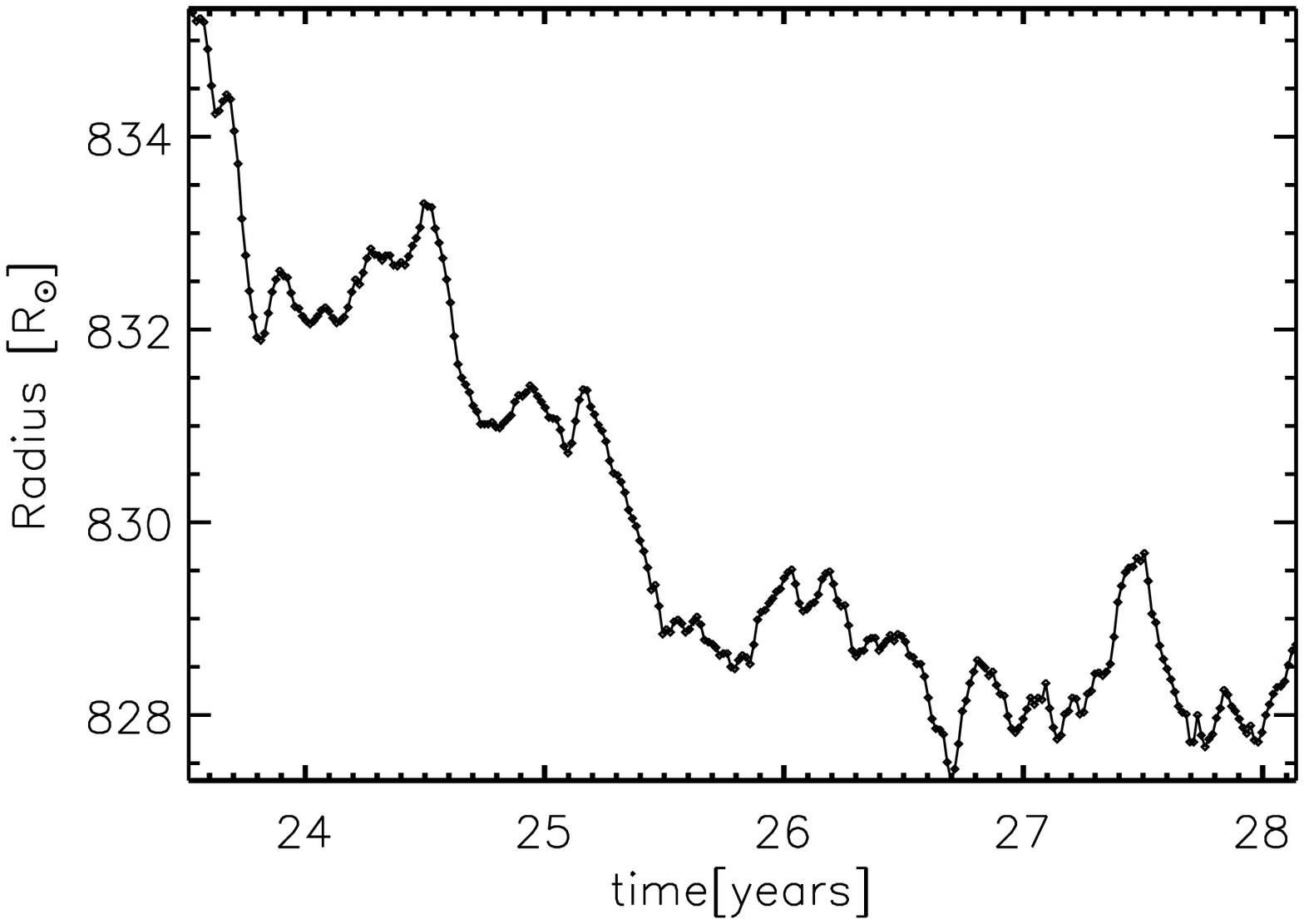} 
    \includegraphics[width=0.25\hsize]{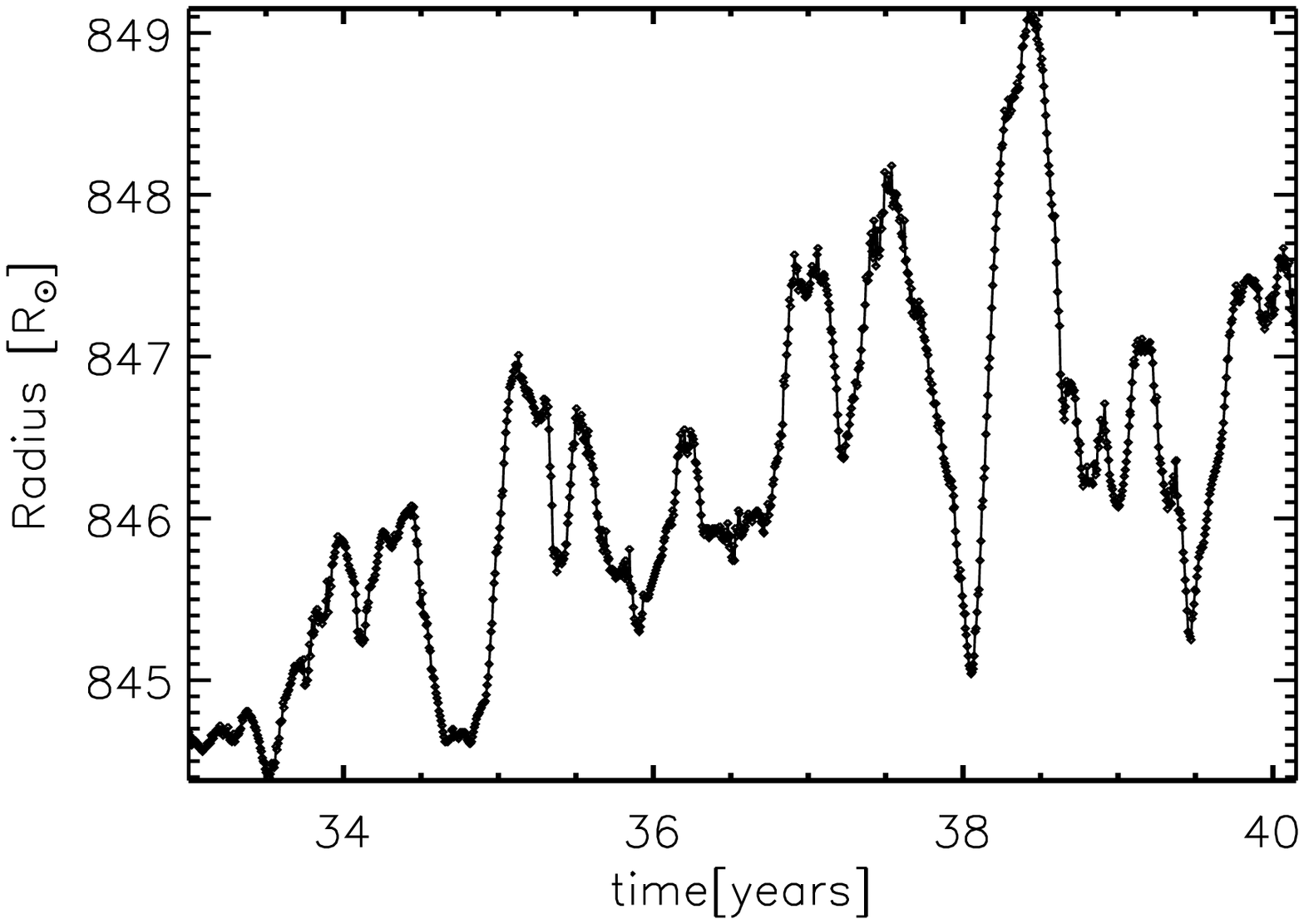} 
    \includegraphics[width=0.25\hsize]{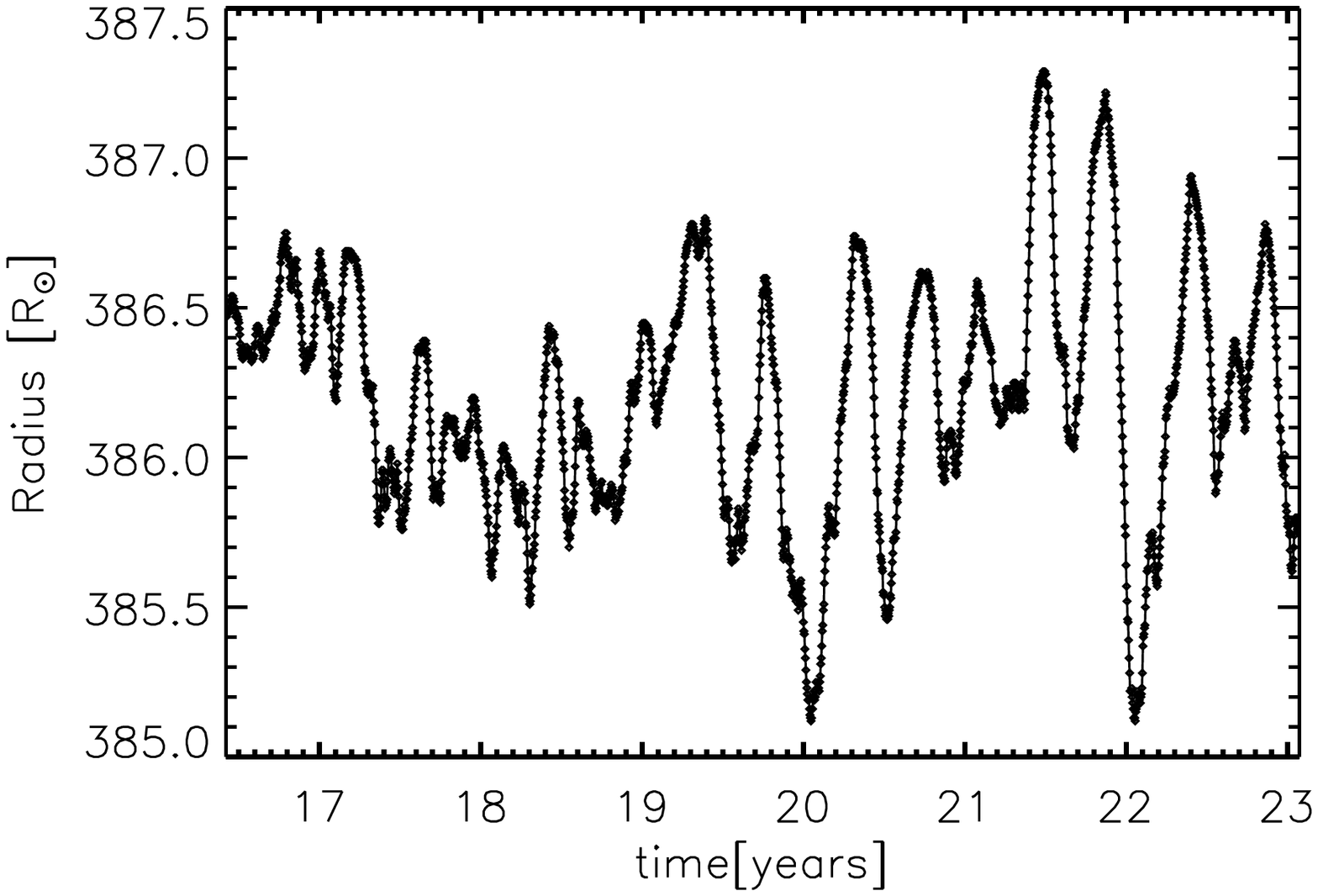} 
      \includegraphics[width=0.25\hsize]{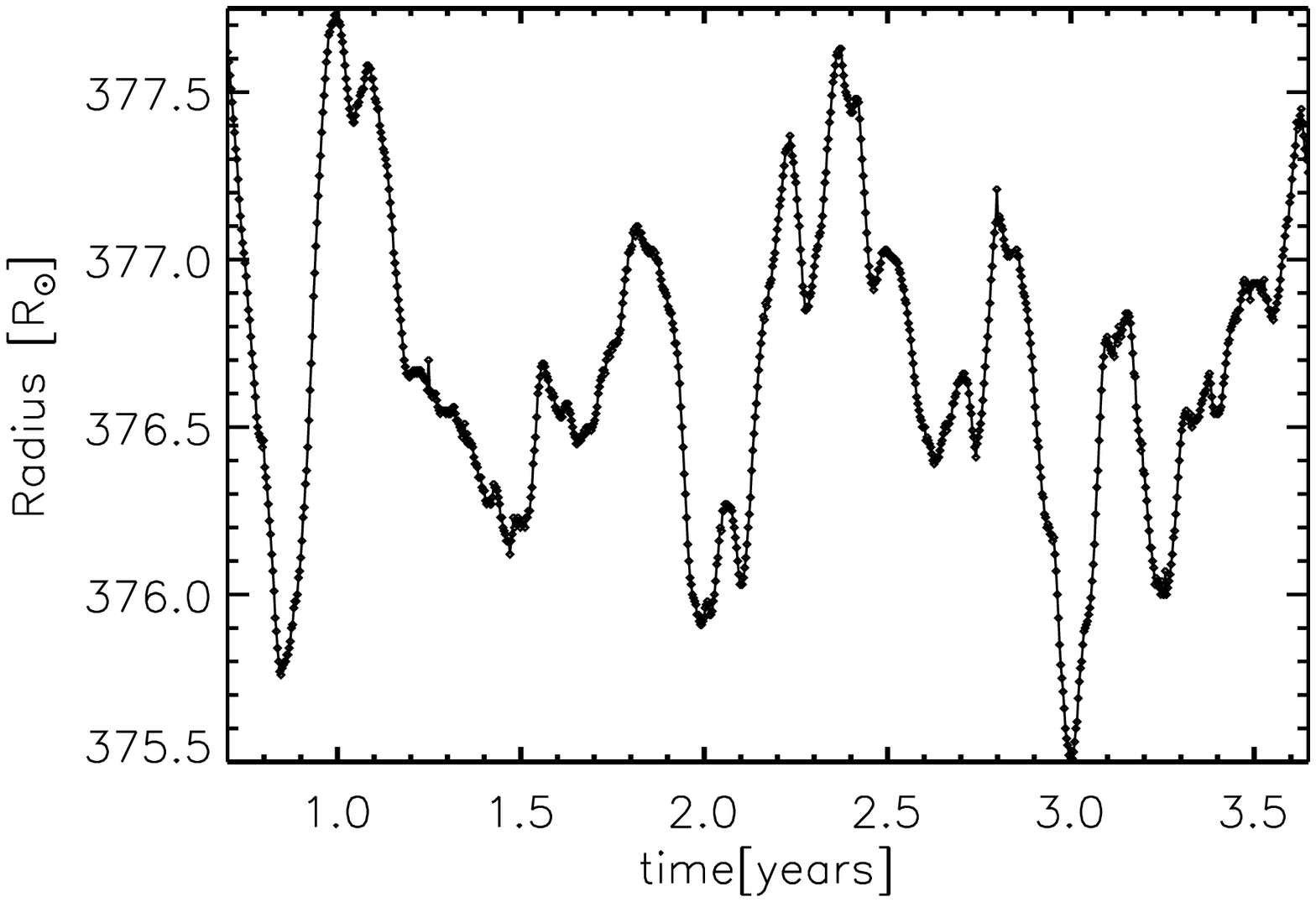} \\
    \includegraphics[width=0.25\hsize]{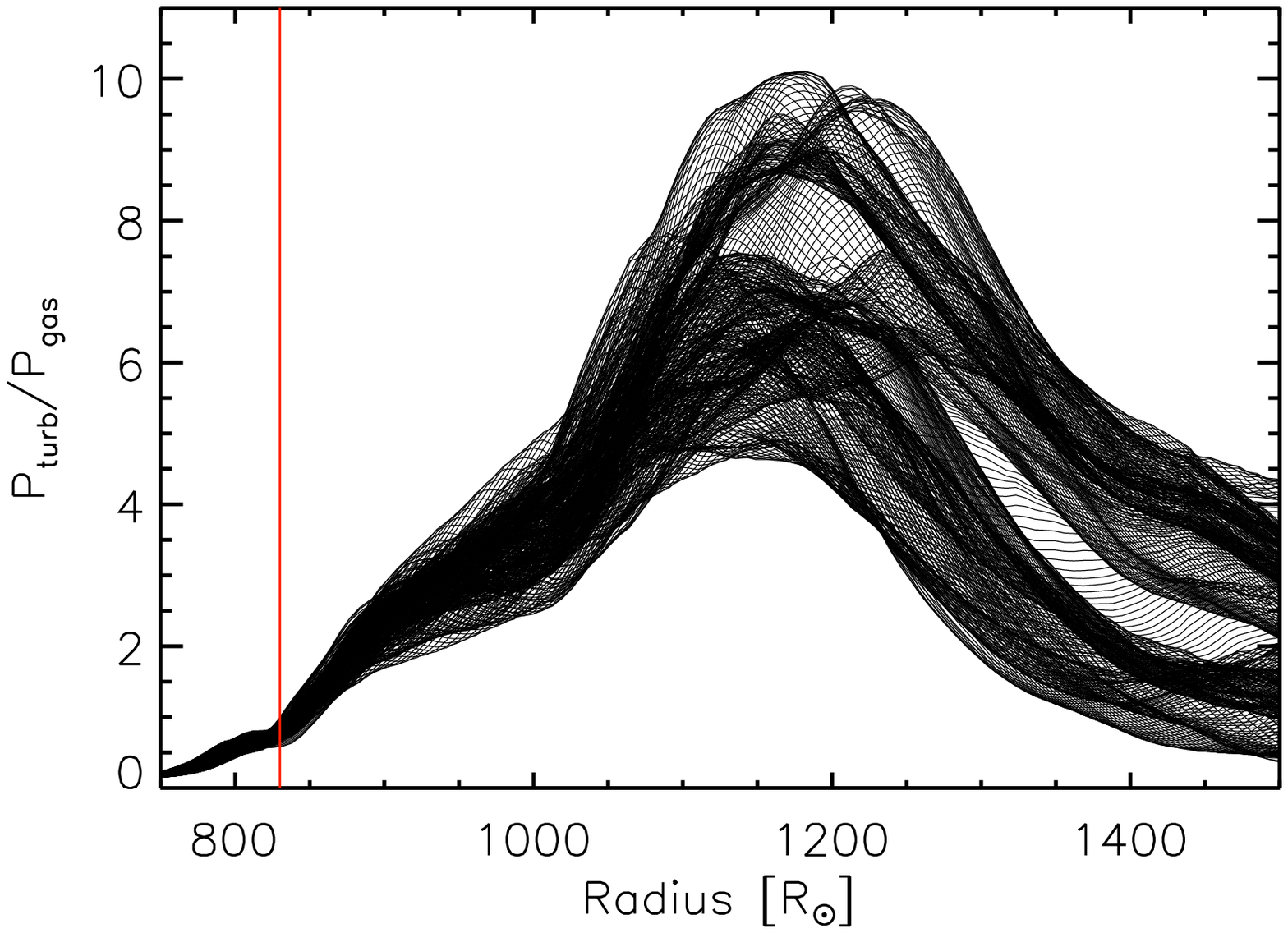} 
    \includegraphics[width=0.25\hsize]{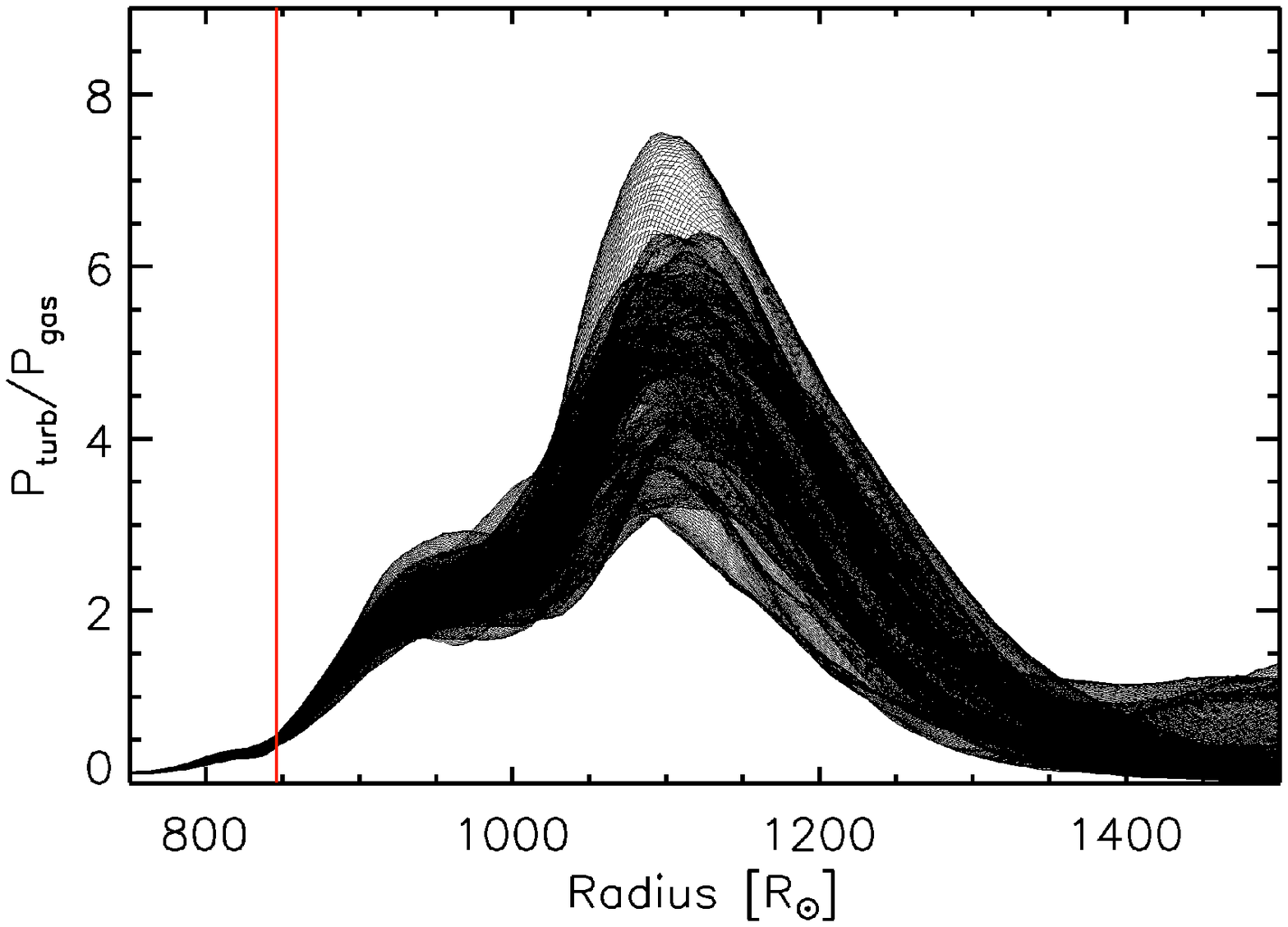} 
    \includegraphics[width=0.25\hsize]{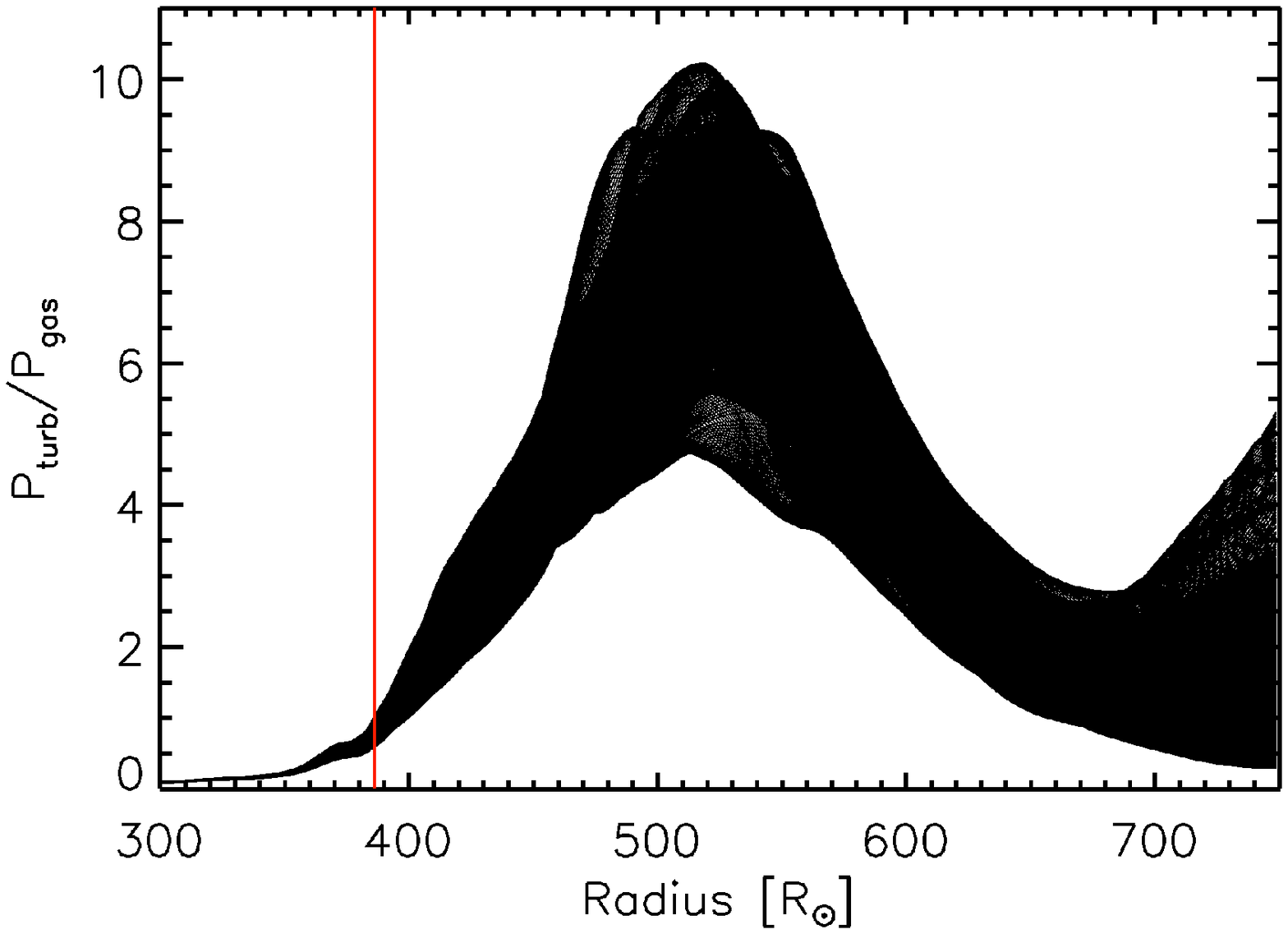} 
      \includegraphics[width=0.25\hsize]{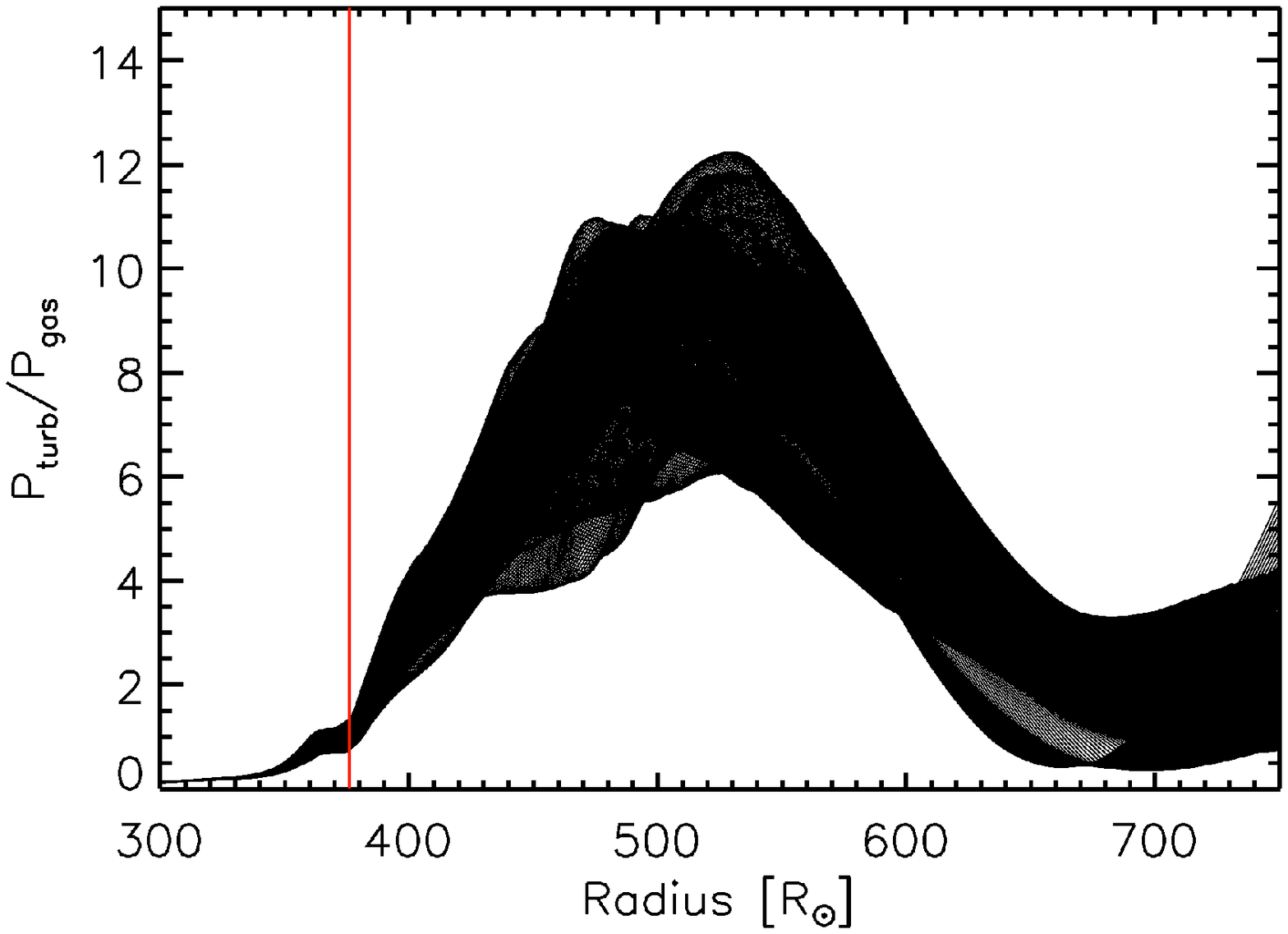} \\
        \end{tabular}
      \caption{Luminosity, temperature, and radius as 
      a function of time for the simulations of Table~\ref{simus}: from left to right columns st35gm03n07, st35gm03n13, st36gm00n04, and st36gm00n05. The bottom panels are the ratio between turbulent pressure and gas pressure for different snapshots. The red vertical lines in all the panels is the approximative position of the radius from Table~\ref{simus}.
           }
        \label{structure}
   \end{figure*}

 \begin{figure*}
   \centering
   \begin{tabular}{cccc}
   \includegraphics[width=0.33\hsize]{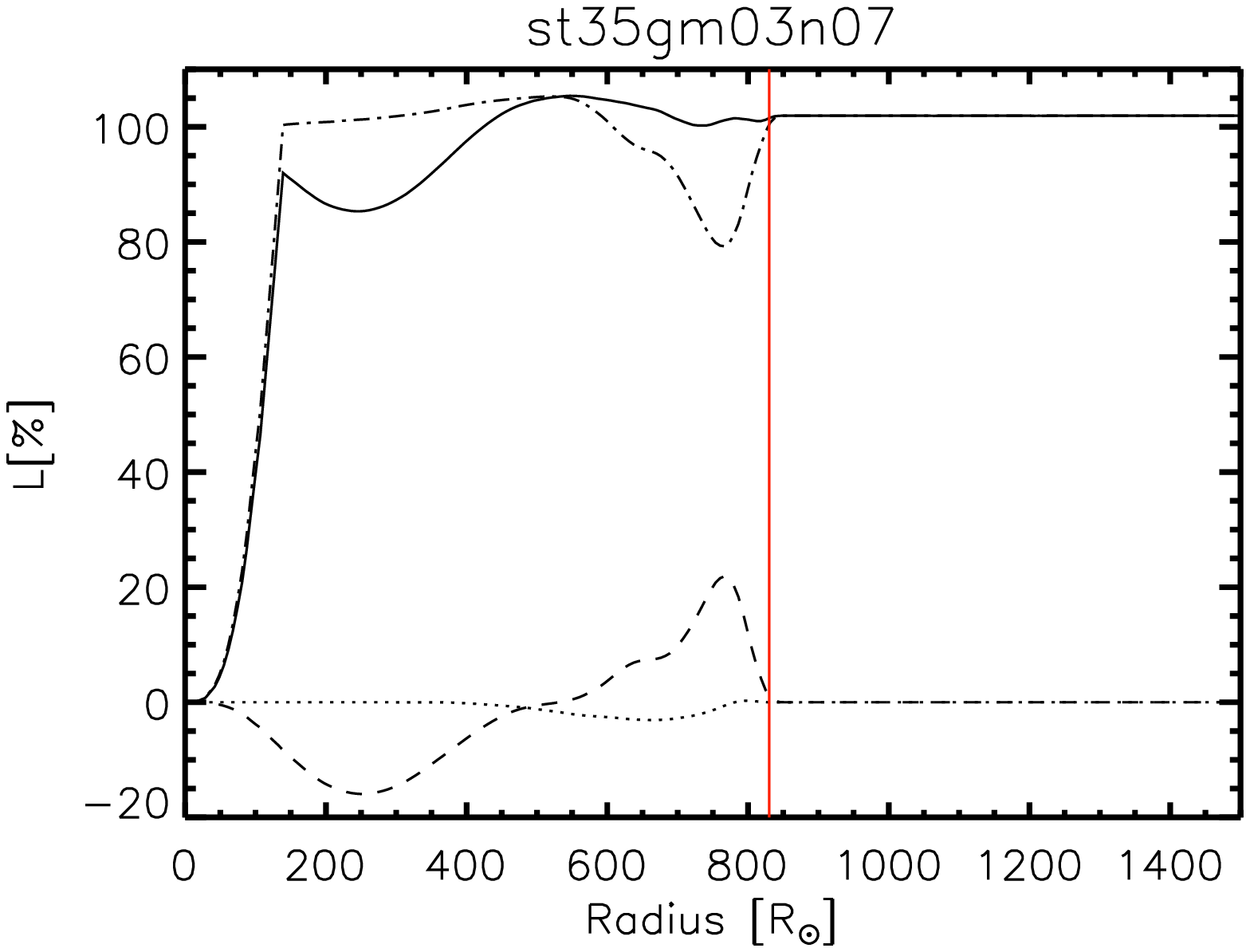}
      \includegraphics[width=0.33\hsize]{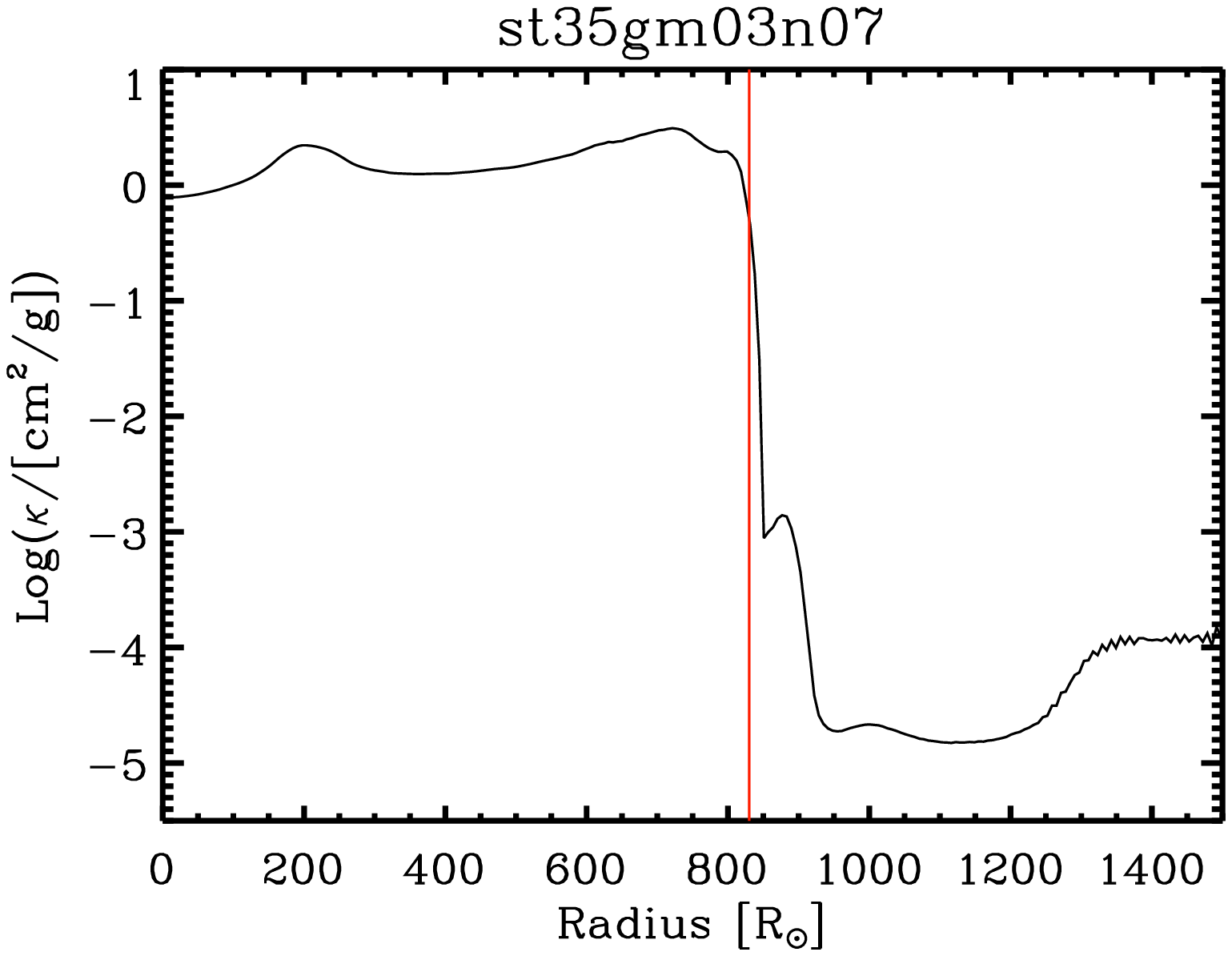}
   \includegraphics[width=0.33\hsize]{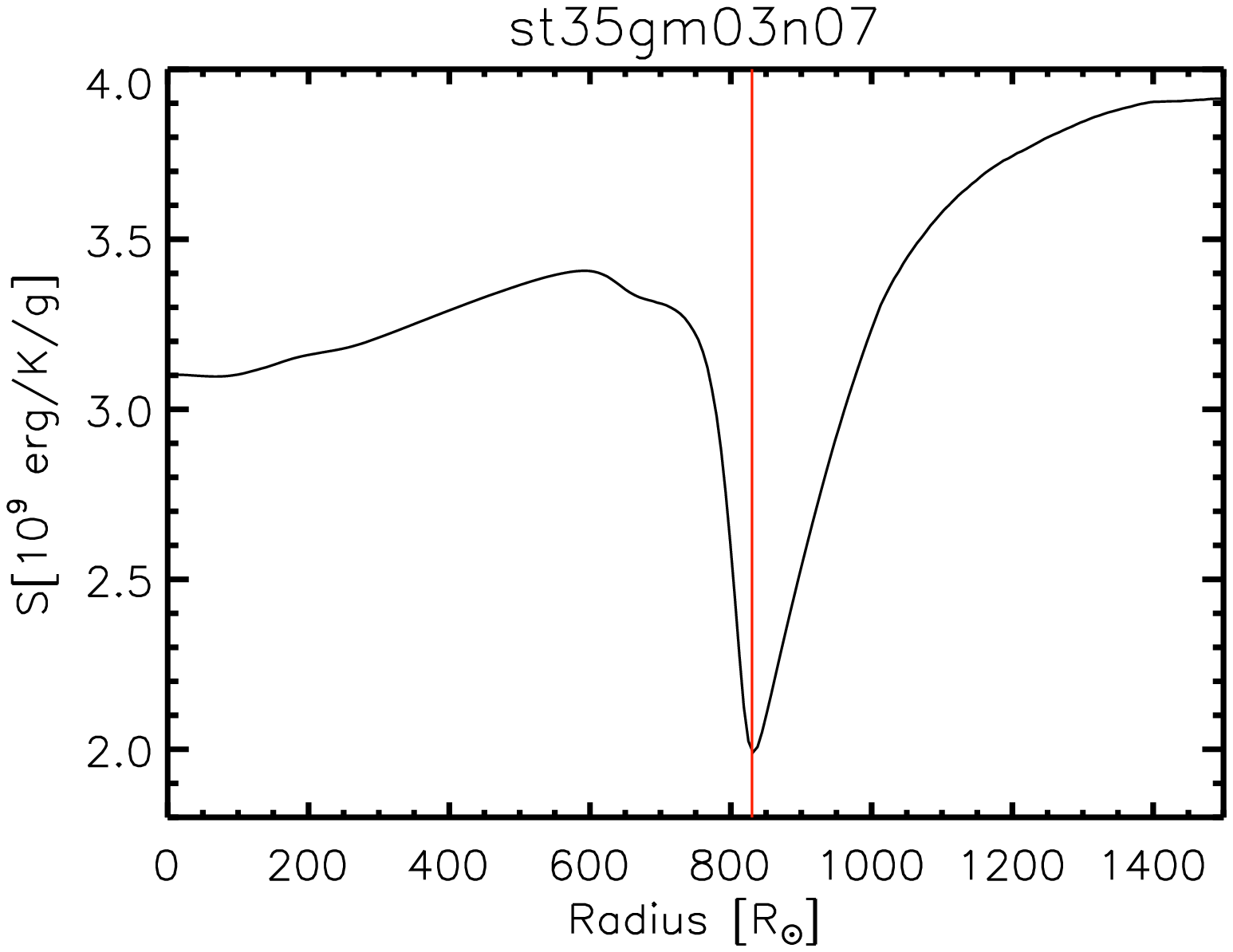}\\
    \includegraphics[width=0.33\hsize]{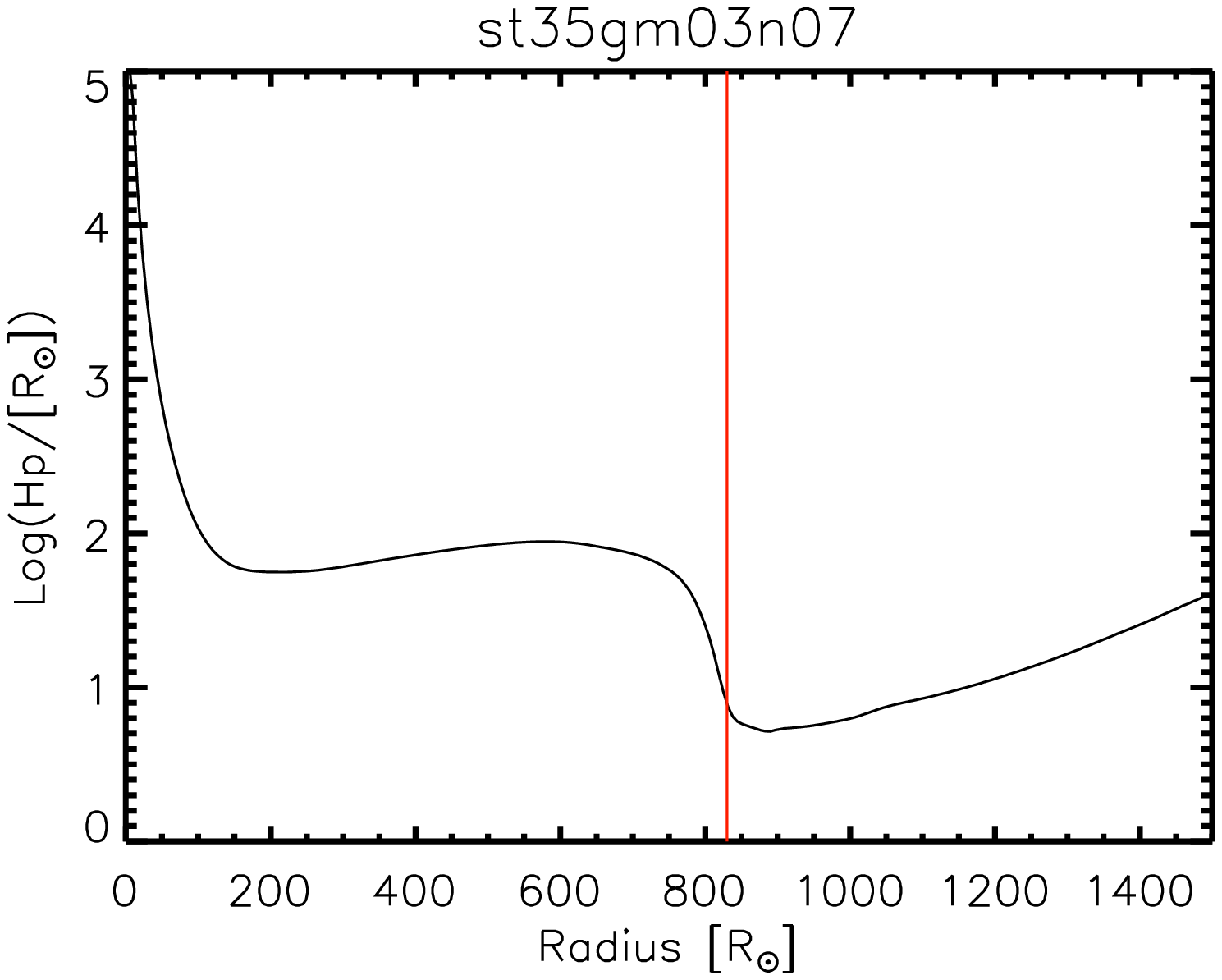}
     \includegraphics[width=0.33\hsize]{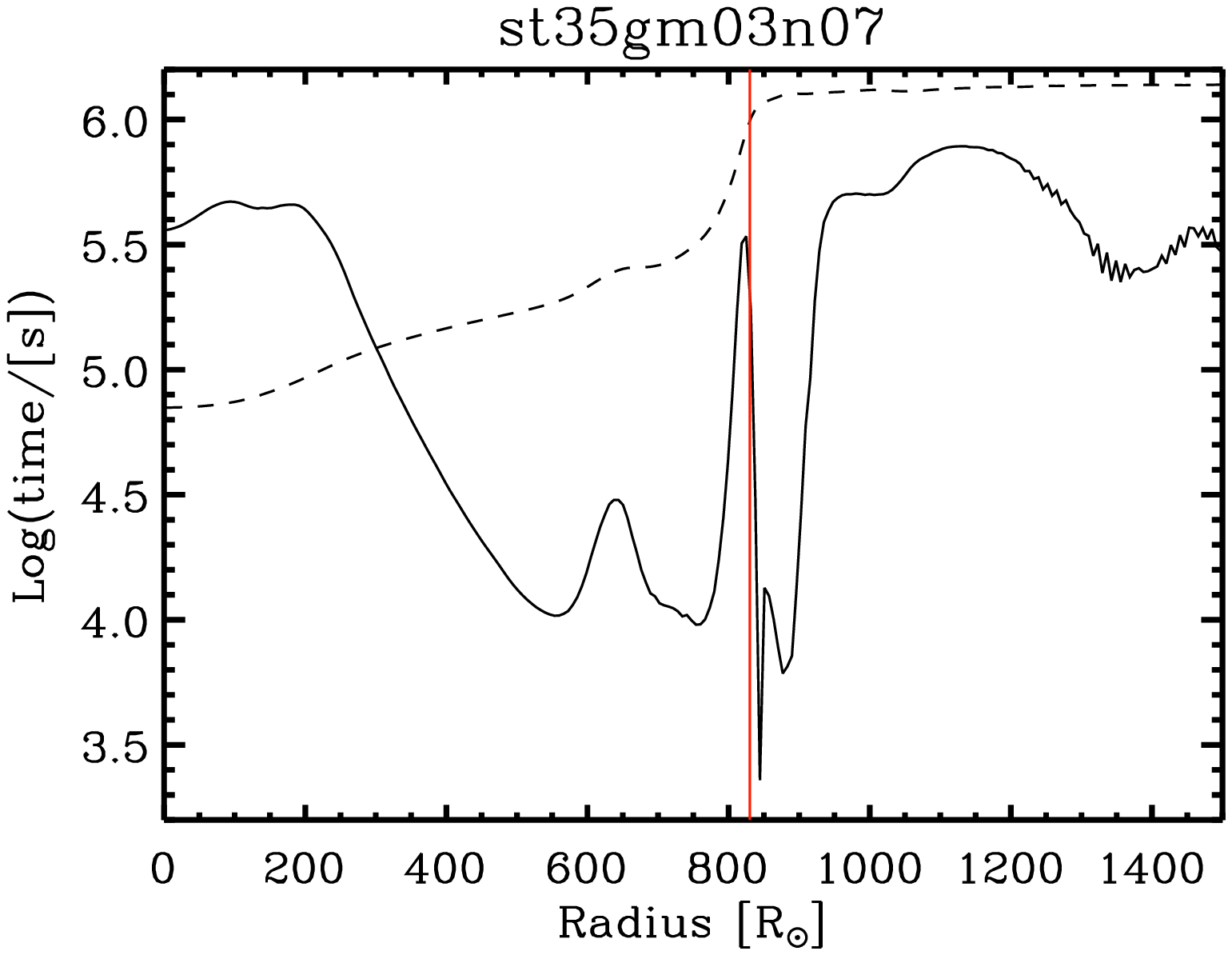}
     \includegraphics[width=0.33\hsize]{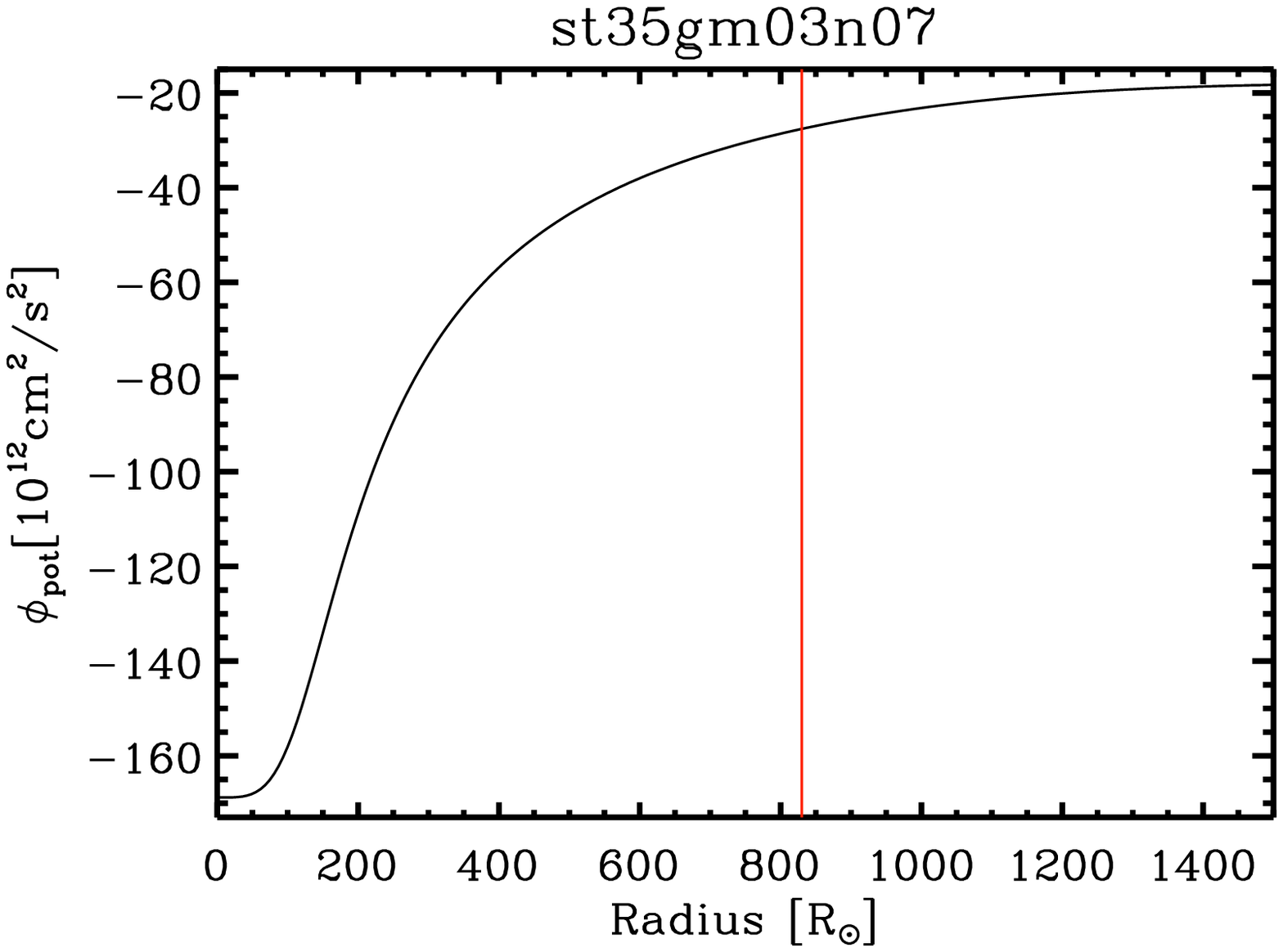}\\
      \includegraphics[width=0.33\hsize]{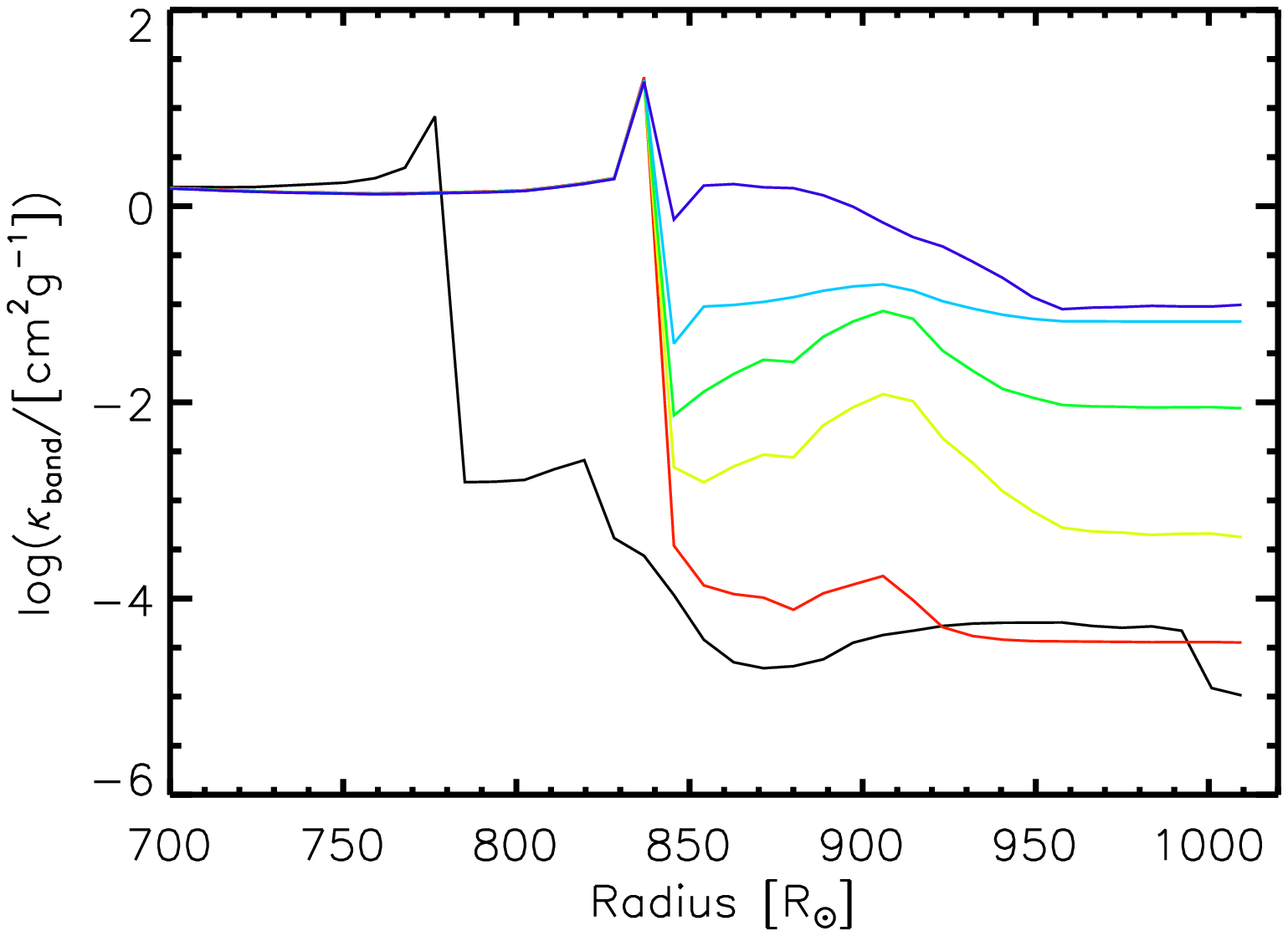}
      \hspace{3 mm}
      \includegraphics[width=0.30\hsize]{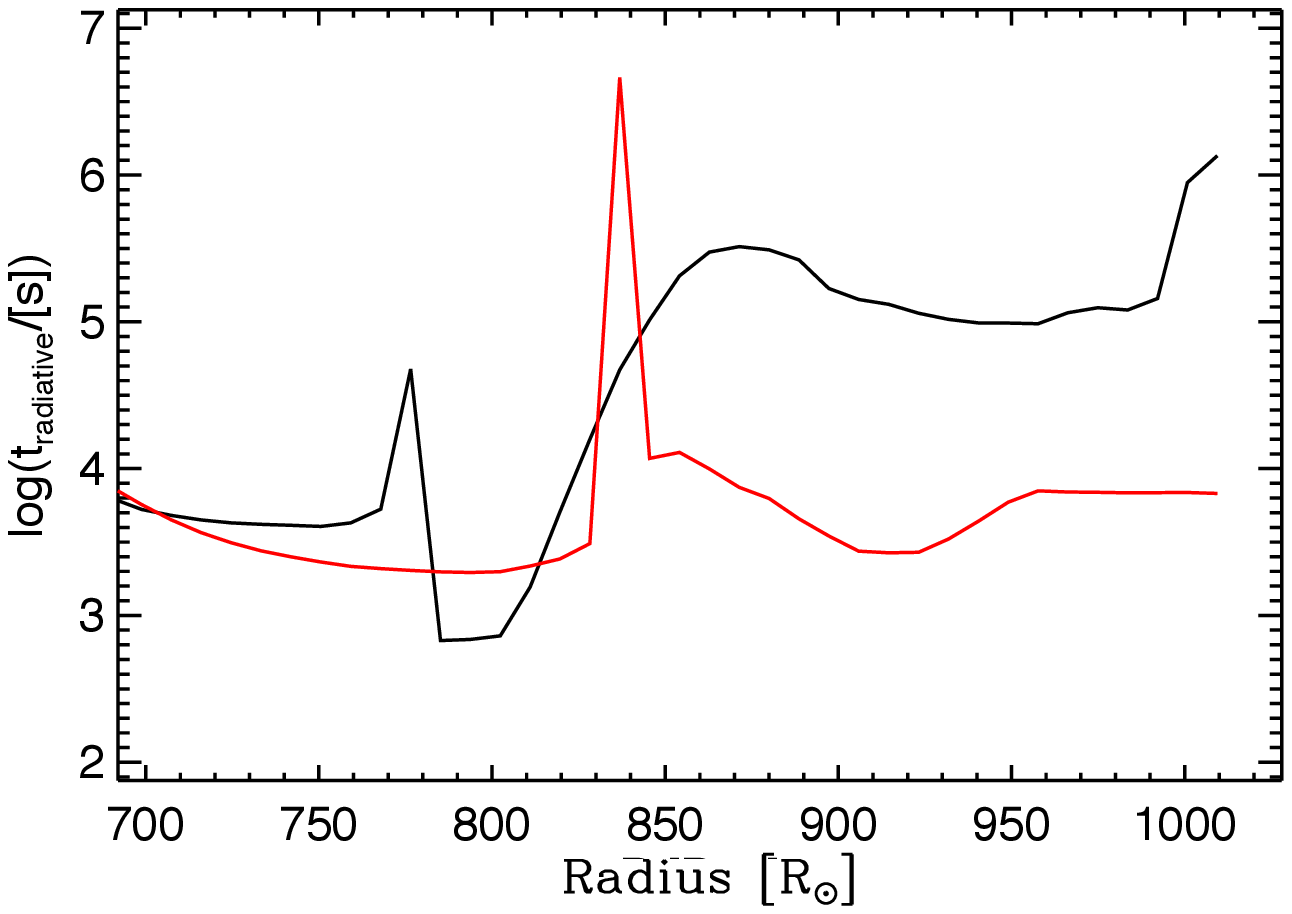}
      \hspace{1 mm}
      \includegraphics[width=0.33\hsize]{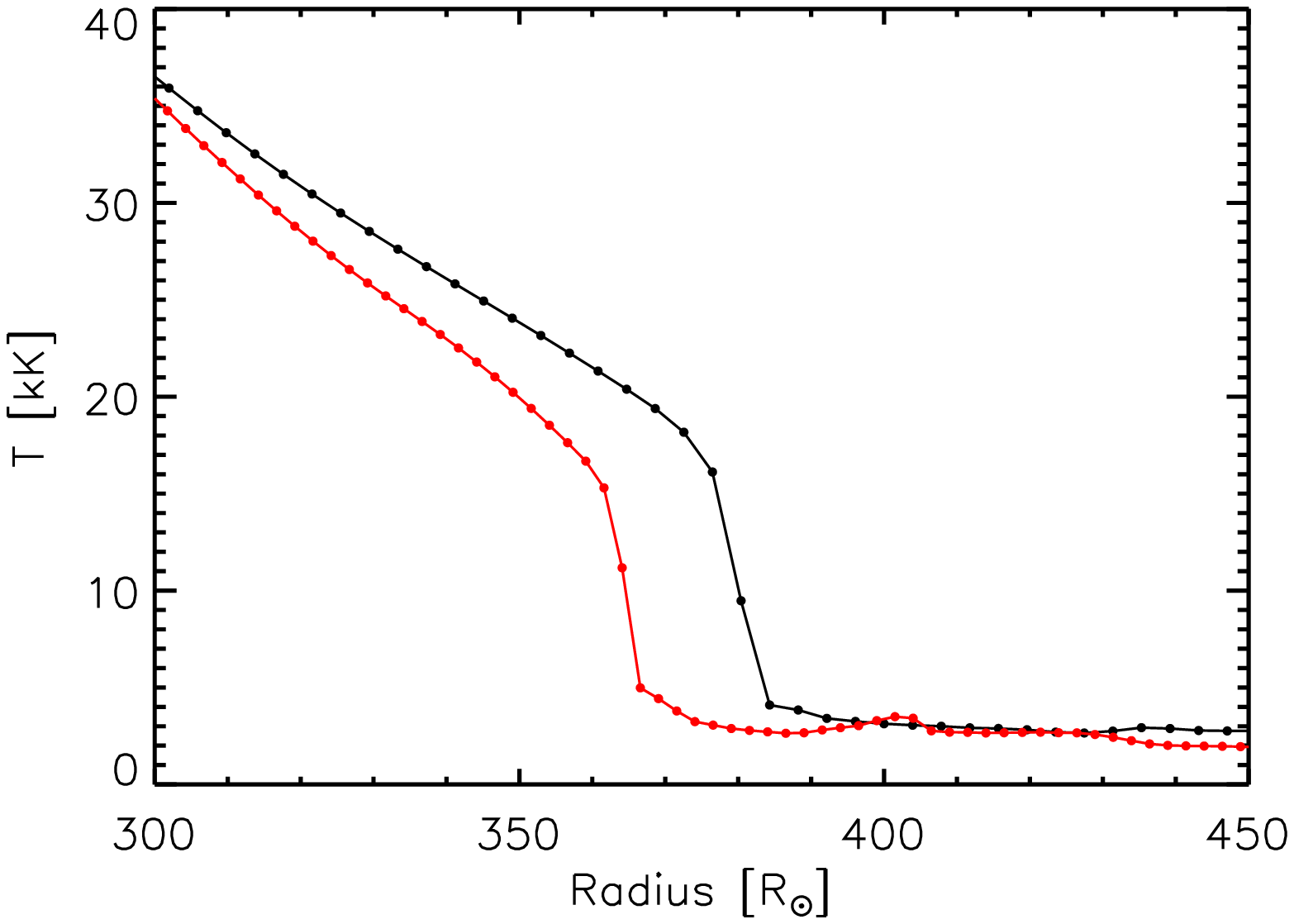}
            \end{tabular}
      \caption{The two rows show some spatially average quantities (i.e., spherical shells) for a snapshot of st35gm03n07 (Table~\ref{simus}). Red vertical line is the location of the radius. \emph{Top row, left panel:} luminosities as function of radius in solar radii: continuous curve is the total luminosity; the dashed curve is the luminosity due to pressure work; the dotted curve is the luminosity of kinetic energy; and the dash-dotted curve is the radiative luminosity. \emph{Top row, center:} opacity. \emph{Top row, right:} specific entropy. \emph{Central row, left panel:} pressure scale height. \emph{Central row, center:} characteristic radiative (continuous line) and hydrodynamical (dotted line) time scales. \emph{Central row, right:} gravitational potential $\Phi_{\rm{pot}}$.        
\emph{Bottom row, left panel:} random columns opacity and radiative timescale for a given snapshot of the gray model st35gm03n07 (black) and non-gray model st35gm03n13 (red). The five coloured curves correspond to the opacity groups. The radiative time scale is estimated
for temperature fluctuations with spatial scales of the order of the grid size. \emph{Bottom row, center:} characteristic radiative timescale for the gray (black) and non-gray (red) models. \emph{Bottom row, right:}  temperature profiles for the st36g00n04 (255$^3$ grid points, black) and st36g00n05 (401$^3$ grid points, red).}
\label{1dquantities}
   \end{figure*}

\section{Model structure}

Table~\ref{simus} reports the simulations analyzed in this work. These models are the result of intensive calculations on parallel computers: the oldest simulation, st35gm03n07, used so far in the other papers of this series, took at least a year (this time may be difficult to estimate due to the queuing system and different machines used) on machines with 8-16 CPUs to provide 9 years of simulated stellar time; more recently, we managed to compute the other simulations of Table~\ref{simus} on a shorter timescale on a machine with 32 CPUs. The model st36gm00n04, which is slightly larger in numerical resolution, needed about 2 months to compute more than 20 stellar years, while st36gm00n05 (the highest numerical resolution so far) about 6 months for about 3 stellar years. Eventually, st35gm03n13 needed about 8 months to compute 10 years of stellar time despite its lower numerical resolution. The CPU time needed for non-gray runs does scale almost linearly with the number of wavelength groups.

\subsection{Global quantities}\label{global_sect}

3D simulations start with an initial model that has an estimated radius,
a certain envelope mass, a certain potential profile, and a prescribed
luminosity as described above. During the run, the internal structure relaxes to something
 not to far away from the initial guess. We needed several trials to
get the right initial model for the latest set. The average final stellar parameters are determined once the
simulation has ended (Fig.~\ref{structure}). For this purpose, we adopted the method reported in \cite{2009A&A...506.1351C} (hereafter Paper~{\sc{I}}). The quantities are
averages over spherical shells and over time, and the errors are one-sigma fluctuations with respect to the average over time. We computed the average temperature and luminosity over spherical shells, $T(r)$,
and $L(r)$. Then, we search for the radius $R$ for which 
\begin{equation}\label{eq_radius_co5bold}
{L(R)}/(4\pi R^2)=\sigma T^4¨,
\end{equation}
where $\sigma$ is the Stefan-Boltzmann constant. Eventually, the effective temperature is $T_{\rm{eff}}=T(R)$. As already pointed out in Paper~{\sc{I}}, the gray model st35gm03n07 (top row of Fig.~\ref{structure}) shows a drift in the first 2 years ($-$0.5$\%$ per year) before stabilisation in the last 2.5 years. This drift in also visible in the radius of the non-gray model st35gm03n13 (second row) even though it is weaker (less than 0.3$\%$). The luminosity fluctuations are of the order of 1$\%$ and the temperature variations $<0.3\%$ for both simulations. It must be noted that the last snapshot of the gray simulation has been used as the initial snapshot for the non-gray model. The two hotter models, st36gm00n04 and st36gm00n05, are also somehow related because they both use a gray treatment of opacities and st36gm00n05 has been started using the last snapshot of st36gm00n04. The temperature (0.5$\%$) and luminosity (2.2$\%$) fluctuations of the higher resolution model are slightly larger than the temperature (0.4$\%$) and luminosity (1.5$\%$) fluctuations of st36gm00n04. None of the simulation shows a drift of the radius in the last years of evolution.

The bottom row in Fig.~\ref{structure} show the ratio between the turbulent pressure and the gas pressure (defined as in Sect.~\ref{temp_sect}). This quantity shows that in the outer layers, just above the stellar radius, the turbulent pressure plays a significant role for the average pressure stratification
and the radial velocities resulting from the vigorous convection are supersonic.

\begin{figure*}
   \centering
   \begin{tabular}{cc}
    \includegraphics[width=0.41\hsize]{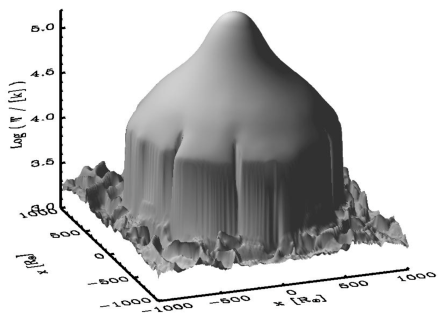}
    \includegraphics[width=0.36\hsize]{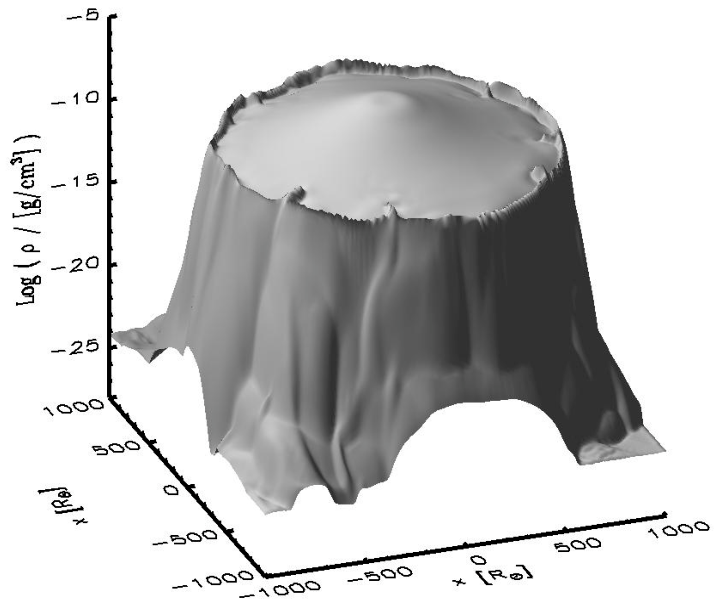} \\
    \includegraphics[width=0.44\hsize]{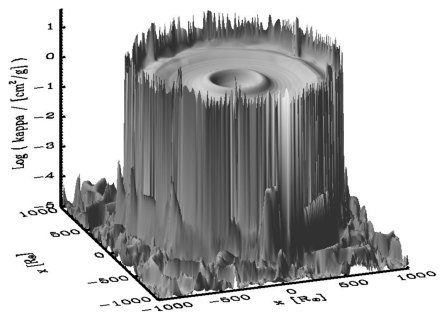}
    \includegraphics[width=0.41\hsize]{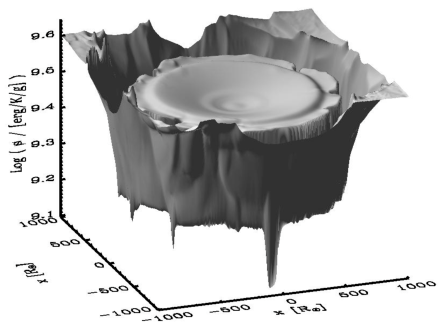} \\
        \end{tabular}
      \caption{Logarithm of temperature (top left panel), density (top right), opacity (bottom left) and entropy (bottom right) of a slice through the center from a snapshot of the RSG simulation st35gm03n07 in Table~\ref{simus}.
           }
        \label{cobold_quantities}
   \end{figure*}

Figure \ref{1dquantities} displays some quantities spatially averaged over spherical shells for a snapshot of the gray simulation st35gm03n07 in Table~\ref{simus}. Figure~\ref{cobold_quantities} shows three-dimensional views of some quantities for the same snapshot and simulation. Radiation is of primary importance for many aspects of convection and the envelope structure in a RSG. It does not only cool the surface to provide a somewhat unsharp
outer boundary for the convective heat transport. It also contributes
significantly to the energy transport in the interior (top left panel of 
Fig.~\ref{1dquantities}), where convection never carries close to $100\%$ of the
luminosity. \\
In the optically thick part the stratification is slightly far
from radiative equilibrium and the entropy jump from the photosphere to the layers below
is fairly large (entropy $s$ in Fig.~\ref{1dquantities} and Fig.~\ref{cobold_quantities}). The He {\sc{II}}/He {\sc{III}} ionization zone is visible in the entropy structure as a small minimum near the surface before the normal steep entropy decrease. The opacity peak at around T=13\ 000K just below the photosphere
causes a very steep temperature jump (temperature $T$ and opacity $\kappa$ in Fig.~\ref{cobold_quantities}) which is very prominent on top of upflow
regions (opacity $\kappa$ in Fig.~\ref{1dquantities}). This causes a density inversion (density $\rho$ in Fig.~\ref{cobold_quantities} and Fig.\ref{temperaturedensity_profile}) which is a sufficient condition of convective instability 
resolved, the entropy drop occurs in a very thin layer, while
the smearing due to the averaging procedure over nonstationary
up- and downflows leads to the large apparent extent of the
mean superadiabatic layer.
  
The local radiative relaxation time scale in the photosphere is much shorter than a typical hydrodynamical time scale (time in Fig.~\ref{1dquantities}). Numerically, the radiative energy exchange terms are
rather stiff and prevent large deviations from a radiative equilibrium
stratification. These terms enforce very short time steps
(about 600 s per individual radiation transport step) and are responsible
for the relatively high computational cost of this type
of simulations. Local fluctuations in opacity, temperature (source function),
and heat capacity pose high demands on the stability of
the radiation transport module. This is true to a lesser degree
for the hydrodynamics module due to shocks, high Mach numbers,
and small scale heights. A side effect of the steep and significant temperature jump
is the increase in pressure scale height from small photospheric
values to values that are a considerable fraction of the radius in
layers just below the photosphere (pressure scale heigh $H_P$ in Fig.~\ref{1dquantities}). The non-gray model has systematically smaller time scales (Fig.~\ref{1dquantities}, bottom row), which causes the increased smoothing efficiency of temperature fluctuations.

\subsection{Temperature and density structures}\label{temp_sect}

\begin{figure*}
   \centering
   \begin{tabular}{cc}
    \includegraphics[width=0.5\hsize]{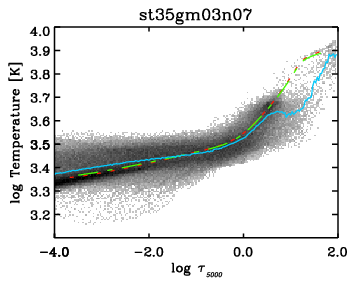}
    \includegraphics[width=0.5\hsize]{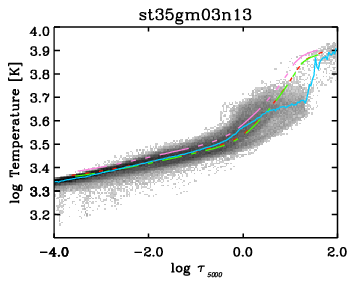} \\
    \includegraphics[width=0.5\hsize]{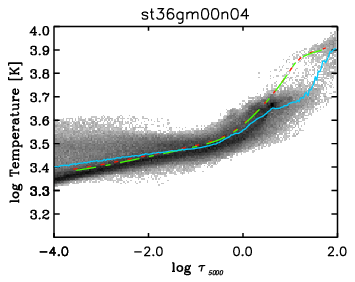}
    \includegraphics[width=0.5\hsize]{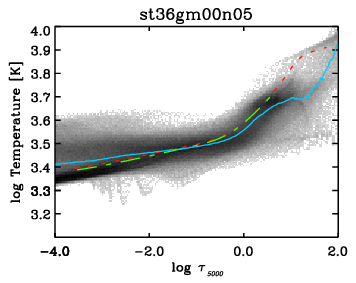} \\
        \end{tabular}
      \caption{Thermal structures of the simulations in Table~\ref{simus} as a function of the optical depth at $\lambda=$ 5000$\rm \AA$. Darker areas indicate temperature values with higher probability. The solid light blue curve is the average temperature, the red dashed line is the 1D MARCS model profile with surface gravity $g$ (Table~\ref{simus1D}) and the green dotted-dashed line is the 1D MARCS model with surface gravity $g_{\rm{eff}}$ (see text). In top right panel, the 3D mean thermal profile is compared to a cool MARCS model at 3430K for the outer layers and a hot one with 3700K (magenta dotted-triple dashed line) for the continuum forming region (see text).
           }
        \label{temperature_profile}
   \end{figure*}

The temperature structure for the 3D simulations is displayed in Fig.~\ref{temperature_profile} as a function of the optical depth at $\lambda=$ 5000$\rm \AA$ for all the simulations used in this work. We have also computed classical 1D, LTE, plane-parallel,
hydrostatic marcs model atmospheres \citep{2008A&A...486..951G} with identical stellar parameters, input data,
and chemical compositions as the 3D simulations (Table~\ref{simus1D}). It should be noted that the MARCS models do not have identical input data and numerical scheme as 3D simulations but they are a good benchmark for a quantitative comparison. The optical depth scale in Fig.~\ref{temperature_profile} has been computed with the radiative transfer code {{\sc Optim3D}} (see Paper~{\sc{I}}, and Sect.~\ref{tio}) for all the rays parallel to the grid axes with a $x$ $-y$ positions $\leq \left(x_{\rm{center}},y_{\rm{center}}\right) \pm \left(G*0.25\right)$ (i.e., a small square at the center of the stellar disk), where $x_{\rm{center}}$ and $y_{\rm{center}}$ are the coordinate of the center of one face of the numerical box, $G$ is the number of grid point from Table~\ref{simus}, and 0.25 has been chosen to consider only the central rays of the box.  

\begin{figure}
   \centering
   \begin{tabular}{cc}
    \includegraphics[width=1.0\hsize]{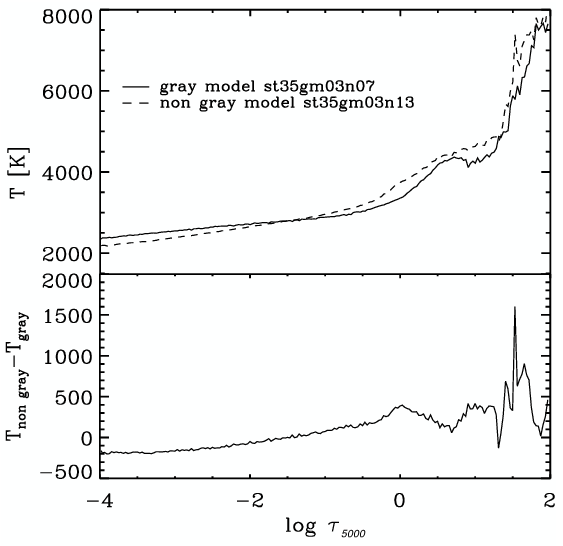}
          \end{tabular}
      \caption{Mean profiles (top) and temperature difference (bottom) of the gray and non-gray simulations from Fig.~\ref{temperature_profile}.
           }
        \label{avg_profiles}
   \end{figure}

The models using a frequency-independent gray treatment show larger temperature fluctuations compared to the non-gray case (top right panel in Fig.~\ref{temperature_profile}), as was already pointed out in \cite{1994A&A...284..105L} for local hydrodynamical models, because the frequency-dependent
radiative transfer causes an intensified heat exchange of a fluid element with its 
environment tending to reduce the temperature differences. Moreover, the temperature structure in the outer layers of the non-gray simulation
tends to remain significantly cooler than the gray case (up to $\sim$200 hundreds K, Fig.~\ref{avg_profiles}), close to the radiative equilibrium. The thermal gradient is then steeper in the non-gray model and this is crucial for the formation of the spectral lines (see Sect.~\ref{tio}). This effect has already been pointed out in the literature and a more recent example is \cite{2004A&A...421..755V} who pointed out that the non-gray treatment of opacities is mandatory to compare solar magneto-convective simulations with {\sc Muram} code to the observations.\\
It is also striking to remark that all the gray simulations largely diverge from 1D MARCS models, while the case of the thermal structure of the non-gray model is more complicated. The outer layers show a very good agreement with a cool 1D MARCS model at 3430K (Fig.~\ref{temperature_profile}, top right panel). At $\lg\tau_{5000}\sim1$ (i.e., where the continuum flux comes from) the mean 3D temperature is warmer than the 1D: a hot 1D MARCS model at 3700K is then necessary to have a better agreement but, however, this 1D profile diverges strongly for $\lg\tau_{5000}<1$. \\

\begin{table}
\centering
\caption{1D MARCS models used in this work.}
\label{simus1D}      
\renewcommand{\footnoterule}{} 
\begin{tabular}{c c c c c}        
\hline\hline                 
    $T_{\rm{eff}}$ & $M$ &  $\log g$ & Corresponding &  Surface  \\
              $[\rm{K}]$ & [$M_\odot$] &  [cgs]  & 3D simulation  & gravity\\
\hline	  
	  3490 & 12 & $-$0.35 & st35gm03n07   & $g$ \\
	  3490 & 12 & $-$0.58 & st35gm03n07  &  $g_{\rm{eff}}$ \\
	  3430 & 12 & $-$0.35 & st35gm03n13  & $g$\\
	  3700 & 12 & $-$0.35 & st35gm03n13  & $g$\\
	  3430 & 12 & $-$0.65 & st35gm03n13 & $g_{\rm{eff}}$ \\
	  3660 & 6   & 0.02 & st36gm00n04 &  $g$ \\
	  3660 & 6   & $-$0.22 & st36gm00n04  &  $g_{\rm{eff}}$\\
	  3710 & 6   & 0.05 & st36gm00n05  &$g$\\
	  3710 & 6   & $-$0.22 & st36gm00n05 & $g_{\rm{eff}}$ \\
\hline

\hline\hline                          
\end{tabular}
\tablefoot{All 1D models are spherical and have been computed with solar metallicity. The effective surface gravity ($g_{\rm{eff}}$) is given at the positions of the radii from Fig.~\ref{geff_figure} and using Eq.~\ref{eqgrav2}. The last column displays if the surface gravity used to compute 1D model is $g_{\rm{eff}}$ or $g$. For $g$ we took the same values as 3D simulations of Table~\ref{simus}. }
\end{table}

While turbulent pressure is naturally included in the RHD simulations, it is modeled in 1D models assuming a parameterisation as 
 
  \begin{equation}\label{eqpturb}
  P_{\rm{turb}}=\beta\rho v_{\rm{t}}^2
  \end{equation}
  
  where $\rho$ is the density, $v_{\rm{t}}$ the characteristic velocity, and $\beta$ is approximatively 1 and depends on whether the motions occur more or less isotropically. This
pressure is measuring the force produced by the kinetic movements
of the gas, whether due to convective or other turbulent
gas motions. In 1D models, assuming spherical symmetry, the equation of hydrostatic equilibrium is solved for 

 \begin{equation}\label{eqgrav1}
 \nabla P_{\rm{tot}}= \nabla P_{\rm{g}} + \nabla P_{\rm{turb}} + \nabla P_{\rm{rad}} = -\rho g
  \end{equation}
  
 where $g\left(r\right)=\frac{GM}{r^{2}}$ (being $r$ the radius, $M$ the mass of the star and $G$ the Newton's constant of gravity), $P_{\rm{g}}=\Re \rho T/\mu_{\rm{mol}}$ the gas pressure ($\Re$ being the gas constant, $\mu_{\rm{mol}}$ the molecular weight, and $T$ the temperature), and $P_{\rm{rad}}$ the radiative pressure. \cite{2008A&A...486..951G} and \cite{1992iesh.conf...86G} showed that, assuming $P_{\rm{rad}}=0$ like in 3D simulations, they could mimic the turbulent pressure on the models by using models with
those effects neglected with an adjusted gravity:
 
 \begin{equation}\label{eqgrav2}
 g_{\rm{eff}}=g\left(\frac{1}{1+\beta\frac{\mu_{\rm{mol}}}{\Re T}v_{\rm{t}}^2}\right)
 \end{equation}
 
\begin{figure}
   \centering
   \begin{tabular}{c}
    \includegraphics[width=1.\hsize]{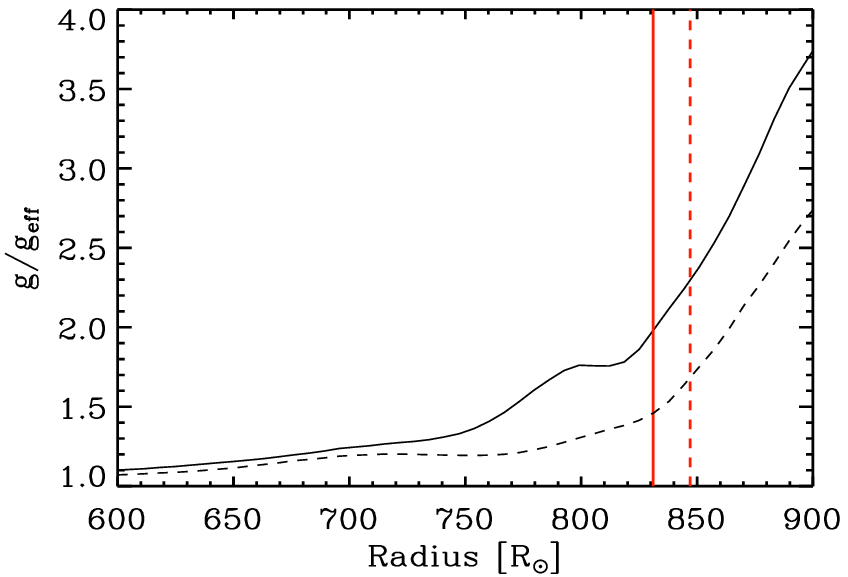}\\
    \includegraphics[width=0.95\hsize]{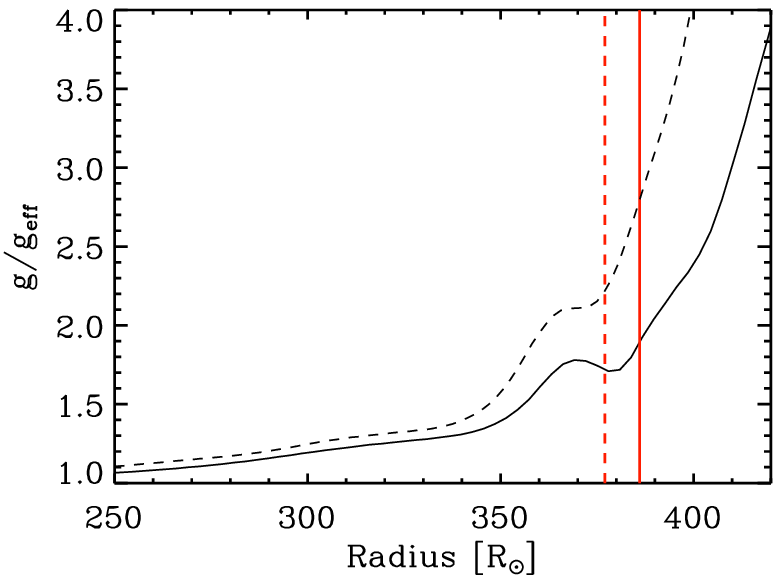} 
        \end{tabular}
      \caption{Ratio between the $g$ and $g_{\rm{eff}}$ from Eq.~\ref{eqgrav2} for simulations st35gm03n07 (solid line) and st35gm03n13 (dotted line) in top panel, and st36gm00n04 (solid line) and st36gm00n05 (dotted line) in bottom panel. We used the same snapshots of Fig.~\ref{temperature_profile} and \ref{temperaturedensity_profile}. The vertical red lines are the approximative positions of the radii from Table~\ref{simus}.}
        \label{geff_figure}
   \end{figure}

\begin{figure*}
      \centering 
   \begin{tabular}{cc}
    \includegraphics[width=0.5\hsize]{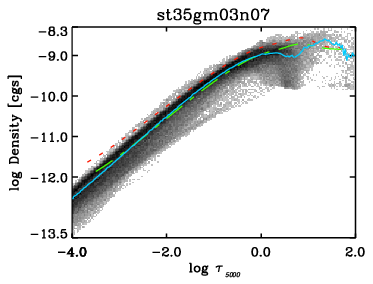}
    \includegraphics[width=0.5\hsize]{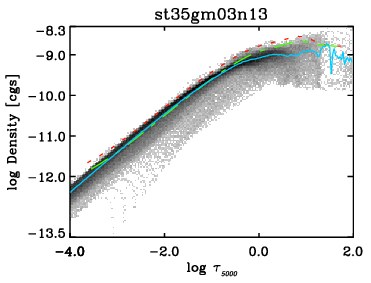} \\
    \includegraphics[width=0.5\hsize]{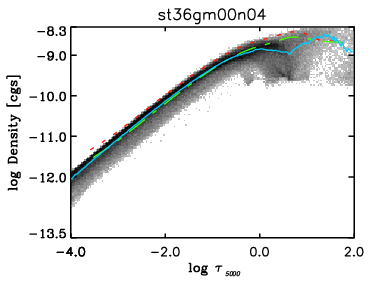}
    \includegraphics[width=0.5\hsize]{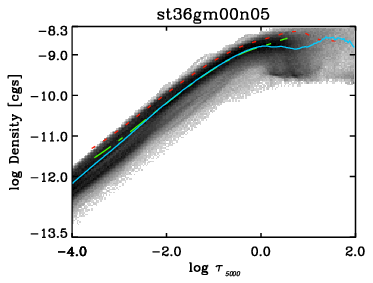} \\
        \end{tabular}
      \caption{Density structures of the simulations in Table~\ref{simus}. Darker areas indicate temperature values with higher probability. The colored curves have the same meaning as in Fig.~\ref{temperature_profile}).}
        \label{temperaturedensity_profile}
   \end{figure*}

 Eventually, \cite{2008A&A...486..951G} chose to set $v_{\rm{t}}$ = 0 for all 1D models in their grid, advicing those who would have liked a different choice to
use models with a different mass or gravity, according to the recipe given in Eq.~\ref{eqgrav2}.\\
Having all the necessary thermodynamical quantities in 3D simulations averaged over spherical shells, we computed $g_{\rm{eff}}$ using Eq.~\ref{eqgrav2} with $\mu_{\rm{mol}}=1.3$ (appropriate value for the atmosphere of RSGs). Figure~\ref{geff_figure} shows the behavior of $g/g_{\rm{eff}}$ for 3D simulations of Table~\ref{simus}. We used the effective surface gravity at the the radius position (red vertical lines in the figure) to compute new 1D models (see Table~\ref{simus1D}). Figure~\ref{temperaturedensity_profile} shows the comparison of the 3D density structures with 1D models. The models with the new effective gravity shows better agreements with 3D mean profiles than the models with $v_{\rm{t}}$ set to zero (i.e., $g_{\rm{eff}}=g$). The effects on the temperature structures of 1D models with different surface gravity and equal effective temperature is negligible as displayed in all the panels of Fig.~\ref{temperature_profile}.\\
We conclude that 1D simulation of red supergiant stars must have turbulent velocity not equal to zero, that accordingly to Eq.~\ref{eqgrav2}, gives the correct value of surface gravity. The effect of the turbulent pressure is a lowering of the gravity that can be calibrated on 3D simulations accordingly to Fig.~\ref{geff_figure}.

\section{Non-gray versus gray simulations}

We used the 3D pure-LTE radiative transfer code {{\sc Optim3D}} described in Paper~{\sc{I}} to compute spectra and intensity maps from the gray simulation st35gm03n07 and the non-gray simulation st35gm03n07 (Table~\ref{simus}). The code takes into account the
Doppler shifts due to convective motions. The radiative
transfer equation is solved monochromatically using extinction
 coefficients pre-tabulated as a function of temperature, density, and
wavelength. The lookup tables were computed for the solar chemical compositions \citep{2006CoAst.147...76A}
using the same extensive atomic and molecular opacity data as the latest generation of
MARCS models \citep{2008A&A...486..951G}. We assumed a zero
micro-turbulence since the velocity fields inherent in 3D models
are expected to self-consistently and adequately account for
non-thermal Doppler broadening of spectral lines.

\subsection{From one to three dimensional models: determination of micro- and macroturbulence}

RHD simulations provide a self-consistent ab-initio
description of the non-thermal velocity field generated by convection,
shock waves and overshoot that are only approximated in 1D
models by empirical calibrations. Thus, the comparison between 1D and 3D
spectra requires the determination of ad-hoc parameters like microturbulence
($\xi_{\rm{mic}}$) and macroturbulence ($\xi_{\rm{mac}}$) velocities that must be
determined for the 1D spectra. \\
For this purpose, we used a similar method as
\cite{2009MmSAI..80..731S}.  Standard 1D radiative transfer codes
usually apply $\xi_{\rm{mic}}$ and $\xi_{\rm{mac}}$ broadening on spectral lines
isotropically, independently of depth, and identically for both atomic
and molecular lines. Hence, to have a good representation of the
conditions throughout the 3D atmosphere, we selected a set of 12 real
Fe~I lines in the 5000-7000 $\rm \AA$ range (see Table~
\ref{tab:micromacrolines}) such that: (i) their equivalent width on the
3D spectrum range between few m$\rm \AA$ to 150m$\rm\AA$; and (ii) with
two excitation potentials, low ($\sim$ 1.0 eV) and high ($\sim$ 5.0 eV). Using
the 1D MARCS models with the same stellar parameters (plus the effective surface gravity, $g_{\rm{eff}}$) of the
corresponding 3D simulations (see Table~\ref{simus1D}), we first
derived the abundance from each line using Turbospectrum
(\citealp{1993ApJ...418..812P}, 
\citealp{1998A&A...330.1109A}, and further improvements by Plez) from the 3D equivalent width. For the non-gray model we used the 1D MARCS model corresponding to the lower temperature (i.e., 3430K) because its  thermal structure is more similar to the 3D mean temperature profile in the outer layers where spectral lines form (see top right panel of Fig.~\ref{temperature_profile}). Note, that we computed the 3D spectra
using {{\sc Optim3D}} with negligible expected differences (less than 5\%; Paper~{\sc{I}}) with respect to Turbospectrum. The
microturbulence velocity was then derived by requiring no trend of the
abundances against the equivalent widths. The error on the
microturbulence velocity is estimated from the uncertainties on the null
trend. Once the microturbulence has
been fixed, the macroturbulence velocity was determined by minimizing
the difference between the 3D and the 1D line profiles, and this for
three profile types: radial-tangential, gaussian and exponential. 
The error on the macroturbulence velocity is calculated from the
dispersion of the macroturbulence from line to line. The reduced $\chi^2$ is also determined on the best-fitting 1D to 3D line
profiles. Hence, the automatic nature of the determination ensures an objective determination.

\begin{table}
\centering
\caption{Parameters of the Fe~I lines used to extract 1D microturbulence
and macroturbulence and equivalent width, $W$, computed for the gray
simulation st35gm03n07 and non-gray st35gm03n13 of Table~\ref{simus}}
\label{tab:micromacrolines}      

\renewcommand{\footnoterule}{} 
\begin{tabular}{c c c c c}        
\hline\hline                 
         Wavelength  & $\chi$   & $\lg\left(gf\right)$ & $W$ gray & $W$ non-gray \\
            $\rm \AA$    & eV      &                                 &  m$\rm \AA$  &  m$\rm \AA$ \\
          \hline
5853.148  &   1.485 & $-$5.280 & 112.7 & 104.0\\
5912.690 & 1.557 & $-$6.605 & 46.1 & 39.8\\
6024.058 & 4.548 & $-$0.120 & 128.4 & 129.8\\
6103.186 & 4.835 & $-$0.770 & 78.6 & 81.1\\
6214.182 & 1.608 & $-$7.250 & 13.9 & 14.6 \\
6364.366 & 4.795 & $-$1.430 & 52.7 & 57.8 \\
6569.216 & 4.733 & $-$0.420 & 106.7 & 107.1 \\
6581.210 & 1.485 & $-$4.679 & 147.3 & 140.7 \\
6604.583 & 4835 & $-$2.473 & 11.4 & 14.5 \\
6712.438 & 4.988 & $-$2.160 &13.6 & 18.4 \\
6717.298 & 5.067 & $-$1.956 & 13.8 & 17.4 \\
6844.667 & 1.557 & $-$6.097 & 78.9 & 71.8 \\
\hline\hline                          
\end{tabular}
\end{table}

\begin{table}
\centering
\caption{1D microturbulence ($\xi_{\rm{mic}}$) and macroturbulence ($\xi_{\rm{mac}}$) velocities determined
to
match the 3D spectra characteristics. }             
\label{tab:micromacro}      
\renewcommand{\footnoterule}{} 
\begin{tabular}{l |c c|c c}        
\hline\hline                 
&\multicolumn{2}{c|}{gray model}        &
\multicolumn{2}{c}{non-gray model}        \\
 &\multicolumn{2}{c|}{st35gm03n07}   &
 \multicolumn{2}{c}{st35gm03n13} \\
\hline
\multicolumn{5}{c}{$\xi_{\rm{mic}}$ (km/s)}    \\
\hline
& \multicolumn{2}{c}{ 1.45    $\pm$ 0.09}
& \multicolumn{2}{c}{1.28 $\pm$ 0.33}  \\
\hline
\multicolumn{5}{c}{Log$\epsilon\left(Fe\right)_{3D}-$Log$\epsilon\left(Fe\right)_{1D}$}    \\
\hline
& \multicolumn{2}{c}{ 0.14   }
& \multicolumn{2}{c}{-0.02 }  \\
\hline\hline
\multicolumn{5}{c}{$\xi_{\rm{mac}}$}   \\
\hline
profile & velocity(km/s) &  $\chi^2$ & velocity (km/s)  &  $\chi^2$  \\
radtan                 & 6.7  $\pm$ 0.9 & 1.15 &   6.4 $\pm$  2.7 &   1.17\\ 
gauss                  & 10.1  $\pm$ 1.2  & 1.19 &   9.7 $\pm$  3.5 &   1.14\\
exp                    & 5.5  $\pm$ 0.9 & 1.21 &   5.2 $\pm$  2.2 &   1.22\\
\hline\hline                          
\end{tabular}
\tablefoot{The 3D$-$1D corrections to Fe abundances, derived for the selected iron lines with the best matching $\xi_{\rm{mic}}$, are also indicated.}
\end{table}

\begin{figure*}
\begin{center} 
\includegraphics[angle=-90, width=0.6\hsize]{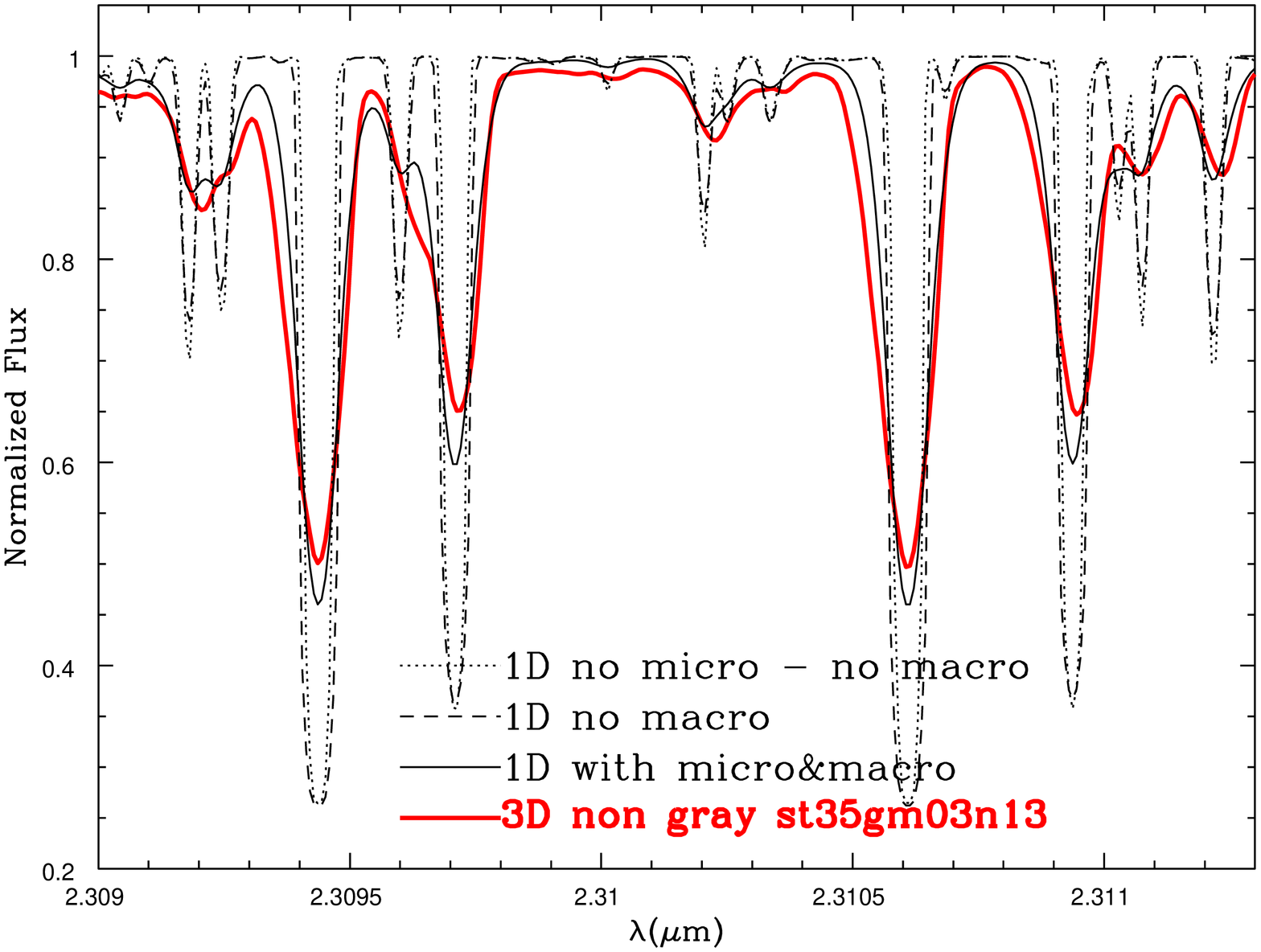}
\caption{Example of CO first overtone lines computed for the 3D non-gray simulation and the corresponding 1D MARCS model of Table~\ref{simus1D}. The $\xi_{\rm{mic}}$ turbulent velocity used is 1.5 km/s and the $\xi_{\rm{mac}}$ is 6.4 km/s with a radial-tangential profile (see Table~\ref{tab:micromacrolines}).
}
\label{micromacro_fig}
\end{center}
\end{figure*}

Table~\ref{tab:micromacro} presents the derived values of the
microturbulence and macroturbulence respectively for the 3D gray and non
gray models. The dispersion of
microturbulent velocities is reasonably small, so that the
depth-independent microturbulence is a fairly good approximation. Note
also that the microturbulence velocity for the gray and non-gray model
are rather close. Although the macroturbulence profile varies from one to another, the reduced $\chi^2$ show that all of them give about the same value.
Note that the micro and macroturbulence standard deviations from the average velocities is systematically larger in the non-gray model. We found that the non-gray spectral lines generally show a more complex profile that affect the line fitting with 1D models causing larger standard deviations. An example for the non-gray model is shown in Fig.~\ref{micromacro_fig}: the spectral lines look shifted and broadened. The line asymmetries are caused by the inhomogeneous velocity field emerging from optical depths where lines form. The lower temperature in the outer layers of non-gray model (Fig.~\ref{avg_profiles}) could cause smaller pressure and density scales heights and consequentially a faster density drops and stronger shocks. These shocks would complex the shape of the non-gray spectral lines. The macroturbulence parameter used for the convolution of 1D spectra,
reproduce only partly the complex profile with larger differences in line wings. 

On a wavelength scale of the size of a few spectral lines, the differences between 3D and 1D models are caused by the velocity field in 3D simulations that affect the spectral
lines in terms of broadening and shift. 1D model do not
take into account velocity fields and ad-hoc parameters for micro- and macroturbulence are used to reproduce this effect. The microturbulence and macroturbulence calibration results are pretty similar for gray and non-gray simulations even if the 3D-1D corrections of the resulting iron abundances are smaller for the non-gray model. The microturbulence values obtained in this work are comparable, albeit a little bit lower, with \cite{2000ApJ...530..307C} who found $1.7\lesssim\xi_{\rm{mic}}\lesssim3.5$ km/s with a different method based on observed CO lines.  Our values are also in general agreement to what \cite{2009MmSAI..80..731S} found for their RHD Sun and Procyon simulations ($0.8\lesssim\xi_{\rm{mic}}\lesssim1.8$).

\subsection{Spectral energy distribution}\label{tio}

\begin{figure*}
\begin{center} 
\includegraphics[angle=-90, width=0.45\hsize]{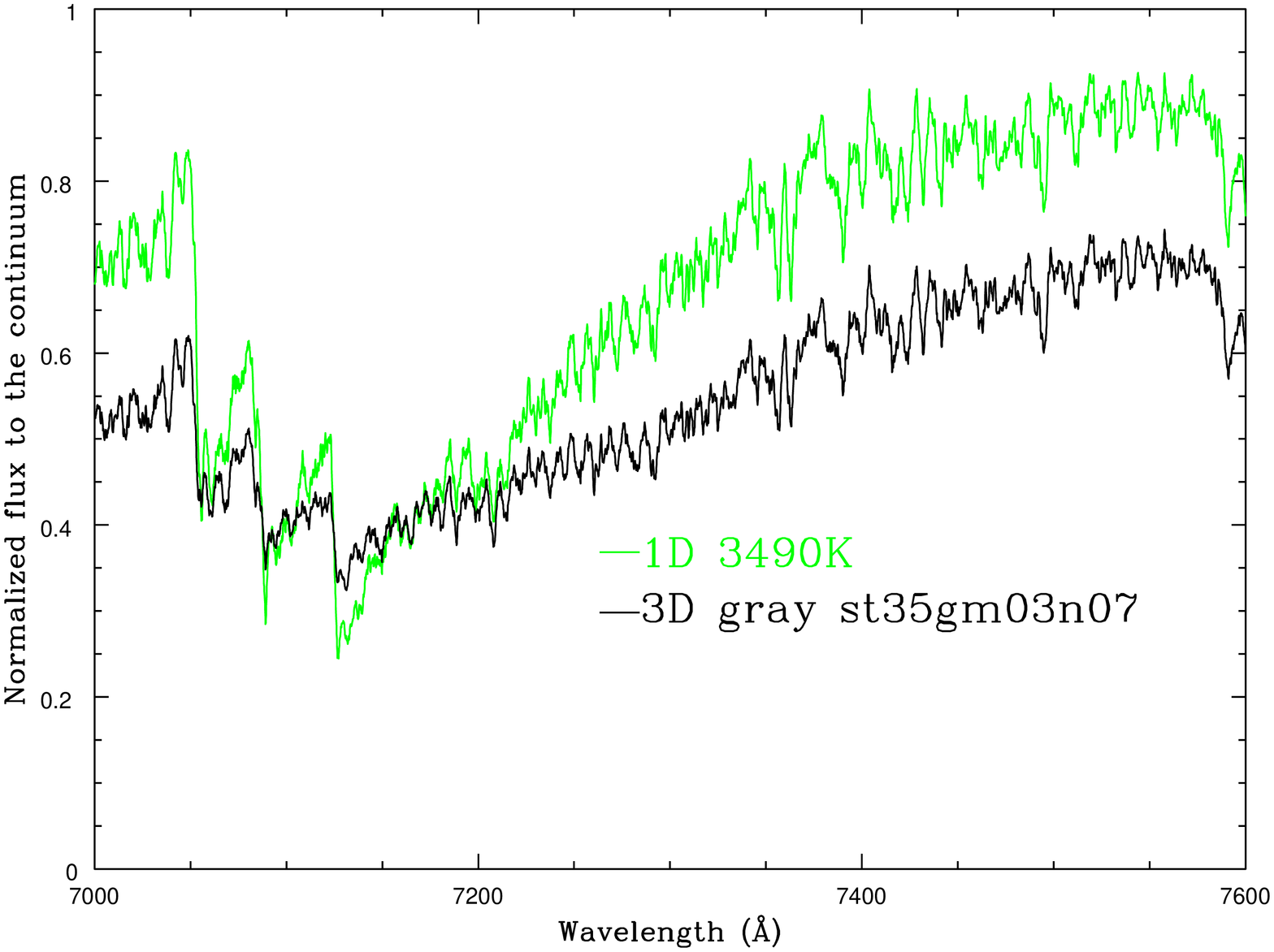}
\includegraphics[angle=-90, width=0.45\hsize]{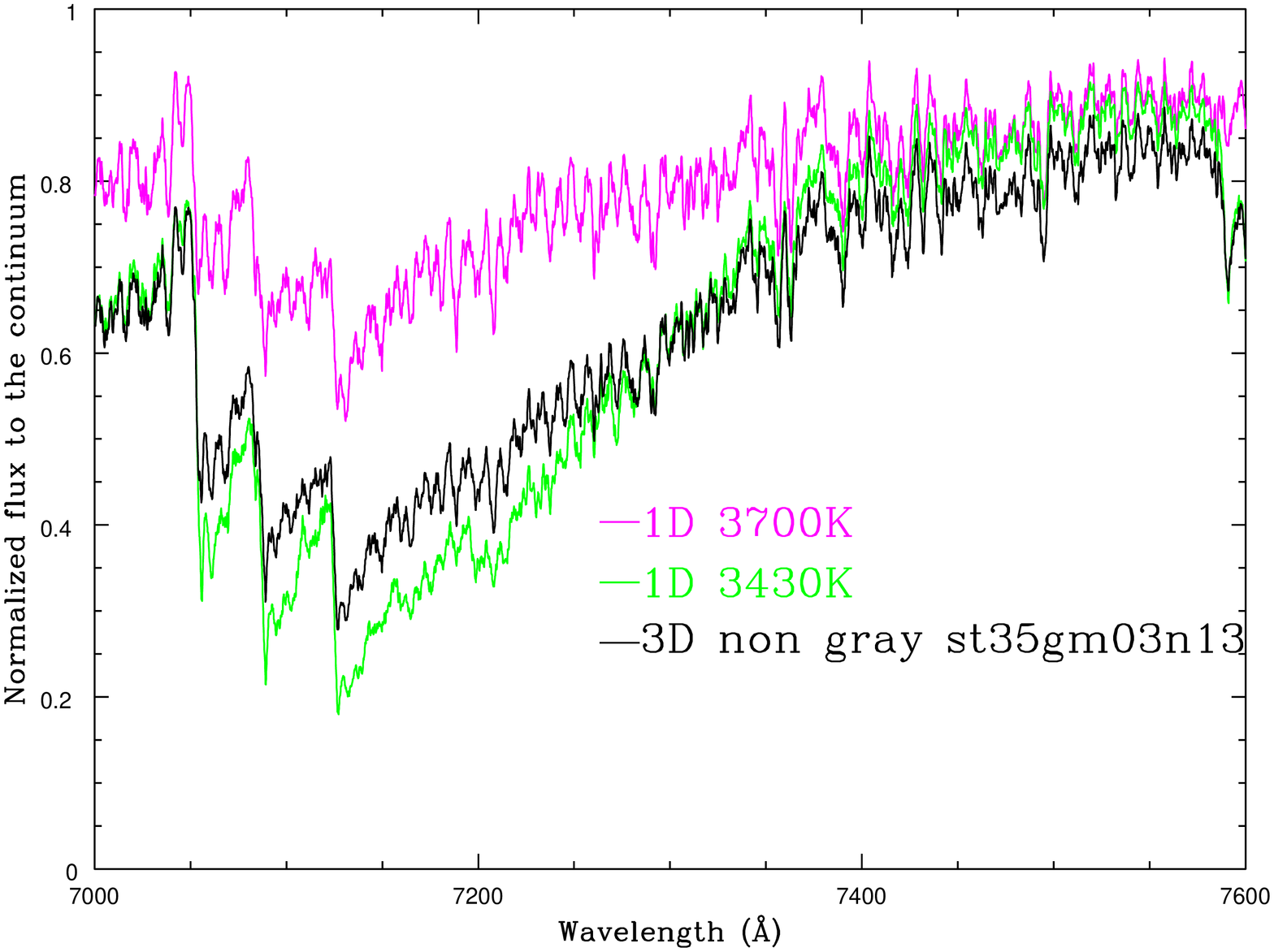}\\
 \includegraphics[angle=-90, width=0.45\hsize]{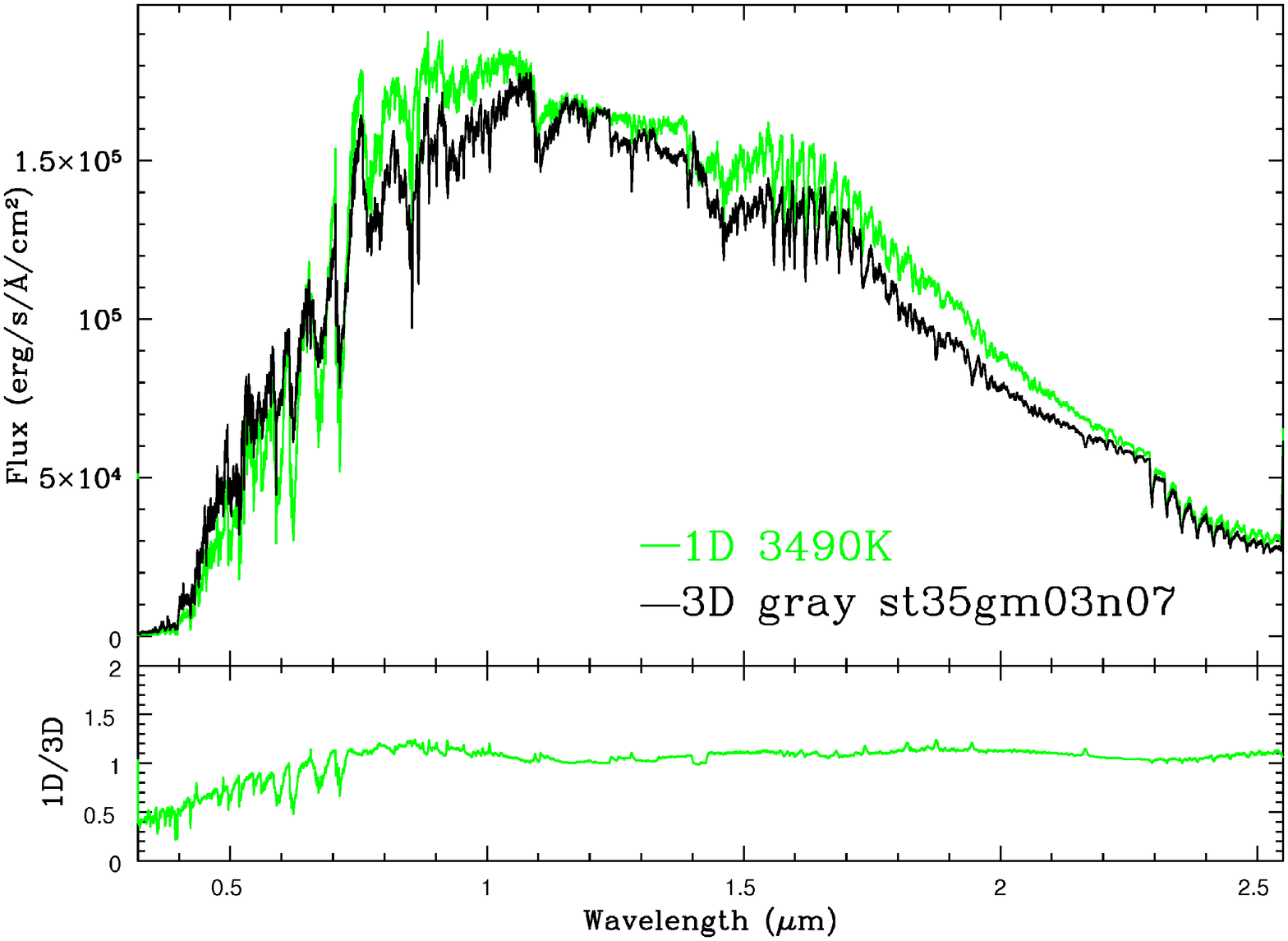}
\includegraphics[angle=-90, width=0.45\hsize]{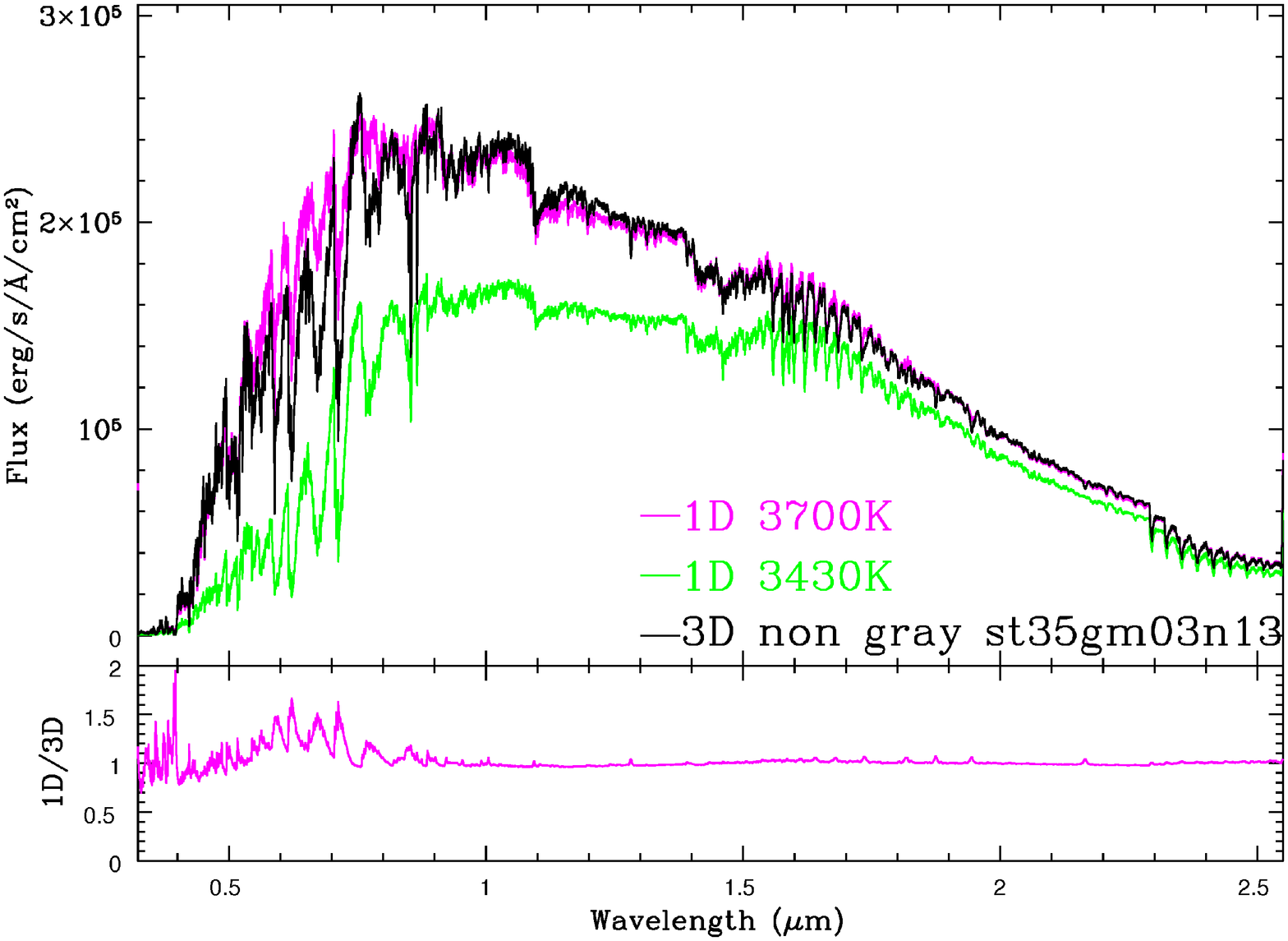}
\caption{\emph{Top panel:} spectral synthesis of TiO band
transition $A^3\Phi-X^3\Delta\left(\gamma\right)$ (top row) and spectral energy distribution (bottom row) for the same snapshots of the gray, st35gm03n07, and non-gray, st35gm03n13, simulations of Fig.~\ref{temperature_profile}. 3D spectra are compared to the corresponding 1D MARCS models (Table~\ref{simus1D}) with $\xi_{\rm{mic}}$ and $\xi_{\rm{mac}}$ from Table~\ref{tab:micromacro} (radial-tangential profile).}
\label{tio_band}
\end{center}
\end{figure*}

The cooler outer layers encountered in the convection simulations are
expected to have a significant impact on temperature sensitive features.
This is the
case of, in particular, molecular lines (e.g., TiO) and strong
low-excitation
lines of neutral metals, the line formation regions of which
are shifted outwards because of the lower photospheric temperatures
\citep{2008A&A...486..951G,1992A&A...256..551P}. The
importance of opacity from molecular electronic transition lines of TiO
are crucial in the spectrum of M-type stars. Figure~\ref{tio_band} (top row)
shows the synthesis of a TiO band. The strength of the transition
depends on the mean thermal gradient in the outer layers ($\tau_{5000}<1$), where TiO has a
large effect, which is larger in the non-gray model with respect to the gray one Fig.~\ref{avg_profiles}. A shallow mean thermal gradient weakens the contrast between strong and weak lines and this is visible in the molecular
band strength which is much stronger for the non-gray model. \\
Top right panel of Fig.~\ref{tio_band} shows also that the TiO band strength of non-gray model is more similar to the cool 1D MARCS model at 3430K than the hot one at 3700K. This reflects the fact that the 3D mean thermal structure in the outer layers is very similar to 1D-3430K model (Fig.~\ref{temperature_profile}, top right panel).\\
The approximation of gray radiative transfer is justified
only in the stellar interior and it is inaccurate in the optically
thin layers. So far, this approximation has been due to the fact that
the RHD simulations have been constrained by execution time.
However, with the advent of more powerful computers, the
frequency-dependent treatment of radiative transfer is now possible and, on a wavelength scale of one or more molecular bands, the use of this method is big step forward for a quantitative analysis of observations.\\

The shape of the spectral energy distribution (SED) reflects the mean
thermal gradient of the simulations. The absolute flux plots in the bottom row of Fig.~\ref{tio_band} display that at lower resolution two important conclusions can be retrieved: (i) the spectrum based on the gray model is largely different from the 1D spectra; (ii) the non-gray model shows that the SED, compared to the 1D MARCS model with the hot temperature (3700K), displays
almost no distinction in infrared region, and weaker in molecular bands in
the visible region and in the near-ultraviolet region. Using the prescriptions by \cite{1998A&A...333..231B} for the filters $BVRJHK$, we computed expected colors
for the gray and non-gray models as well as the corresponding MARCS models (Table~\ref{tablecolors}). The non-gray 3D model shows smaller differences to the 1D model than the gray one. The stronger
radiative efficiency in the non-gray case forces the temperature stratification to be closer to the radiative equilibrium than in the gray case, where convection and waves have a larger impact on it.\\
\cite{2000A&A...357..225J} probed the usefulness of the index $V-K$ as a temperature indicator for Galactic RSGs, but \cite{2006ApJ...645.1102L}, fitting TiO band depths, showed that $V-K$ and $V-R$ provides systematic higher effective temperatures in Galactic and Magellan clouds RSGs (in average +170K and +30K, respectively) than the spectral fitting with 1D MARCS models. \cite{2006ApJ...645.1102L} concluded that the systematic offset was probably due to the limitations of static 1D models. Using the radius definition of Eq.~\ref{eq_radius_co5bold}, we find that the effective temperature from the SED of 3D non-gray spectrum is $T_{\rm eff}$=3700 K. However, in the optical the spectrum looks more like a 3500K 1D model (judging from the TiO bandheads of Fig.~\ref{tio_band}, top right panel). Assuming that the 3D non-gray model spectrum is close to real stars spectra and using $V-K$, this leads to a $T_{\rm eff}$  that is higher by about 200K than the $T_{\rm eff}$ from TiO bands. This goes in the right direction given what \cite{2006ApJ...645.1102L} found.

\begin{table*}
\centering
\caption{Photometric colors for RHD simulations of Table~\ref{simus} and for the corresponding 1D models of Table~\ref{simus1D}. $\Delta\left(\rm{3D}-\rm{1D}\right)$ shows the 3D/1D difference for each index.}
\label{tablecolors}      
\renewcommand{\footnoterule}{} 
\begin{tabular}{c c c c c c c}        
\hline\hline                 
Model    &    $B-V$   &  $V-R$  &  $V-K$ & $J-H$   &   $J-K$ &  $H-K$  \\ 
\hline
3D gray, st35gm03n07   &  1.680 &  0.955   &  4.751  &   0.859  & 1.153  & 0.293        \\  
1D,  $T_{\rm{eff}}=3490$K, $\log g=-0.58$ & 1.881 &   1.125   &  5.070  &   0.931 &  1.200 &  0.269 \\
$\Delta\left(\rm{3D}-\rm{1D}\right)$  &    $-$0.201       & $-$0.170 & $-$0.319  &  $-$0.072  & $-$0.047 & 0.024  \\
\hline
3D non-gray, st35gm03n13              &   1.661 &  0.886  &   4.356 &    0.807 &  1.064 &  0.258 \\  
1D,  $T_{\rm{eff}}=3700$K, $\log g=-0.35$ &   1.850 &  0.962   & 4.202  &  0.854 &  1.080 &  0.226  \\  
$\Delta\left(\rm{3D}-\rm{1D}\right)$   & $-$0.189 & $-$0.076 &  0.154 & $-$0.047 &  $-$0.016 & 0.032\\
\hline	  

\hline\hline                          
\end{tabular}
\end{table*}

\subsection{Interferometry: visibility fluctuations}\label{interfero1}

\begin{figure*}
   \centering
   \begin{tabular}{cc}
    \includegraphics[width=0.36\hsize]{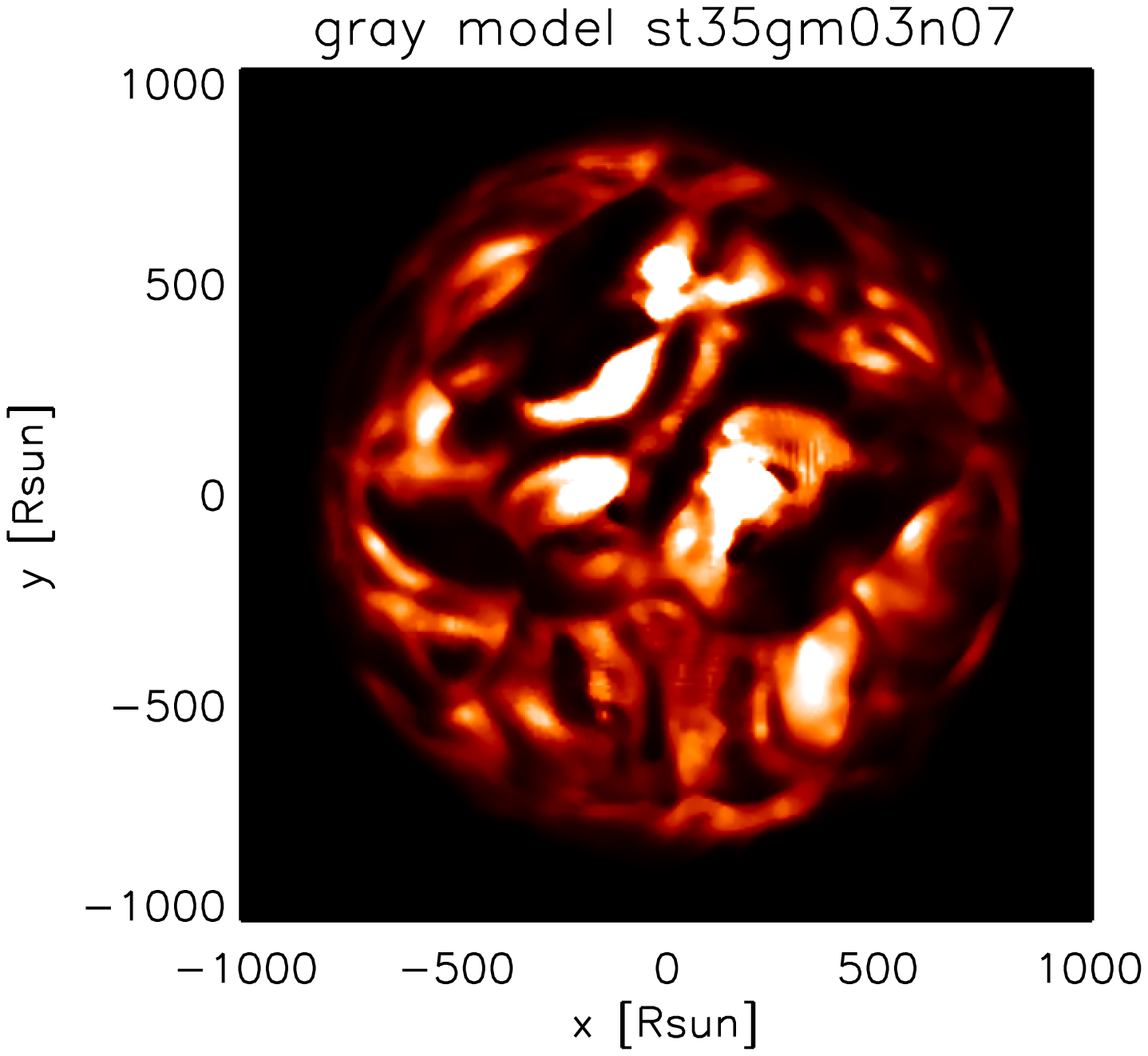}
        \includegraphics[width=0.36\hsize]{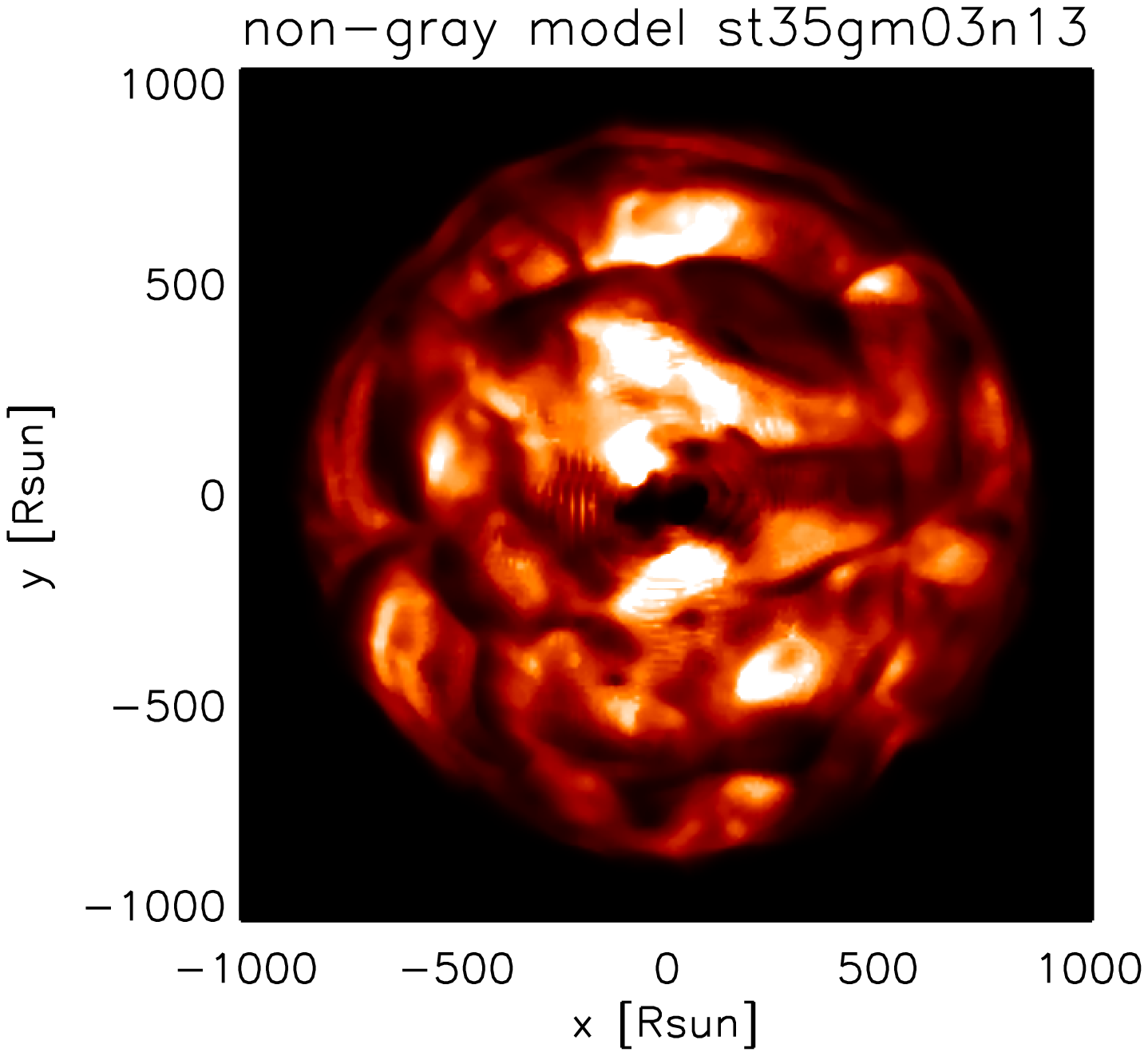}\\
    \includegraphics[width=0.325\hsize]{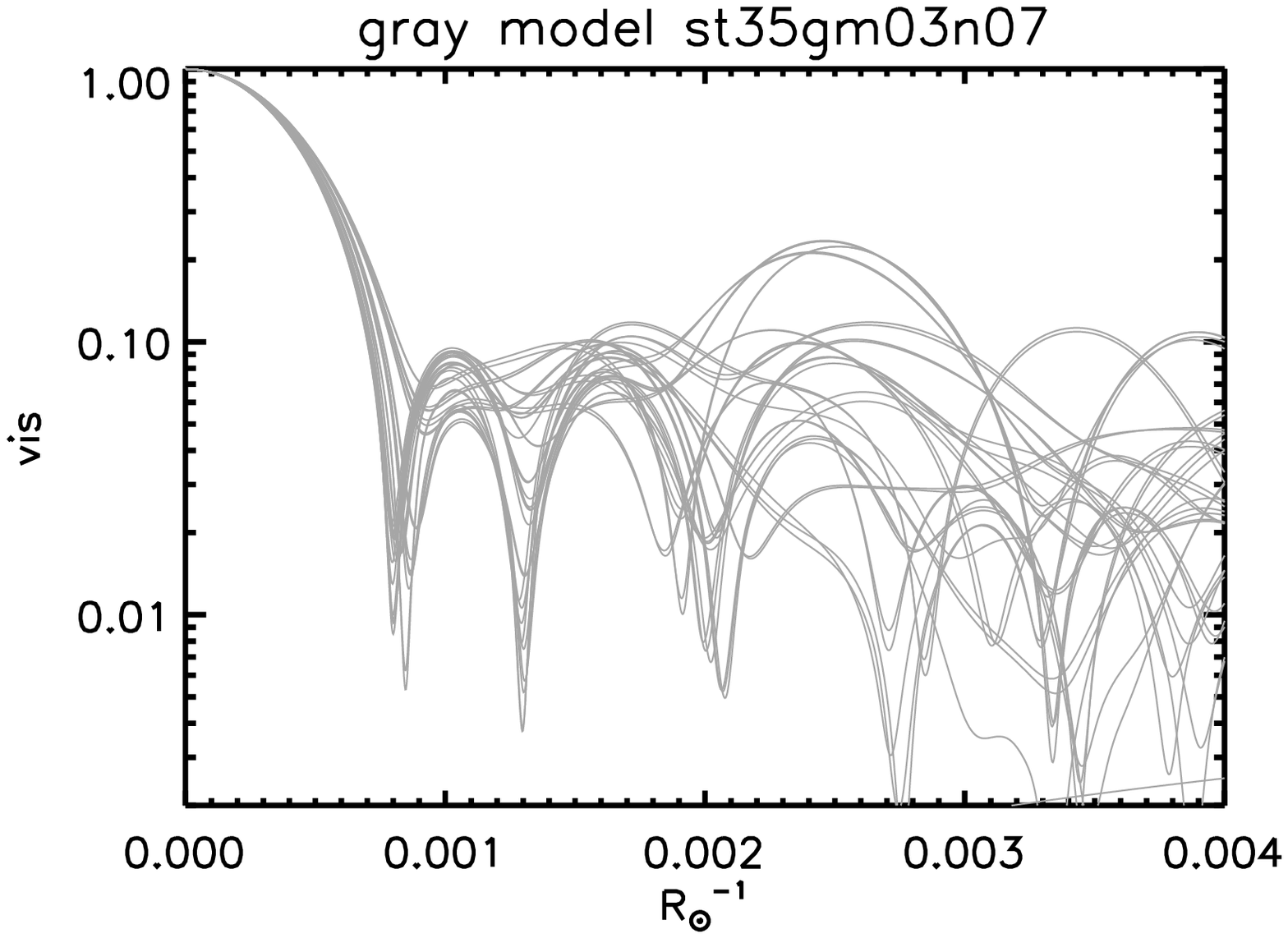} 
    \includegraphics[width=0.325\hsize]{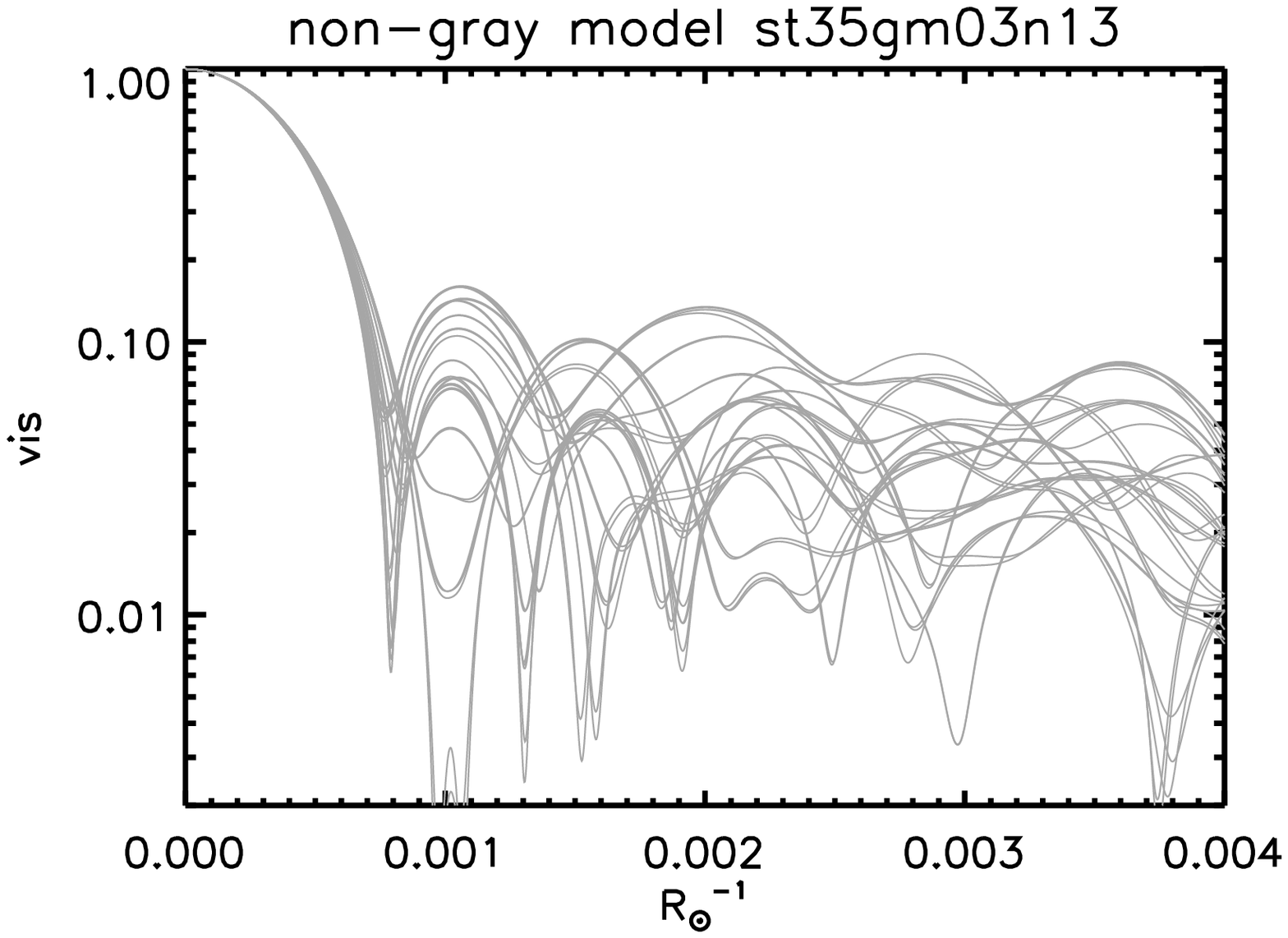} 
    \includegraphics[width=0.325\hsize]{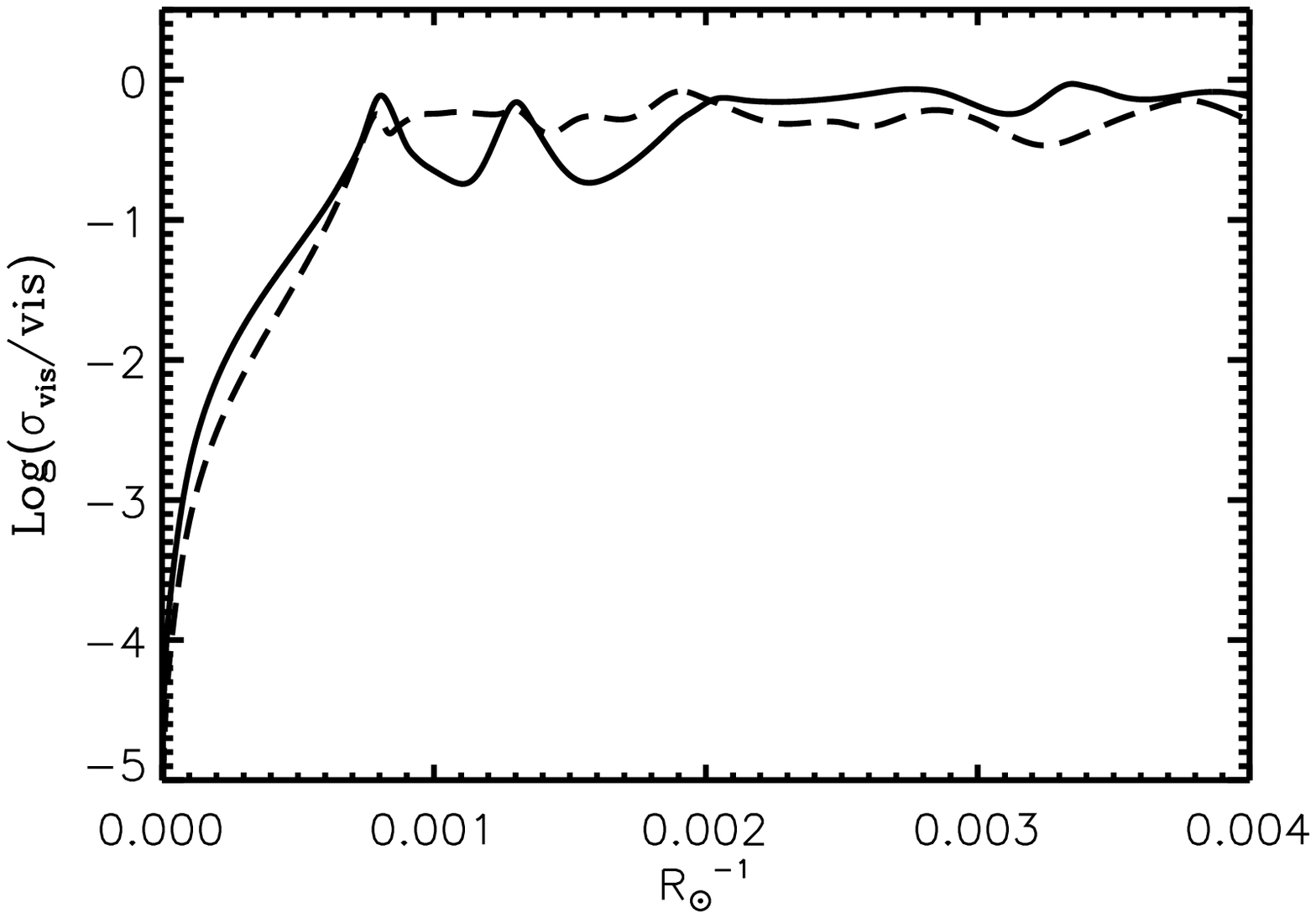} 
        \end{tabular}
      \caption{\emph{Top row:} map of the linear intensity in the
  TiO band transition $A^3\Phi-X^3\Delta\left(\gamma\right)$ of Fig.~\ref{tio_band}. The range is
  [0; $3.5\times10^5$] for the gray simulation st35gm03n07 of Table~\ref{simus} and [0; $4.5\times10^5$]\,erg\,cm$^{-2}$\,s$^{-1}$\,{\AA}$^{-1}$ for the non-gray simulation st35gm03n13. \emph{Bottom row:} visibility curves obtained from the maps above. The visibilities are computed for 36 different azimuth angles 5$^\circ$ apart. \emph{Bottom right panel:} visibility fluctuations with respect to the average value for the gray (solid line) and non-gray (dashed line).}
        \label{gray_interfero}
   \end{figure*}

\cite{2010A&A...515A..12C,2011A&A...528A.120C} showed that in the visible spectral region, the gray approximation in our modelling may significantly affect the intensity contrast of the granulation. Figure~\ref{gray_interfero} (top row) shows the resulting intensity maps computed in a TiO electronic transition. The resulting surface pattern is connected, throughout the granulation, to dynamical effects. The emerging intensity is related to layers where waves and shocks dominate together with the variation in opacity 
through the atmosphere and veiling by TiO molecules. 





The intensity fluctuations are linked with the temperature inhomogeneities, which are weaker in the thermal structure of the non-gray simulation (see Sect.~\ref{temp_sect} and Fig.~\ref{temperature_profile}). 
We analyzed the impact on the interferometric observables using the method described in Paper~{\sc{I}}. We computed visibility
curves for 36 different rotation angles with a step of
5$^\circ$ from the intensity maps in Fig.~\ref{gray_interfero}. In the plots, we introduced a theoretical spatial frequency scale
expressed in units of inverse solar radii ($R_\odot^{-1}$). The conversion between visibilities expressed in the latter scale and in the more usual scale is given by:  $[\arcsec]^{-1}=[R_\odot^{-1}]\cdot d~[{\rm pc}]\cdot214.9$. The resulting visibility curves are plotted in bottom row of the Fig.~\ref{gray_interfero} together with the one $\sigma$ visibility
fluctuations, $F$, with respect to the average value $\overline{{vis}}$ ($F=\sigma/\overline{{vis}}$). The visibility fluctuations of the non-gray model are lower than the gray model ones for spatial frequencies lower than the first null point (approximately 0.0075 $R_\odot^{-1}$), as a consequence the uncertainty on the radius determination \citep[the first null point of the visibility curves is sensitive to the stellar radius, ][]{1974MNRAS.167..475H} is smaller: $\sim$10$\%$ in gray model versus $\sim$4$\%$ in the non-gray one. At higher frequencies, the visibility fluctuations are larger in the non-gray model between 0.0075 and 0.002 $R_\odot^{-1}$, then the non-gray model fluctuations are systematically lower between $\sim$0.002 and 0.004 $R_\odot^{-1}$. However, it must be noted that after 0.03 $R_\odot^{-1}$ (corresponding to 33 $R_\odot$, i.e., $\sim$4 pixels), the numerical resolution limit is reached and visibility curves can be affected by numerical artifacts (Paper~{\sc{I}} and Sect.~\ref{high_res_interfero}).

\section{Study on the increase of the numerical resolution of simulations}\label{high_res_interfero}

In this work, we presented two simulations with approximatively the same stellar parameters but with an increase of the numerical resolution. We analyze the effect of increasing the resolution from $255^3$ grid points of simulation st36gm00n04 to $401^3$ grid points of st36gm00n05 (Table~\ref{simus}). Due to the large computer time needed for the high-resolution model, the better way to proceed is to compute a gray model with the desired stellar parameters first, then the numerical resolution can be increased, and eventually the frequency-dependent non-gray treatment can be applied. With the actual computer power and the use of the OpenMP method of parallelization in CO$^5$BOLD, this process may take several months.

\begin{figure}
   \centering
   \begin{tabular}{cc}
    \includegraphics[width=0.68\hsize]{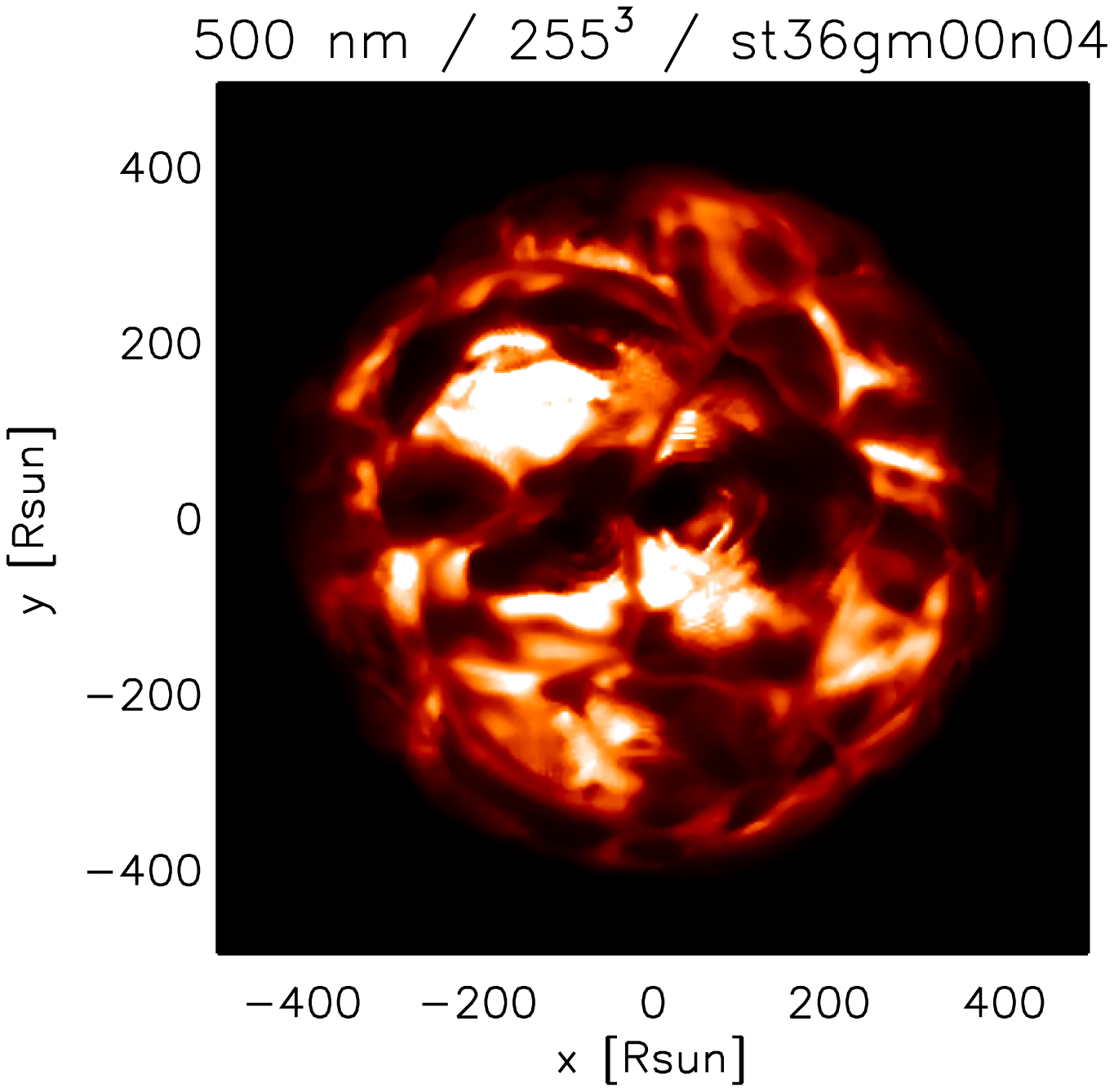}\\
        \includegraphics[width=0.78\hsize]{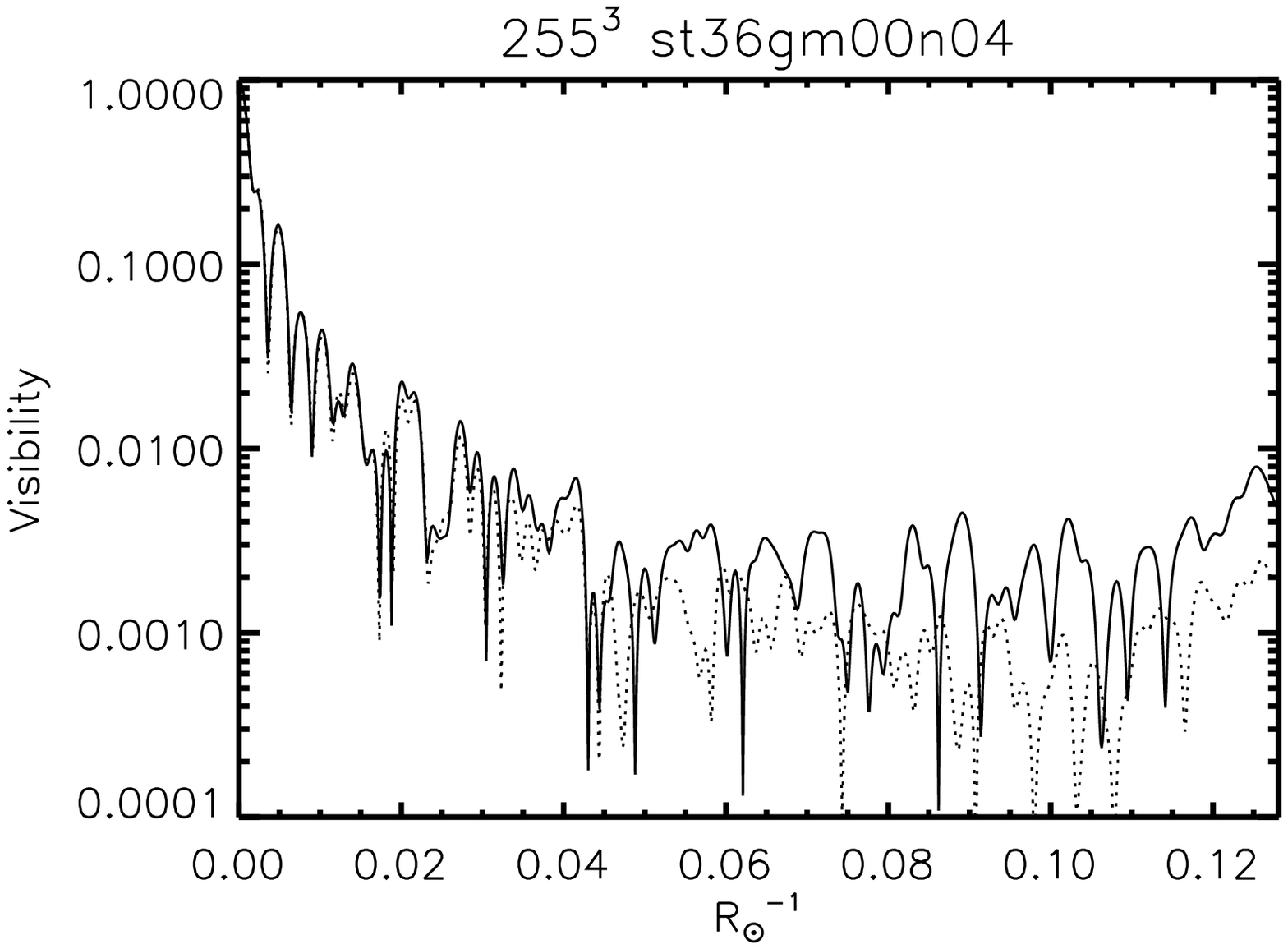}\\
    \includegraphics[width=0.68\hsize]{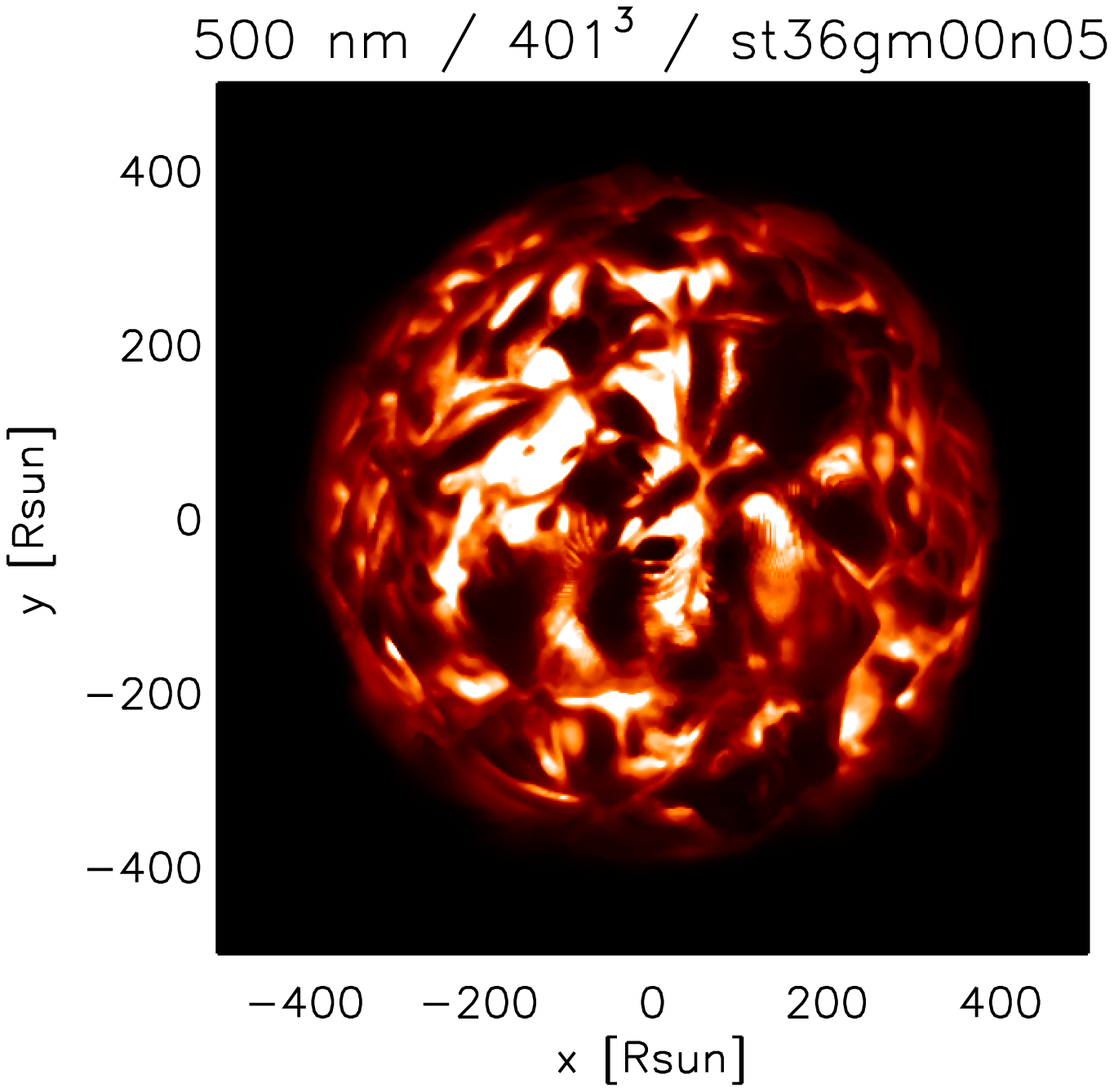} \\
    \includegraphics[width=0.78\hsize]{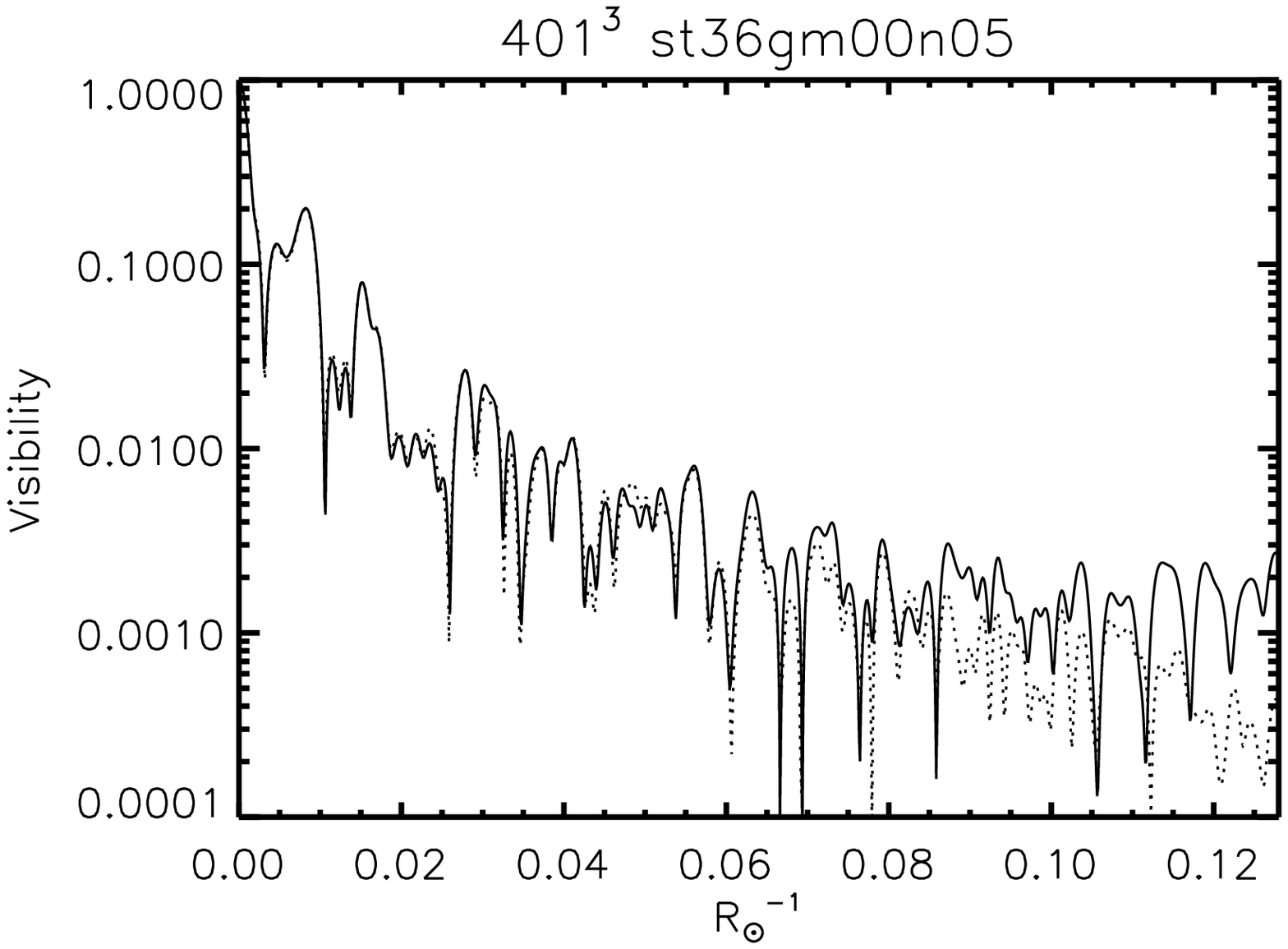}
        \end{tabular}
      \caption{Maps of the square root intensity (to better show the structure in the higher resolution simulation) of st36gm00n04 and st36gm00n05 (Table~\ref{simus}) at 5000$\rm \AA$ together with a visibility curve extracted for a particular angle (solid curve) and the the same visibility's angle after applying a $\left[3\times3\right]$ median smoothing (dashed curve). The intensity range in the maps is [0; $547.7$]\,erg\,cm$^{-2}$\,s$^{-1}$\,{\AA}$^{-1}$.
              }
        \label{high_res_interfero_fig}
   \end{figure}

We computed intensity maps at 5000$\rm \AA$. This spectral region, as for the TiO band, is interested by a granulation pattern with high intensity contrast. Figure~\ref{high_res_interfero_fig} displays the intensity maps permeated by a variety of convection-related structures; while the large structures are visible in both images, the higher resolution model resolves much better the details and  smaller scale structures appear within the large ones \citep[as already shown in a previous work,][]{2006EAS....18..177C}. Artifacts due to Moire-like patterns are show up in the maps of the higher numerical resolution simulation. They are composed of thin ripples up to 40$\%$ brighter than the surrounding points and they are already been pointed out in Paper~{\sc{I}}. They are caused by large opacities, large variations in opacity, and large changes in optical depth between 2 successive columns in the numerical box (see $\kappa$ in Fig.~\ref{1dquantities}). To overcome this problem, we computed visibility curves from these images and compared the visibility curves for one particular projected baseline from the raw image, and after applying a median $\left[3\times3\right]$ smoothing that effectively erases the artifacts. The visibility curves in Fig.~\ref{high_res_interfero_fig} show that the signal from the low-resolution st36g00n04 simulation is larger than st36g00n05 for spatial frequencies higher than about 0.07 $R_\odot^{-1}$. A lower signal at a particular frequency means  better spatial resolution: thus very small structures (corresponding to roughly $\leq$ 15-20 $R_\odot$) are better resolved in st36g00n05. \\
Moreover, the visibility curves are mostly affected by the numerical artifacts for frequencies higher than $\sim$0.05 $R_\odot^{-1}$ (corresponding to 25 $R_\odot$, i.e., $\sim$5 pixels) for st36g00n04 and 0.09 $R_\odot^{-1}$ (corresponding to 11 $R_\odot$, i.e., $\sim$4 pixels) for st36g00n05. The increase of numerical resolution reduces only partly the problem in the intensity maps and therefore the cosmetic median filter must be still applied. This will not affect the visibilities at lower frequencies, which are the
only ones observable in practice with modern interferometers. 

Apart for the problems on the Moire-like patterns of intensity maps, the number of convection related surface structures seems to visually increase and possible change size passing from $255^3$ to $401^3$ grid points (Fig.~\ref{high_res_interfero_fig}). In addition to this, the bottom right panel of Fig.~\ref{1dquantities} shows that going from $255^3$ to $401^3$ grid points the resolution of the temperature jump does not change so much. Thus, the numerical resolution limit has not been reached yet. The principal limitation to the computation of larger numerical simulation is the computer power. A more complete study concerning the impact of numerical resolution on the number and size of granules is necessary and will be addressed in a forthcoming paper where a series of simulations with the same stellar parameters and increasing numerical resolution will be analyzed.
   
\section{Conclusions}\label{conclusion}

We presented the radiation hydrodynamical simulations of red supergiants with the code CO$^5$BOLD together with the new generation of models based on a more sophisticated opacity treatment, and higher effective temperature and surface gravity.

The actual pattern and the mean temperature structure of RSGsÕ simulations are influenced: (i) by differences in the relative efficiency of convective
and radiative energy transport and by the efficiency of radiation
to smooth out temperature fluctuations; (ii) by the optical depth of the upper
boundary of the convection zone; (iii) by the extent of convective
overshoot.

The main difference between RSG and Sun-like granulation (except for geometrical scales) comes from the altered role of radiation: it is much more efficient in transporting energy in a RSG than in the Sun. 
This strongly influences the stratification and keeps it closer
to the case of radiative equilibrium than its inefficient counterpart
in the deeper layers of the Sun. Moreover, the strong entropy
gradient causes a large buoyancy and render the convective motions,
that compete with the radiative energy transport, more
violent in a relatively sharper region
over which the sub-photospheric entropy jump occurs (compared
to a typical size of a convective element). Eventually, higher velocities are accompanied by larger pressure
fluctuations and a stronger influence of shockwaves on the photosphere.

The most important improvement we brought with this work is the relaxation of the assumption of gray opacities through multigroup frequency-dependent opacities. The non-gray simulation shows: (i) a steeper mean thermal gradient in the outer layers that affect strongly the molecular and line strengths that are deeper in the non-gray case; (ii) that the general shape of the spectrum of 3D non-gray simulation are similar to 1D model while the 3D gray simulation is largely different. Hence, we concluded that the frequency-dependent treatment of radiative transfer with 5 bins represents a good approximation of the more complex profile of 1D based on $\sim$110\,000 frequency points. Moreover, the smaller temperature fluctuations of the non-gray model, due to the intensified heat exchange of a fluid element with its  environment, affect the surface intensity contrast and consequentially the interferometric observables reducing the uncertainty on the stellar radius determination. 

We also showed that 1D models of RSGs must include a turbulent velocity calibrated on 3D simulations to obtain the effective surface gravity that mimic the effect of turbulent pressure on the stellar atmosphere.

We provided an calibration of the ad-hoc micro- and macroturbulence parameters for 1D models using the 3D simulations, which have a self-consistent ab initio description of the non-thermal velocity field generated by convection, shock waves and overshoot. We found that the microturbulence velocity for the gray and non-gray model are rather close and the depth-independent microturbulence assumption in 1D models is a fairly good approximation, even if the 3D-1D corrections of the resulting iron abundances are smaller for the non-gray model. We also assessed that there is not a clear distinction between the different macroturbulent profiles needed in 1D models to fit 3D synthetic lines; nevertheless, we noticed that micro and macroturbulence standard deviations on the average velocities are systematically larger in the non-gray model, that shows more complex line profiles than the gray simulation. This may be caused by stronger shocks in the outer layers of non-gray model where the pressure and density scales heights are smaller.

While simulations with more sophisticated opacity treatment are essential for the forthcoming analysis, models with a better numerical resolution are desirable to study the impact of numerical resolution on the number and size of granules. We provide a first investigation showing that the number of convection related surface structures seems to increase and change size passing from $255^3$ to $401^3$ grid points, thus the numerical resolution limit has not been reached yet. Future work will focus on a series of simulations with the same stellar parameters and increasing numerical resolution.

\begin{acknowledgements}
A.C. thanks the Rechenzentrum Garching (RZG) and the CINES for providing the computational resources necessary for this
work. A.C. thanks also Pieter Neyskens and Sophie Van Eck for enlightening discussions. B.F. acknowledges financial support from
the {\sl Agence Nationale de la Recherche} (ANR),
the {\sl ``Programme National de Physique Stellaire''} (PNPS) of CNRS/INSU,
and the {\sl ``{\'E}cole Normale Sup{\'e}rieure''} (ENS) of Lyon, France,
and the {\sl Istituto Nazionale di Astrofisica / Osservatorio Astronomico di Capodimonte}
(INAF/OAC) in Naples, Italy.
\end{acknowledgements}

\bibliographystyle{aa}
\bibliography{biblio}

\end{document}